\documentclass[12pt,preprint]{aastex}

\usepackage{local_commands}
\usepackage{arydshln}

\usepackage{natbib}
\usepackage{aas_macros}
\slugcomment{\today}
\shorttitle{The structure of gas-accreting protoplanets and the condition of the critical core mass}
\shortauthors{KANAGAWA K. D. \& FUJIMOTO M.Y.}

\begin{document}
\title{The structure of gas-accreting protoplanets and the condition of the critical core mass}
\author{Kazuhiro D. KANAGAWA\altaffilmark{1} and Masayuki Y. FUJIMOTO\altaffilmark{2,3}}
\altaffiltext{1}{Department of Cosmosciences, Hokkaido University, Kita 10 Nishi 8, Kita-ku, Sapporo 060-0810, Japan}
\altaffiltext{2}{Nuclear reaction data center, Graduate School of Science, Hokkaido University, Kita 10 Nishi 8, Kita-ku, Sapporo 060-0810, Japan}
\altaffiltext{3}{visiting researcher, Faculty of Engineering Hokkai-Gakuen University, 4-1-40, Asahimachi, Toyohira, Sapporo, Japan}\email{kanagawa@astro1.sci.hokudai.ac.jp}

\begin{abstract}
In the core accretion model for the formation of gas giant planets, runaway gas accretion onto a core is the primary requisite, triggered when the core mass reaches a critical value.
The recently revealed wide diversity of the extrasolar giant planets suggests the necessity to further the understanding of the conditions resulting in the critical core mass that initiates runaway accretion.
We study the internal structure of protoplanets under hydrostatic and thermal equilibria represented in terms of a polytropic equation of state to investigate what factors determine and affect the critical core mass.
We find that the protoplanets, embedded in protoplanetary disks, have the same configuration as red giants, characterized by the envelope of the centrally-condensed type solution.
Applying the theory of stellar structure with homology invariants, we demonstrate that there are three types of criteria for the critical core mass depending on the stiffness of polytrope and the nature of outer boundary condition.
For the stiff polytropes of index $N \le 3$ with the Bondi radius as the outer boundary, the criterion governing the critical core mass occurs at the surface.
For stiff polytropes with the Hill outer boundary and for soft polytropes of $N>3$, this criterion acts at the bottom of gaseous envelope.
Further, we elucidate the roles and effects of coexistent radiative and convective zones in the envelope of critical core mass.
Based on the results, we discuss the relevance of Bondi and Hill surface conditions and explore the parameter dependences of critical core mass.
\end{abstract}

\section{Introduction} \label{sec:introduction}
The core accretion model \citep[e.g.,][]{Perri_Cameron1974,Mizuno1980,Bodenheimer_Pollack1986} is the most widely-accepted scenario for the formation of gas giant planets, such as Jupiter and Saturn.
In this scenario, the solid core grows by the collision and merger of planetesimals and attracts surrounding gas by its own gravity from the protoplanetary disk.
Once the core of a protoplanet grows in mass sufficiently to reach a critical core mass, gas accretion onto the protoplanet is accelerated rapidly and so-called runaway gas accretion occurs.
Eventually, the core acquires a massive enough gaseous envelope to be a new-born gas giant planet.
This scenario has been discussed extensively with regard to the formation of planets in the solar system with the minimum mass solar nebula (MMSN) model proposed by \cite{Hayashi1981}.
However, recent discovery of more than 3000 exoplanets, including Kepler candidates, has revealed the existence of a variety of planetary systems \citep[e.g.,][]{Udry_Santos2007,Borucki_etal2011,Batalha_etal2012}.
This implies a variety of disks distinct from the MMSN model, e.g., their dependence on the masses of circumstellar disks \citep{Beckwith_Sargent1996}.
Therefore, in order to understand formation of extrasolar planets, we have to know the conditions to form gas giant planets in various disks.

According to the core accretion model, runaway gas accretion is indispensable to form the gas giant planets.
It is thought to be initiated when the core mass reaches the critical core mass.
Quasi-static calculations have demonstrated that the runaway gas accretion occurs. \citep{Bodenheimer_Pollack1986,Pollack_etal1996,Ikoma_Nakazawa_Emori2000,Hubickyj_Bodenheimer_Lissauer2005,Hori_Ikoma2010}.
In these works, however, it is difficult to identify the onset of the runaway gas accretion exactly since the accretion rate changes continuously and increases with the growth of the core and with the cooling of the envelope even before the runaway gas accretion.
This led \cite{Bodenheimer_Pollack1986} to propose an expedient of defining the critical core mass as the core mass when the gaseous envelope has the same mass as the core, but this is too ambiguous to predict when the runaway gas accretion sets in.
An alternative definition of the critical core mass is discussed based on the structure of the gaseous envelope under hydrostatic and thermal equilibria where time-dependent effects, such as gravitational contraction, are ignored.
Previous works have demonstrated that there exists an upper limit to the core mass, above which no static envelope solution is possible \citep{Perri_Cameron1974,Mizuno1980,Stevenson1982,Wuchterl1993,Ikoma_Emori_Nakazawa2001,Rafikov2006,Rafikov2011}, which is regarded as implying the onset of a time-dependent process which may trigger rapid gas accretion.
They have studied what kinds of parameters are important for determining the critical core mass, such as the growth rates of core, opacities in the gaseous envelopes, and physical properties of the ambient gas (temperature and density of accreted gas).
Such attempts have been made to derive the critical core mass and the mass ratio between the gaseous envelope and solid core in some analytical ways \citep[e.g.][]{Stevenson1982,Wuchterl1993,Rafikov2006,Rafikov2011}.
They are concerned mostly with the thermal properties of planetary envelopes.
Previous works have little discussion on the mechanical structure of gaseous envelope other than the predominance of its self-gravity over the gravity of the core.
Further studies are, therefore, needed to reveal why the critical core mass exists for static envelopes and what determines its mass and to derive a more general criterion for the critical core mass so as to understand the formation of gas giant planets under various conditions.

In consideration of these problems, we rely upon the similarity in interior structures between the gas giant planets and the red giant stars, as self-gravitating systems.
Both have basically the same internal configurations, composed of core and envelope.
The red giant branch (RGB) stars are known to take a peculiar form, characterized by the envelope structure described by the so-called centrally-condensed type solution \citep{Schwarzchild1958,Hayashi_Hoshi_Sugimoto1962}, distinct from the dwarf stars.
The comparative research between planets and stars may be properly pursued only by using the dimensionless, homology invariants because of large differences in physical dimensions.
So far, these variables have been referred to cursorily mainly in relation to the isothermal cases \citep{Pecnik_Wuchterl2005,Schonke2007}.
We exploit the results of the theory of stellar structure to study the similarity and disparity in the structures between the protoplanets and the RGB stars, which may give a better understanding of the reason why static systems have the critical core masses and why the protoplanetary envelope undergoes an instability leading to runaway gas accretion.

In this paper, we discuss the structure of protoplanetary envelope in thermal equilibrium, by using a polytropic equation of state, to elucidate the conditions under which systems form critical core masses and also the relation between the runaway gas accretion and the critical core mass.
The next section is devoted to the description of the basic equations, the assumptions, and the boundary conditions, and to the mathematical preparations for our models to analyze the features in the structure of protoplanets.
In \S \ref{sec:single_polytropic_model}, based on the interior structures of protoplanets in terms of homology invariants, we discuss the relation between the total mass and core mass of protoplanets for the model with a single polytropic envelope.
Application to realistic structure of envelope is investigated in \S \ref{sec:realistic_model} with the model of composite polytropes.
The comparison with the existent works and implications for formation of gas giants planets are discussed in \S \ref{sec:Discussion}.
Conclusions follow in \S \ref{sec:Conclusion}.

\section{Model and Assumptions} \label{sec:Model_and_Assumptions}
\subsection{Basic equations and assumptions}
We discuss the hydrostatic structure of systems consisting of two components, solid core and gaseous, where the solid component is highly concentrated in the center and immersed in a gaseous component.
\cite{Fujimoto_Tomisaka1992} show that the RGB stars are described as gravitating systems composed on two components, each of which obeys a different equation of state and for which the hydrostatic equations hold separately.
In this paper, we adopt their formulation with the two components taken to be the gaseous and solid matter, which form the envelope and core, respectively, in the protoplanet configuration.
Hence, the pressure and density are expressed as the combinations of those of solid and gaseous components as $P=\psol+\pgas$ and $\rho=\rhos+\rhog$.
Here and following, subscripts, $\solid$ and $\gas$, denote the solid and gaseous components.
   The mass, $M_r$, interior to the sphere of radius, $r$, is also given by the sum of mass of the two components as $M_r = \qrsol + \qrgas$, where $\qrsol$ and $\qrgas$ are defined by
\begin{eqnarray}
	\qrsol&=&\int^{r}_0 4\pi r^2 \rhos dr,\label{eq:cont_sol}\\
	\qrgas&=&\int^{r}_0 4\pi r^2 \rhog dr.\label{eq:cont_gas}
\end{eqnarray}
   Then, the equation of mechanical balance is divided into two parts as
\begin{eqnarray}
	\dpar{\psol}{r}&=&-\rhos\frac{GM_r}{r^2},\label{eq:balance_sol}\\
	\dpar{\pgas}{r}&=&-\rhog\frac{GM_r}{r^2},\label{eq:balance_gas}
\end{eqnarray}
   for the sold and gaseous components, respectively, where $G$ is the gravitational constant.
In the present problem, we may assume the density of solid component to be constant for simplicity.
Given the total mass, $\qsol$, of the solid component, the core radius, $\rcore$, is defined as
\begin{eqnarray}
	&&\rcore=\bfracb{3\qsol}{4\pi \rhos}{1/3}, \label{eq:core_radius}
\end{eqnarray}
and the gravity of the solid component can be replaced by a potential.

Since a planet is embedded and grows in a protoplanetary disk, the outer boundary condition during their formation should be different from those of stars.
We adopt the same outer boundary condition as previous works \citep[e.g.,][]{Bodenheimer_Pollack1986};
at the outer edge of planet, density and pressure are assumed to connect continuously with the disk density, $\rhodisk$, and pressure, $\pdisk$, i.e.,
\begin{eqnarray}
	&&\rho=\rhodisk,P=\pdisk\; \hbox{ at } r=\rtot,\label{eq:outer_boundary_conditions}
\end{eqnarray}
where the radius, $\rtot$, of the planet is taken to be the smaller of the two, Bondi radius and Hill radius, denoted by $\rbondi$ and $\rhill$ in the following and defined by
\begin{eqnarray}
	\rbondi&=&{G \qtot}/{\sonic^2},\label{eq:bondi_radius}\\
	\rhill&=&\sma\left[ {\qtot}/{3(\qhost+\qtot)} \right]^{1/3},\label{eq:hill_radius}
\end{eqnarray}
respectively,
where $\qtot$ is the total (solid plus gas) mass of planet:
$\sonic$ the sonic speed in the protoplanetary disk $(= \sqrt{\gammad \pdisk/\rhodisk}$,
$\gammad$ being the ratio of specific heats):
$\sma$ the semi-major axis of planet orbit:
and $\qhost$ the mass of a parent star.

The inner boundary conditions are taken to be the same as those for stars, i.e., $\rho$ and $P$ are finite at $r=0$.
This is different from previous works, which set the inner boundary at the surface of solid core \citep[e.g.,][]{Mizuno1980,Bodenheimer_Pollack1986,Hubickyj_Bodenheimer_Lissauer2005}.
In this work, the gaseous component is assumed to permeate through the core, without influencing the solid component.
We define the core mass, $\qcore$, and core density, $\rhocore$, as the sum of the mass and the sum of the mean density of solid and gaseous components inside the sphere of $r = \rcore$, i.e., $\qcore \equiv \qsol + \qrgas (\rcore)$ and $\rhocore \equiv \rhos + \qrgas (\rcore) /(4 \pi \rcore^3/3)$, respectively.
Since $\rhos \gg \rhog$ in the present study, differences between $\qcore$ and $\qsol$ and between $\rhocore$ and $\rhos$ remain very small for most cases, in particular during the early phase of planet formation.

We may summarize some quantities relevant to the core accretion model \citep[e.g.,][]{Rafikov2006}.
First we introduce the characteristic mass, $\qchar$, related to the physical conditions in the protoplanetary disk by
\begin{eqnarray}
	&&\qchar = \left[ \left( {1}/{4\pi G^3}  \right)  \left( {\pdisk^3} / {\rhodisk^4} \right)\right]^{1/2}.\label{eq:qchar}
\end{eqnarray}
This mass corresponds to the mass scale of the Emden solution with $\rhodisk$ and $\pdisk$ as the central density and pressure, respectively.
We may also define mean density, $\dmean$, of matter in the parent star, averaged over the sphere of radius equal to the separation as
\begin{eqnarray}
	&&\dmean={\qhost}/{(4\pi \sma^3/3)}.\label{eq:mean_density_hill}
\end{eqnarray}
   The above two outer boundary radii are distinct in the nature;
    e.g., Bondi radius gives the mean density of protoplanet, including the core, decreasing with the total mass as $4 \pi \rhodisk (\qtot / \qchar)^{-2}$, while Hill radius designates a constant mean density of $4 \pi \dmean$ for $\qtot \ll \qhost$.
With use of these quantities, the ratios of the Bondi and Hill radii to the core radius are written as
\begin{eqnarray}
	\frac{\rbondi}{\rcore}&=&\frac{1}{\gammad}\bfracb{\rhocore}{3\rhodisk}{1/3}\bfracb{\qcore}{\qchar}{2/3}\bfraca{\qtot}{\qcore},
\label{eq:ratio_rb_rc}\\
	\frac{\rhill}{\rcore}&=&\bfracb{\rhocore}{3\dmean}{1/3}\bfracb{\qtot}{\qcore}{1/3},
\label{eq:ratio_rh_rc}
\end{eqnarray}
respectively.
Here, we assume that $\qtot \ll \qhost$.
The Bondi radius grows larger than the core radius when
\begin{eqnarray}
	\qcore &>& \qcoremin \nonumber \\
	&\equiv& \gammad^{3/2}\left( 3\rhodisk/\rhocore \right)^{1/2} \qchar,\label{eq:qcoremin}
\end{eqnarray}
and the solid core is thought to attract surrounding gas.  
On the other hand, the Hill radius is larger than the core radius regardless of the core mass, and yet, is surpassed by $\rbondi$ when the planet mass grows larger than
\begin{eqnarray}
	\qtot = \gammad^{3/2} \left( \rhodisk/\dmean \right)^{1/2} \qchar = \qhost \left( H/\sma \right)^3/\sqrt{3},\label{eq:changeover_m}
\end{eqnarray}
where $H$ is vertical thickness of protoplanetary disk [$= \sma \sonic / (G \qhost / \sma)^{1/2}$].
As the protoplanetary disk is thicker, therefore, the changeover from the Bondi radius to Hill radius will be postponed until the planet grows more massive.
Both the Bondi and Hill radii may, however, exceed thickness of protoplanetary disk when planets grow more massive than $\qtot \simeq ( 1 \hbox{ and }3 ) \times (3 \gammad^3 \rhodisk / \dmean)^{1/2} \qchar $, respectively \citep{Rafikov2006}.
For the planet of mass beyond this bound, its structure may deviate from spherical symmetry, in particular near its surface.
We will not consider such deviations from spherically symmetry in this paper since we are here interested in basic properties of hydrostatic and thermal structure of protoplanets.

\subsection{Description with Homology Invariants} \label{sec:uvplane}

In studying structures of gravitating systems in spherical symmetry, it is convenient to introduce homology invariants, $U$ and $V$, defined by
\begin{eqnarray}
	U&\equiv& \dpar{\log M_r}{\log r}=\frac{\rho}{M_r/(4\pi r^3)},\label{eq:u}\\
	V&\equiv& -\dpar{\log P}{\log r}=\frac{GM_r/r}{P/\rho},\label{eq:v}
\end{eqnarray}
\citep[e.g.,][]{Chandrasekhar1939,Schwarzchild1958,Hayashi_Hoshi_Sugimoto1962}.
With the aid of local polytropic index, defined by
\begin{eqnarray}
	\frac{N}{N+1}&=&\frac{d\log \rho/d\log r}{d\log P/d\log r},\label{eq:polytropic_eos}
\end{eqnarray}
   the equations of hydrostatic equilibrium reduce to the first-order differential equation;
\begin{eqnarray}
	\dpar{\log U}{\log V}&=&-\frac{U+VN/(N+1)-3}{U+V/(N+1)-1}.\label{eq:hydroeq_uv}
\end{eqnarray}
Accordingly, structures are described as curves on the plane of these variables.

Merits of using these variables lie in the fact that properties of gravitating systems in spherical symmetry can be fully expounded on the characteristic plain of $\log U$ - $\log V$ diagram, instead of the three-dimensional space of the mass, radius and pressure.
Indeed, distributions of the latter quantities through the interior are given by the integration along the structure line on the characteristic plane as
\begin{eqnarray}
	d\log r&=&\frac{d\log M_r}{U}=\frac{d\log P}{-V} \nonumber\\
	&=& \frac{d\log(V/U)}{2U+V-4} \label{eq:struct_uv}
\end{eqnarray}
without referring to other quantities \citep{Sugimoto_Nomoto1980}.
   Here, the quantity, $U/V$, denotes the ratio between the mass, $|dM_r/d\log P|$, contained within a pressure scale height to the mass, $M_r$, inside the sphere of radius, $r$.
This may afford a better perspective on the analysis and understanding of the properties of structures.
The denominator in the last member of eq.~(\ref{eq:struct_uv}) defines the critical line of
\begin{eqnarray}
	2U+V-4=0 & & \hbox{   (critical line)}.\label{eq:critical_line}
\end{eqnarray}
This line divides the characteristic plane into two domains;
the upper domain of $2U+V-4>0$, where $U/V$ increases from outer to inner shells along the structure line, and the lower domain of $2U+V-4 <0$, where it increases from inner to outer shells.
   The structure line has to traverse this line diagonally with $d \log (U/V) = 0$.
In addition, two more characteristic lines are defined from the differentials of $U$ and $V$, respectively, though dependent on the polytropic index;
\begin{eqnarray}
	&U+NV/(N+1)-3=0 &\hbox{ (vertical line)},\label{eq:vertical_line}\\
	&U+V/(N+1)-1=0 &\hbox{ (horizontal line)}.  \label{eq:horizontal_line}
\end{eqnarray}
For $N \ge 3$, these three lines intersect with each other at a point
\begin{eqnarray}
	U&=&(N-3)/(N-1),\nonumber\\
	V&=&2(N+1)/(N-1),\label{eq:singular_point}
\end{eqnarray}
which corresponds to singular solutions of eq.~(\ref{eq:hydroeq_uv}) such that $\rho \propto r^{-2N/(N-1)}$ and $M_r \propto r^{(N-3)/(N-1)}$.
For $N<3$, these three lines have no intersections in the first quadrant of $U, \ V \ge 0$.

\begin{figure}
	\resizebox{\textwidth}{!}{\plotone{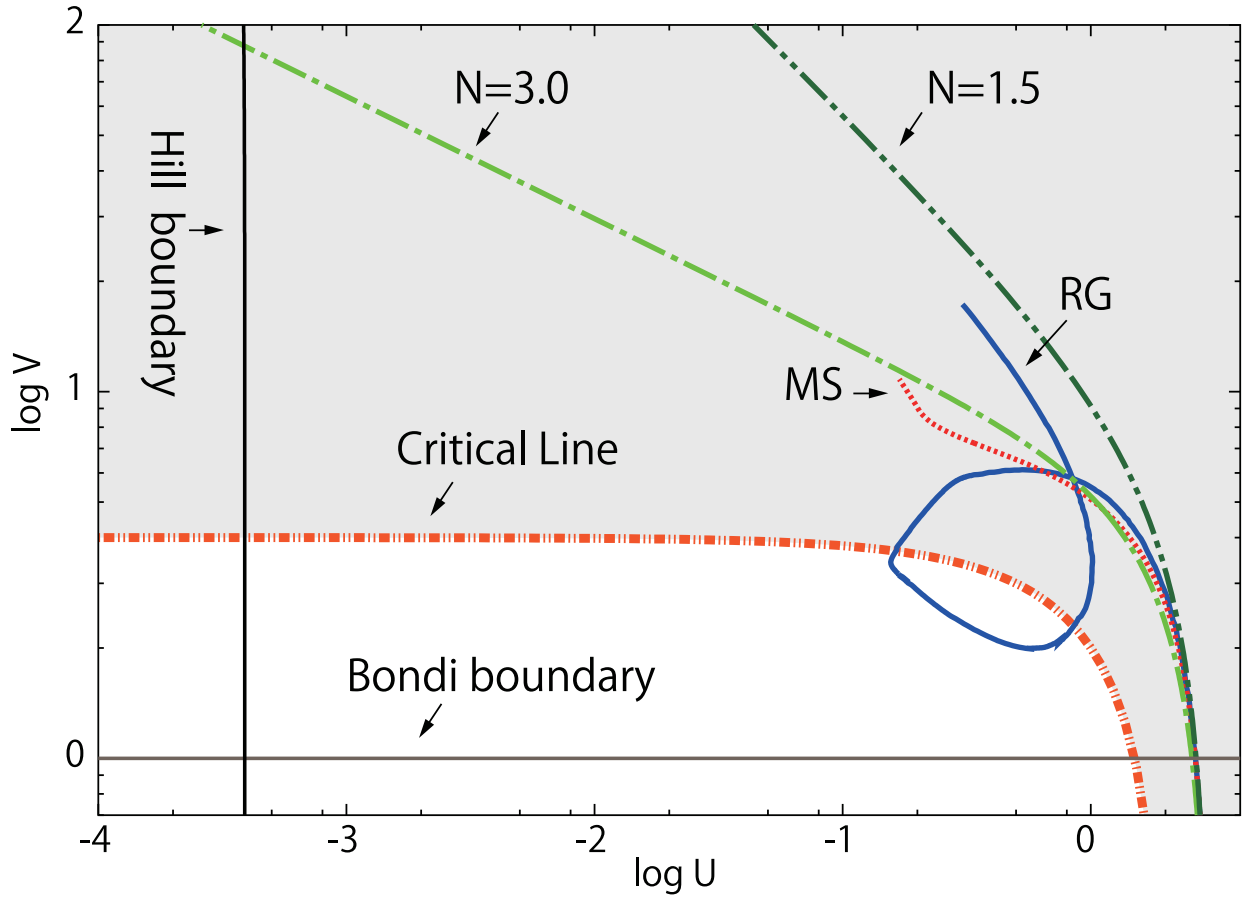}}
	\caption{
	Examples of stellar structure lines on $\log U $-$ \log V$ diagram; a main sequence star (long dash line), a RGB star (solid line), and the Emden solutions of polytropes with indices $N= 1.5$ and 3 (dotted lines).
	Also plotted is the critical line of $2U + V - 4 = 0$ (dash-two-dotted line) and loci of outer boundary conditions with the Bondi radius and Hill radius (horizontal and vertical straight lines, respectively) for parameters, listed in Table~\ref{tab:input_parameters}.
	}
	\label{fig:behavior_curves}
\end{figure}

Figure~\ref{fig:behavior_curves} shows some typical examples of structure lines of stars on the $\log U $-$ \log V$ diagram.
For stars, the boundary conditions are given in the upper domain of the critical line, i.e., $U = 3$ and $V = 0$ at the center and $U \ll 1$ and $V \gg 1$ at the surface.
Dwarf stars draw structure lines in the upper domain with the value of $V/U$ monotonically increasing all the way from the center to the surface.
Their structures well approximate to Emden solutions.
On the other hand, RGB stars are characterized by the structure lines forming a loop (or loops) on the way;
starting from the center in the upper domain, the structure lines enter once into the lower domain at the edge of core, decrease $V/U$ in the bottom of the envelope, and then, return again into the upper domain to reach the surface.
Accordingly, these two types of structures are distinguished by configurations on this characteristic plane, one connecting the center and surface directly and the other forming a loop (loops) on the way \citep{Sugimoto_Nomoto1980,Fujimoto_Tomisaka1992,Sugimoto_Fujimoto2000}.
The loop is attended with an increase in thermal energy relative to gravitational energy and brings about a large variation particularly in the radius.
For the configuration of the RGB stars, the envelope structure is known to be described in terms of the centrally-condensed type solutions, which converge to $U \rightarrow 0$ and $V=N+1$ for $N\leq 3$ or spiral in the singular point of eq.~(\ref{eq:singular_point}) for $N> 3$, depending on the polytropic index.
The radial expansion occurs near the bottom of envelope where the structure line comes close to the critical line and runs along it during the loop, as seen from eq.~(\ref{eq:struct_uv}).
Among two intersections with the critical line, the mass increase is larger near the outer crossing of larger $U$, while the pressure change is larger near the inner crossing of large $V$, although there is no contribution at the very crossing point since $d \log V/U = 0$.

With the homology invariants, the outer boundary conditions in eqs.~(\ref{eq:outer_boundary_conditions}) are written as
\begin{eqnarray}
	\usurf&=&\gammad^{-3}[\qtot/\qchar]^2, \nonumber \\
	\vsurf &=& \gammad,
\label{eq:surf_uv_bondi} \\
\noalign{\noindent for the Bondi radius, and}
	\usurf &=& \rhodisk/\dmean,\nonumber\\
	\vsurf &=&[\dmean/\rhodisk]^{1/3} [\qtot/\qchar]^{2/3}.
\label{eq:surf_uv_hill}
\end{eqnarray}
for the Hill radius.
The distinctive feature of these outer boundary conditions is that they pass below the critical and characteristic lines in eqs.~ (\ref{eq:critical_line}) , (\ref{eq:vertical_line}) and (\ref{eq:horizontal_line}) different from those of stars.
These boundary conditions are both expressed by straight lines in this characteristic plane but in different directions, i.e., they run along the lines of a fixed value of $V$ and $U$, respectively, for the Bondi and Hill radii, and intersect each other in the domain below the critical and characteristic lines for the planet mass in eq.~(\ref{eq:changeover_m}).

In the core-accretion model, there is a discontinuity of density between the top of core and the bottom of gaseous envelope.
Since other quantities are continuous across it, the structure line of envelope is connected to that of the core on the $U $-$ V$ plane through the following jump condition;
\begin{eqnarray}
	&&\uoe/\uoi=\voe/\voi=\rhooe/\rhooi,\label{eq:jump_condition}
\end{eqnarray}
where subscript, 1, denotes the interface between the core and the envelope, and subscripts, $i$ and $e$, denote the interior and exterior to the interface, respectively.
Thus, the values of $\uoe$ and $\voe$ at the bottom of envelope are given by
\begin{eqnarray}
	\uoe&=&3\rhooe/\rhocore,\label{eq:jump_uoe} \\
	\voe&=&(3\rhodisk / \rhocore)^{(3-N)/3N}(\qcore / \qchar)^{2/3} (3\rhooe / \rhocore) ^{-1/N}
\label{eq:jump_voe}
\end{eqnarray}
If the properties of the core are specified, therefore, there is the one-to-one correspondence between the location on the characteristic plane and the physical conditions of gas in the bottom of envelope.

In addition to the jump condition, the inner edge of the envelope should satisfy the condition of radius ratio between the core and surface in eqs.~(\ref{eq:ratio_rb_rc}) or (\ref{eq:ratio_rh_rc}) for the Bondi and Hill radius, respectively.
The latter condition is written in the form of an integral along the structure line on the characteristic plane for the radius ratio as
\begin{eqnarray}
	\oint_{\be}^{\surf}\frac{d\log(V/U)}{2U+V-4} &=& \log\left( \frac{\rtot}{\rcore} \right).
\label{eq:radius_condition}
\end{eqnarray}
or equivalently for the mass ratio as
\begin{eqnarray}
	\oint_{\be}^{\surf} U \frac{d\log(V/U)}{2U+V-4} &=& \log\left( \frac{\qtot}{\qcore} \right)
\label{eq:mass_condition}
\end{eqnarray}
For a given total mass, we have $\usurf$ and $\vsurf$ from the surface boundary condition of eq.~(\ref{eq:surf_uv_bondi}) or eq.~(\ref{eq:surf_uv_hill}), corresponding either to the Bondi or Hill radius.
Then, the values of $\uoe$ and $\voe$ at the core-envelope interface ensue from this equation with the jump conditions, and so does the core mass.

\section{Single Polytrope Models} \label{sec:single_polytropic_model}

If the barotropic relation, or the entropy distribution, is specified, the structure is determined by hydrostatic equilibrium.
It is known that the polytropic equation of state, which is written in the form
\begin{eqnarray}
	&&P = K \rho{}^{1+1/N}
\label{eq:polytrope_relation}
\end{eqnarray}
with the polytropic index $N$ and the polytropic constant $K$ as parameters, gives a good approximation to thermal structure of stars with properly chosen parameters.
The polytropic index may take different values and even vary locally in the interior according to the physical conditions of protoplanets, as discussed in \S~\ref{sec:realistic_model}.
In order to discuss the basic characteristics of structure of protoplanets, in this section we will assume a single polytropic equation of state for the gaseous component.
We call the model prescribed by a single polytrope 'single polytrope model' in the following.
A more realistic structure will be discussed by using models of composite polytropes with different parameters in \S \ref{sec:realistic_model}.

\begin{table}
	\begin{center}
	\caption{Input Parameters and Characteristic Quantities}
	\label{tab:input_parameters}
		\begin{tabular}{cl}
		\tableline\tableline
		Parameter & Value\\
		\tableline
		$\rhodisk$ & $5.5 \times 10^{-11} \gcm$\\
		$\tdisk$ & $150 \hbox{ K}$\\
		$\pdisk$ & $0.343 \hbox{ dyn cm}^{-2}$\\
		$\mu_{\disk}$ & $2.0$ \\
		$1 - \beta_{\disk}$ & $3.7 \times 10^{-6}$\\
		$\rhos$ & $5.5 \ \gcm$\\
		$\sma$ & $1 \hbox{ AU (for Hill radius)}$\\
		$\qhost$ & $1 \ \msun \hbox{ (for Hill radius)}$\\
		\hdashline
		$\dmean$ & $1.41 \times 10^{-7} \gcm$\\
		$\qchar$ & $179 \ \mear$\\
		$\qcoremin$ & $ 9.84\times 10^{-4} \ \mear$\\
		\tableline
		\end{tabular}
	\end{center}
\end{table}

\begin{table*}
	\begin{center}
	\caption{Properties of single polytrope models with the maximum core mass \label{tab:properties_single_model_BH}}
		\begin{tabular}{lccclcc}
		\tableline\tableline
		$N$ & $\log(\ucrit)$ & $\log(\vcrit)$ & $\qtot^{\crit}(\mear)$ &
		\multicolumn{1}{c}{$\qcore^{\crit}(\mear)$\tablenotemark{a}} & $\log(\usurf)$ & $ \toe/\mu $ ($10^{4} $K) \\
		\tableline
		\multicolumn{7}{c}{Bondi boundary condition}\\
		\tableline
		1.5       & -6.10 & 0.40 & 162.7 & 106.2(0.65)& -0.09 & $ 6.94$ \\
		2.0       & -4.79 & 0.48 & 159.4 & 103.7(0.65)& -0.10 & $ 5.71$ \\
		2.5       & -3.53 & 0.54 & 158.3 & 101.2(0.64)& -0.11 & $ 4.71$ \\
		3.0       & -2.33 & 0.60 & 156.2 & 97.2 (0.62)& -0.12 & $ 4.05$ \\
		\hdashline
		4.0 (1st) & -0.68 & 0.65 & 60.2  & 37.7 (0.63)& -0.95 & $ 2.29$ \\
		4.0 (2nd) & -0.69 & 0.65 & 240.1 & 37.6 (0.16)&  0.25 & $ 2.25$ \\
		5.0       & -0.79 & 0.75 & 11.53 & 8.96 (0.78)& -2.39 & $ 0.68$ \\
		\tableline
		\multicolumn{7}{c}{Hill boundary condition}\\
		\tableline
		$N$ & $\log(\ucrit)$ & $\log(\vcrit)$ & $\qtot^{\crit}(\mear)$ &
		\multicolumn{1}{c}{$\qcore^{\crit}(\mear)$\tablenotemark{a}} & $\log(\vsurf)$ & $ \toe/\mu $ ($10^{4} $K) \\
		\tableline
		1.5 & -3.19 & 0.39 & $2.48 \times 10^{5}$& $8.15\times 10^{4}$ (0.33)& 3.31 & $ 579 $ \\
		2.0 & -2.40 & 0.47 & $1.39 \times 10^{4}$& $6.20\times 10^{3}$ (0.44)& 2.40 & $ 89.5$ \\
		2.5 & -1.64 & 0.52 & $2.71\times 10^{3}$ & $1.27\times 10^{3}$ (0.47)& 1.92 & $ 27.7$ \\
		3.0 & -1.04 & 0.55 & 715.1               & 344.0               (0.48)& 1.56 & $ 11.1$ \\
		\hdashline
		4.0 & -0.67 & 0.65 & 56.3                & 38.0                (0.68)& 0.80 & $ 2.29$ \\
		5.0 & -0.77 & 0.75 & 11.54               & 8.97                (0.78)& 0.34 & $ 0.69$ \\
		\tableline
		\end{tabular}
	\tablenotetext{a}{the value in brackets denotes a ratio of $\qcore$ to $\qtot$}
	\end{center}
\end{table*}

\begin{figure}
	\resizebox{\textwidth}{!}{\plotone{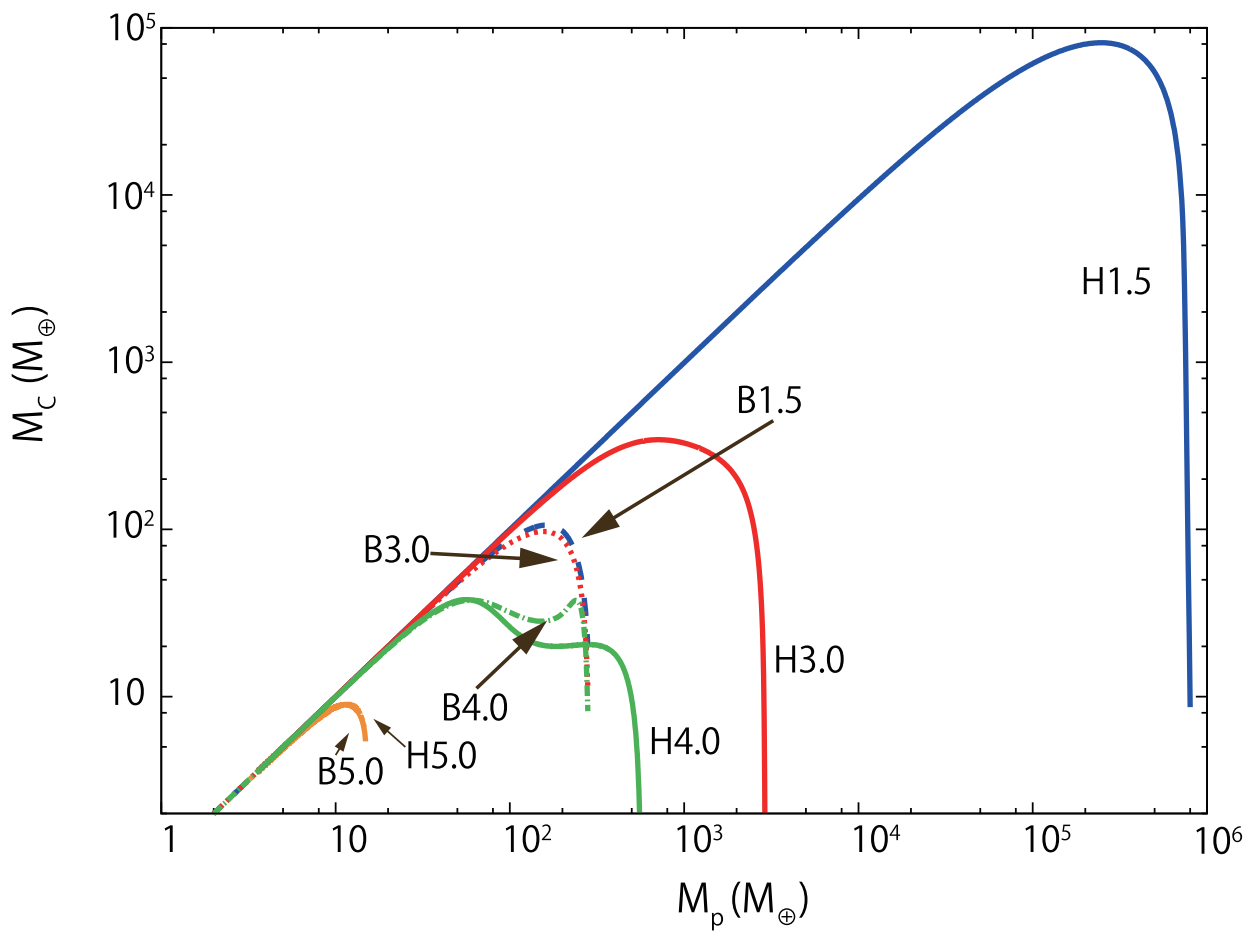}}
	\caption{
	Relation between a total mass, $\qtot$, and a core mass, $\qcore$, for single polytrope models of indices, $N = 1.5$ (blue), 3 (red), 4 (green), and 5 (orange), and with Bondi model and Hill model.
	Solid lines indicate curves for Hill model, and dashed lines indicate curves for Bondi model.
	}
	\label{fig:solutions_property_mass}
\end{figure}

\begin{figure}
	\resizebox{\textwidth}{!}{\plotone{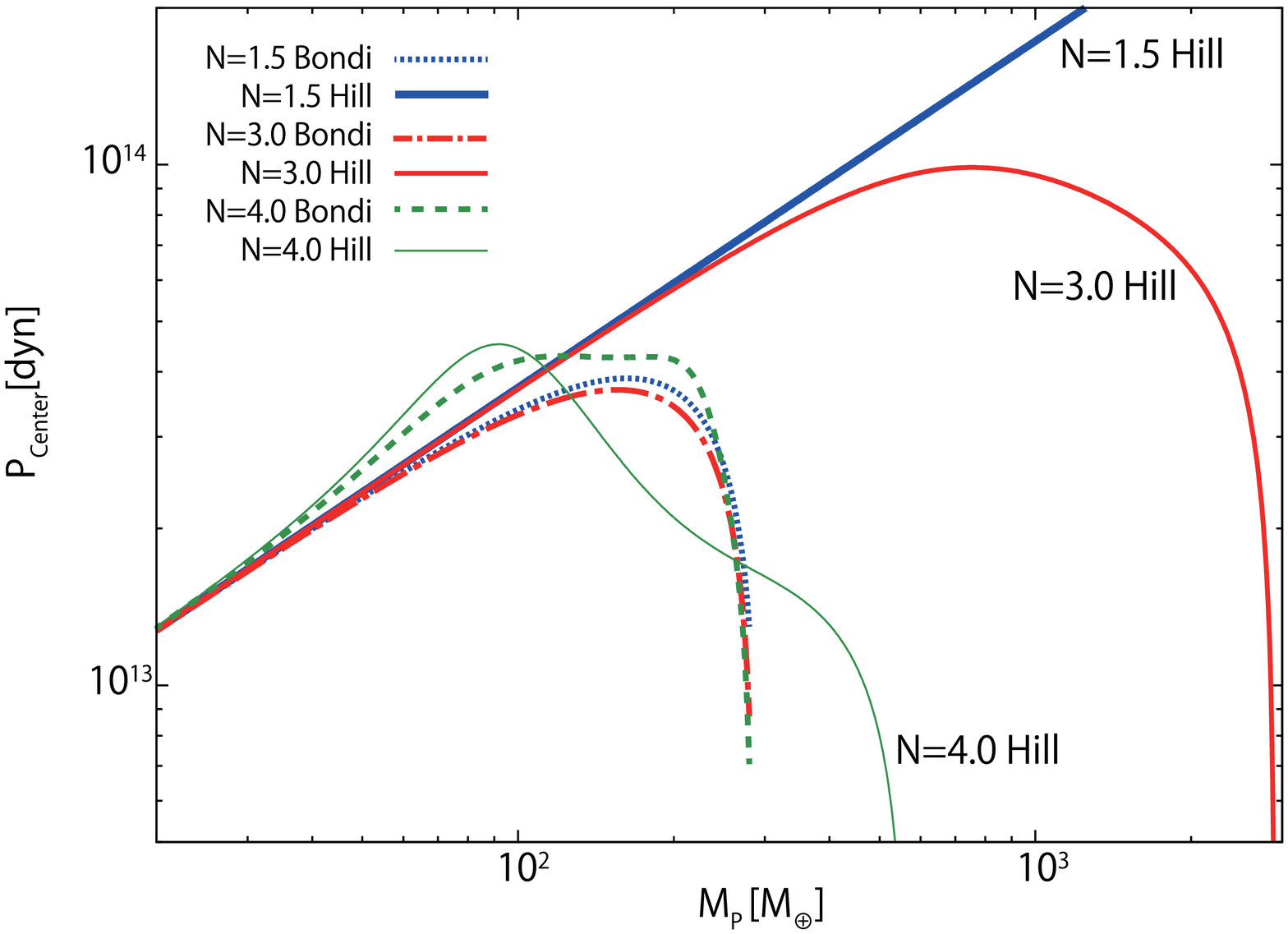}}
	\caption{
	Variations of central total pressure, i.e., the sum of those of gas and solid components, with the total mass, $\qtot$, for single polytrope models of indices, $N = 1.5$ (blue), 3 (red), and 4 (green) with Bondi model (dashed line) and Hill model (solid line).
	}
	\label{fig:solutions_property_pressure}
\end{figure}

We study the behaviors of the single polytrope models with various polytropic indices;
the polytropic constant is given by
\begin{equation}
K= \kd = \pdisk / \rhodisk^{1+1/N}
\label{eq:poly_const_disk}
\end{equation}
from the outer boundary condition.
In the previous studies, the radius of the planet is usually switched from the Bondi radius to the Hill radius as the planet grows in mass.
However, the nature of the boundary conditions is very different, as discussed above, and we are interested in their effects on the structures of protoplanets.
In this paper, we take either of Bondi radius and Hill radius as outer boundary regardless of the total mass, and label solutions as Bondi or Hill models according to their outer boundary radius.
For other parameters, we adopt values similar to previous works \citep[e.g.,][]{Ikoma_Emori_Nakazawa2001}, as summarized in Table~\ref{tab:input_parameters} with characteristic values.
We set $\gammad =1$, and we take $\sma=1$ au instead of $5.2$ au in order to allow a larger range of the ratio between the mass of the gaseous envelope and the mass of the solid core for the Hill radius [$\qgas/\qcore > \rhodisk/\dmean = 3.9 \times 10^{-4}(\sma/1\au)^3$].

In our computations, we give the total mass and obtain the structure of gaseous envelope by searching the mass of solid core which satisfies the jump conditions at the inner edge of envelope.
We use the Henyey method \citep{Henyey_Forbes_Gould1964,NR} to solve the two-point boundary value and eigenvalue problem.
Figure~\ref{fig:solutions_property_mass} shows the resultant relationship between the core and total mass for the polytropic indices of $N = 1.5$ - $5$.
Our models demonstrate that the core mass has a maximum (or maxima) as a function of the total mass, as reported by previous works with a realistic equations of state \citep[e.g.,][]{Mizuno1980,Ikoma_Emori_Nakazawa2001}.
Table~\ref{tab:properties_single_model_BH} summarizes characteristics of the models with the maximum cores.
In the following, we call the first maximum core mass the critical core mass and the critical model with it the critical model.

Behaviors of models differ by the polytropic index as well as by the surface boundary conditions.
In general, the critical core mass increases for models with the stiffer equation of state (the smaller polytropic index).
The model characteristics are divided according to whether the polytropic index is larger than $N = 3$ or not.
In the case of the polytropes of $N \le 3$, Hill models have much larger critical core masses than the Bondi models.
For the Hill models, the critical core mass increases greatly as the polytropic index decreases, while it depends only weakly for the Bondi models.
In the case of the polytropes of $N > 3$, the critical core mass turns out similar both for the Hill and Bondi radius and decreases significantly with the polytropic index.
The behaviors also diverge after the critical core mass is reached.
In the case of $N \le 3$, the core mass monotonically decreases as the total mass increases irrespective of the outer boundary conditions.
In the case of $N > 3$, Bondi model has the second maximum of core mass with almost the same value as the first maximum, while the Hill model has a plateau on the way of continuous decline of core mass.
In the following, we refer to the polytrope of $N \le 3$ and $N>3$ as 'stiff polytrope' and 'soft polytrope', respectively.
As compared with the previous works with the realistic thermal structures, the critical core masses of the single polytrope model are larger, and especially, for the polytropes of smaller polytropic indices.
This is due to larger entropy, or more precisely, larger polytropic constant, at the bottom of envelope, implying that the effects of radiative cooling and/or the effects of ionization and molecular dissociation are important in the actual envelope, as will be discussed in \S \ref{sec:realistic_model}.

The occurrence of the maximum core mass are related to the general property of gravitating systems with two components of different thermal states in spherical symmetry \citep[e.g.,][]{Sugimoto_Nomoto1980,Fujimoto_Tomisaka1992}.
There are two different ways of responding to variations in the total mass.
When the envelope is sufficiently less massive than the solid core and the structure of the whole envelope is governed by the external gravity of the core of smaller radius, the variation of pressure with the density ($\propto \rho^{1+1/N}$) is larger than that of gravity ($\propto \rho$) so that the increase of the weight of envelope with the envelope mass can be balanced with the increase in the pressure due to compression.
In this case, it demands higher gravity, and hence, a larger mass of core to retain the larger amount of envelope which requires larger gravity of core, and hence, a larger core mass.
When the envelope mass grows larger than the core mass, on the other hand, the self-gravity of envelope comes to be effective as compared with the gravity of core except for the innermost shells near to the core.
In this case, the compression can no longer sustain the larger mass of envelope in thermal equilibrium as designated by polytropic equation of state in our models, for the enhanced self-gravity ($\propto \rho^2$) outweighs the increase in the pressure.
Instead, such a system can accommodate more envelope mass by expelling the envelope mass outward to diminish the weight of overlying layers.
This demands a reduction of the core gravity, and hence, a decrease in the core mass.
The maximum core mass is realized during
The transition of these two envelope configurations may occur when the envelope mass increases to be comparable with the core mass, which may realize the maximum core mass.

The opposite responses manifest themselves in the variations of the thermal properties of protoplanets, also.
Figure~\ref{fig:solutions_property_pressure} shows the central total pressure ($P_{\rm center} = \pgas + \psol$) exhibits the similar behaviors with the maximum and the diversity;
$\pgas$ and $\psol$ increase with the total mass as well as the core mass for a total mass smaller than the critical model, whereas both turn to decrease after the models pass through the critical models.
The peak of $P_{\rm center}$ occurs very close to the critical model in the case of the stiff polytropes, but they may not necessarily coincide since the weight of envelope results from the competition between the decrease of the gravity of core and the increase of envelop mass.
In the case of the soft polytrope, $P_{\rm center}$ attains at the maximum appreciably beyond the critical model, and for the Bondi model, it remains with the maximum value until the second maximum of core mass is reached.
Such distinct behaviors indicate that different mechanisms operate to bring about the critical models.

\begin{figure}
	\begin{center}
		\resizebox{\textwidth}{!}{\plotone{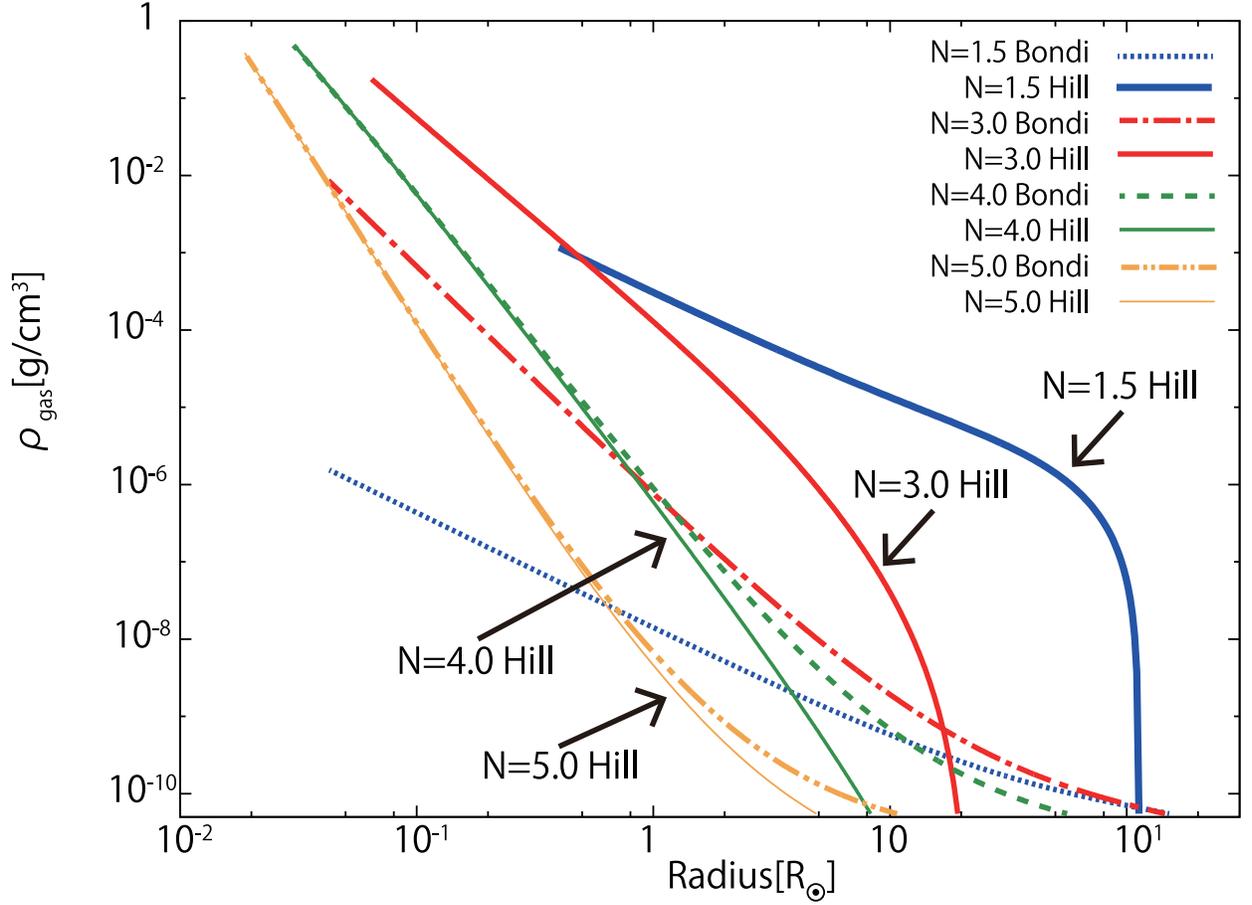}}
	\end{center}
	\caption{
	Density distribution of gaseous components against the radius of $r \ge \rcore$ in the critical models, with the polytropes of $N=1.5$, 3, 4 and 5 and for Bondi model and Hill model, respectively.
For a description of models, see the legend at top right corner.
	}
	\label{fig:density_distribution_crit_model}
\end{figure}

Figure \ref{fig:density_distribution_crit_model} shows the internal structure of critical models, where the density is plotted against the radial distance from the center.
The density distribution differs also according to the polytropic index and the outer boundaries.
The density reaches much larger in the interior for softer polytrope.
In the case of stiff polytropes, the internal density is greatly larger for Hill models than for Bondi models, whereas there is little difference except the outermost layers in the case of the soft polytropes.
For Hill models, there is a steep density drop near the surface particularly in the case of stiff polytropes, while the slope grows small as the surface approaches for Bondi models.
Nevertheless, all the models share a common feature that a power-law distribution of density develops in the bottom of envelope, but with the slope steeper for softer polytropes ($\propto r^{-N}$).
This is characteristic of the centrally-condensed type solution, which develops under the dominance of the gravity of core over the self-gravity of envelope \citep[][]{Chandrasekhar1939,Hayashi_Hoshi_Sugimoto1962}.
Further, close examination discloses that the slope grows flatter in the outer part of the power-law distribution in the case of the stiff polytropes of $N \le 3$, while in the case of the soft polytropes, it occurs in the inner part between the core and the power-law-distribution.
These differences in the internal structure are closely related to the condition of the critical core mass.

In the following, we discuss the characteristics of the internal structure of envelope and elucidate the occurrence of the maximum core mass and the origins of their differences.
Below we first deal with the case of small envelope mass, and then, the models of the full range of envelope mass by using the homology invariants.

\subsection{The Models of Small Envelope Mass}
\label{ss:small_envelope,result_of_polytrope} \label{sec:small_envelope}

During the earlier stages, when the envelope mass, $\qenv$, remains much smaller than the mass in the core, i.e.,
\begin{eqnarray}
	&&\qenv =\qtot-\qcore \ll \qcore,
\label{eq:qenv}
\end{eqnarray}
the self-gravity of the envelope mass can be ignored as compare with the gravity of core in the whole envelope.
Then, the equation of mechanical balance is integrated to yield the density distribution;
\begin{eqnarray}
	\rhog(r)&=& \rhodisk \left[ 1-\frac{\vsurf}{N+1}\left( 1-\frac{\rtot}{r} \right) \right]^{N}
\label{eq:env_dens_dist}
\end{eqnarray}
\citep{Fujimoto_Tomisaka1992,Rafikov2006}.
Around the core where $r / \rtot \ll \min [1, \vsurf/(N+1)]$, therefore, the power-law distribution $\rho \propto r^{-N}$ develops, which is characteristic of the centrally-condensed type solution and associated with the expansion of radius of red giants.
The density at the inner edge of envelope is given by
\begin{eqnarray}
	&&{\rhooe}/{\rhodisk} \simeq (N+1)^{-N}\left[ \rhocore/( 3\rhodisk) \right]^{N/3}\left( \qcore/\qchar \right)^{2N/3}. \label{eq:density_ratio_env}
\end{eqnarray}
It is larger for a softer polytropes of larger $N$, which is the consequence of greater decrease in the entropy, or, more precisely, slower increase in the thermal energy, $P / \rho$, with the density.
In terms of homology invariants, the above distribution is written as;
\begin{eqnarray}
	V&=& (N+1)/ \left[ 1 + (r/\rtot) \{  (N+1)/\vsurf -1\} \right].
\label{eq:v_eq_nplus1}
\end{eqnarray}
Accordingly, the power-law part is featured by the constancy of $V = N+1$, indicative of the fact that the balance is maintained between the gravitational energy and the thermal energy.
On the characteristic plane, it is described as a straight line, along which $U$ changes the direction of movement according to the soft and stiff polytropes, as given by
\begin{eqnarray}
	U&=&\left( \frac{1}{N+1} \right)^{N}\left( \frac{3\rhodisk}{\rhocore} \right)^{(3-N)/3} \left( \frac{\qcore}{\qchar} \right)^{2N/3} \left( \frac{r}{\rcore} \right)^{3-N}.  \label{eq:u_at_v_eq_nplus1}
\end{eqnarray}

The density distribution in eq.~(\ref{eq:env_dens_dist}) bifurcates in the outer part according to the surface value of $\vsurf$.
For $\vsurf / (N+1) < 1$ and $ > 1$, the density distribution grows flatter and steeper, respectively, as $r \rightarrow \rtot$.
In particular, there is a large drop near the surface for $\vsurf / (N+1) \gg 1$.
The two groups are clearly distinguished on the characteristic plane;
one with $V > (N+1)$ and the other with $V < (N+1)$ in the whole envelope, as seen in eq.~(\ref{eq:v_eq_nplus1}).
As a corollary, only the latter distribution is applicable to the Bondi model, while the Hill model can take the both distributions.

Given the density distribution, the mass ratio between the envelope and the core is written in the form;
\begin{eqnarray}
	&&\frac{\qenv}{\qcore}=\frac{4\pi \rtot^3 \rhodisk}{\qcore} \int_{\rcore/\rtot}^{1} \left[ \xi+\frac{\vsurf}{N+1}\left( 1-\xi \right) \right]^{N} \xi^{3-N}d \log \xi.
\label{eq:env_mass}
\end{eqnarray}
In the case of the stiff polytropes, the contribution to the integrand is larger in the outer shells because of slow outward decrease of density in the power-law distribution, and the envelope mass also concentrates in the outer shells.
For Bondi model, the density distribution, normalized with respect to the power-law distribution, remains the same, and the largest mass concentration occurs in the outermost shell, regardless of total mass as long as $\rcore \ll \rtot$.
Since the integral remains an order of unity for $\vsurf < N+1$ so that the envelope mass is simply proportional to the volume of sphere with the radius equal to Bondi radius in the multiplier in front of the integral;
\begin{eqnarray}
	&&\qenv / \qcore \simeq (\qcore/\qchar)^2/\gammad^3.\label{eq:env_mass_bondi_stiff}
\end{eqnarray}
Consequently, the envelope mass is kept smaller than the core mass until the core mass approaches to the characteristic mass.

For Hill models, on the other hand, the multiplier is constant and the mass ratio augments through the increase of density with $\vsurf / (N+1)$;
for $\vsurf \gg N+1$, the envelope mass is evaluated as
\begin{eqnarray}
	&&\qenv/\qcore \sim (N+1)^{-N}(\rhodisk/\dmean)^{(3-N)/3}(\qcore/\qchar)^{2N/3}  \label{eq:env_mass_hill_stiff}
\end{eqnarray}
with the largest mass concentration in the middle of envelope.
The increase in the envelope with the core mass is slower as compared with Bondi models because of weaker dependence of the Hill radius on the total mass.
It demands higher density, and thus larger gravity of core, to stuff a given mass within the sphere of Hill radius since the Hill radius grows smaller than the Bondi radius for $\qcore / \qchar > (\rhodisk / \dmean)^{1/2}$.
This effect is greater for smaller polytropic index owing to larger thermal energy for a given density, as represented by the term of $(\rhodisk/\dmean)^{(3-N)/3}$, and explains much larger critical mass for the model of smaller $N < 3$.
For $N = 3$, the mass ratio no longer depends on the surface radius since the homologous transformation holds so that the difference from Bondi model arises solely from the difference in the outer boundary condition.

In the case of the soft polytropes, the integral is dominated by the contribution from the innermost envelope in contrast to the stiff polytropes because of steeper decrease of density outwards.
Hence, the mass ratio is insensitive to the outer boundary radius and results similar both for Bondi and Hill models;
\begin{eqnarray}
	&&\qenv/\qcore \sim (N+1)^{-N} (\rhocore/\rhodisk)^{(N-3)/3}(\qcore/\qchar)^{2N/3}.\label{eq:env_mass_soft}
\end{eqnarray}
For a given core mass, the ratio is much larger, and the dependence on the core mass is stronger than the models of the stiff polytropes.
The envelope mass catches up with the core mass at the core mass smaller than the characteristic mass by $\qcore / \qchar \sim (N + 1)^{3/2} (\rhodisk / \rhocore)^{(N - 3) /2N}$.
This is mostly attributable to slower increase of the thermal energy, $P / \rho$, or steeper decrease of entropy, with the density in the interior.

\subsection{Structural Characteristics and the conditions of Critical Core Mass}

\begin{figure}
	\includegraphics[width=0.45\textwidth]{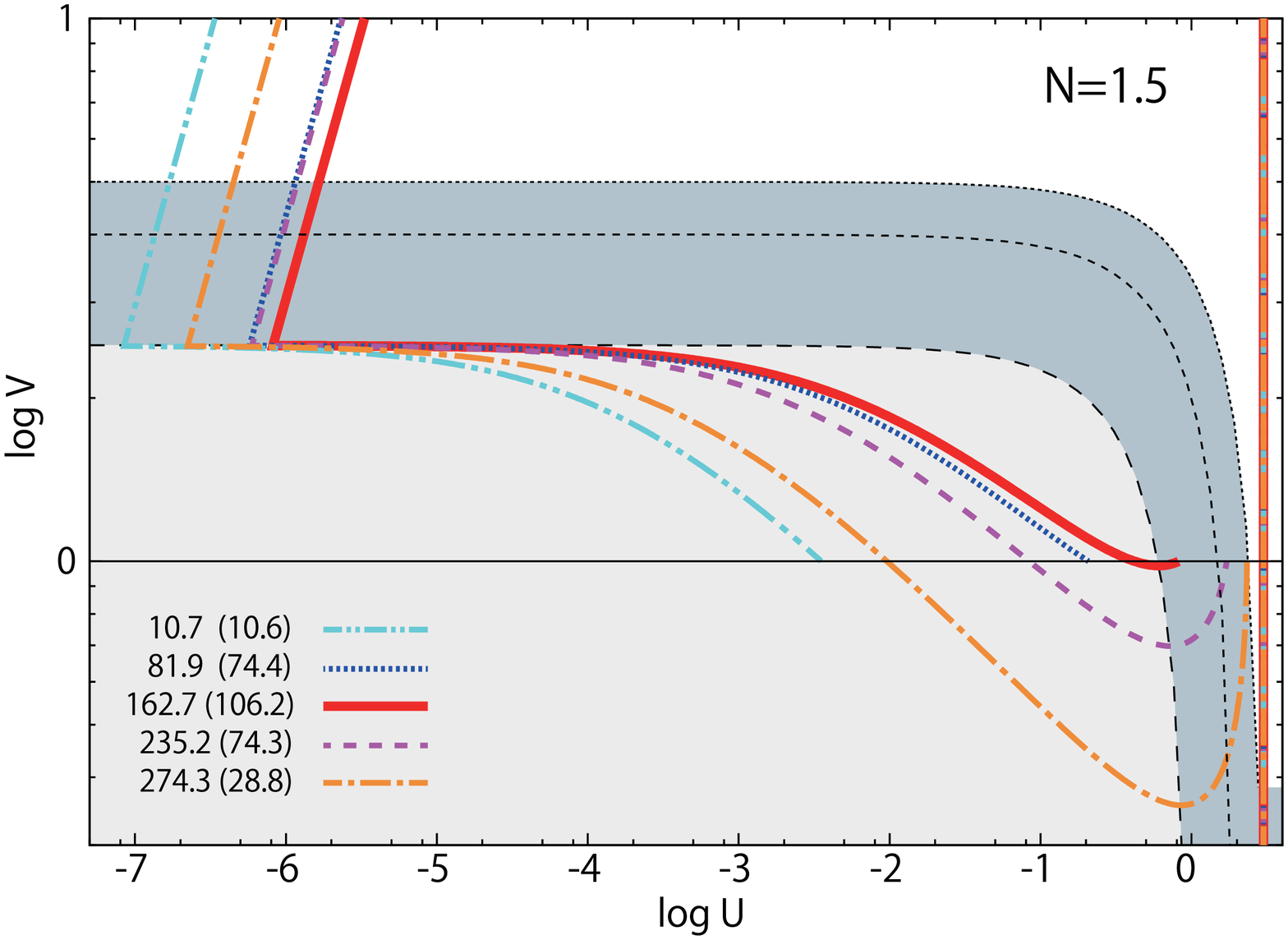}
	\hfil
	\includegraphics[width=0.45\textwidth]{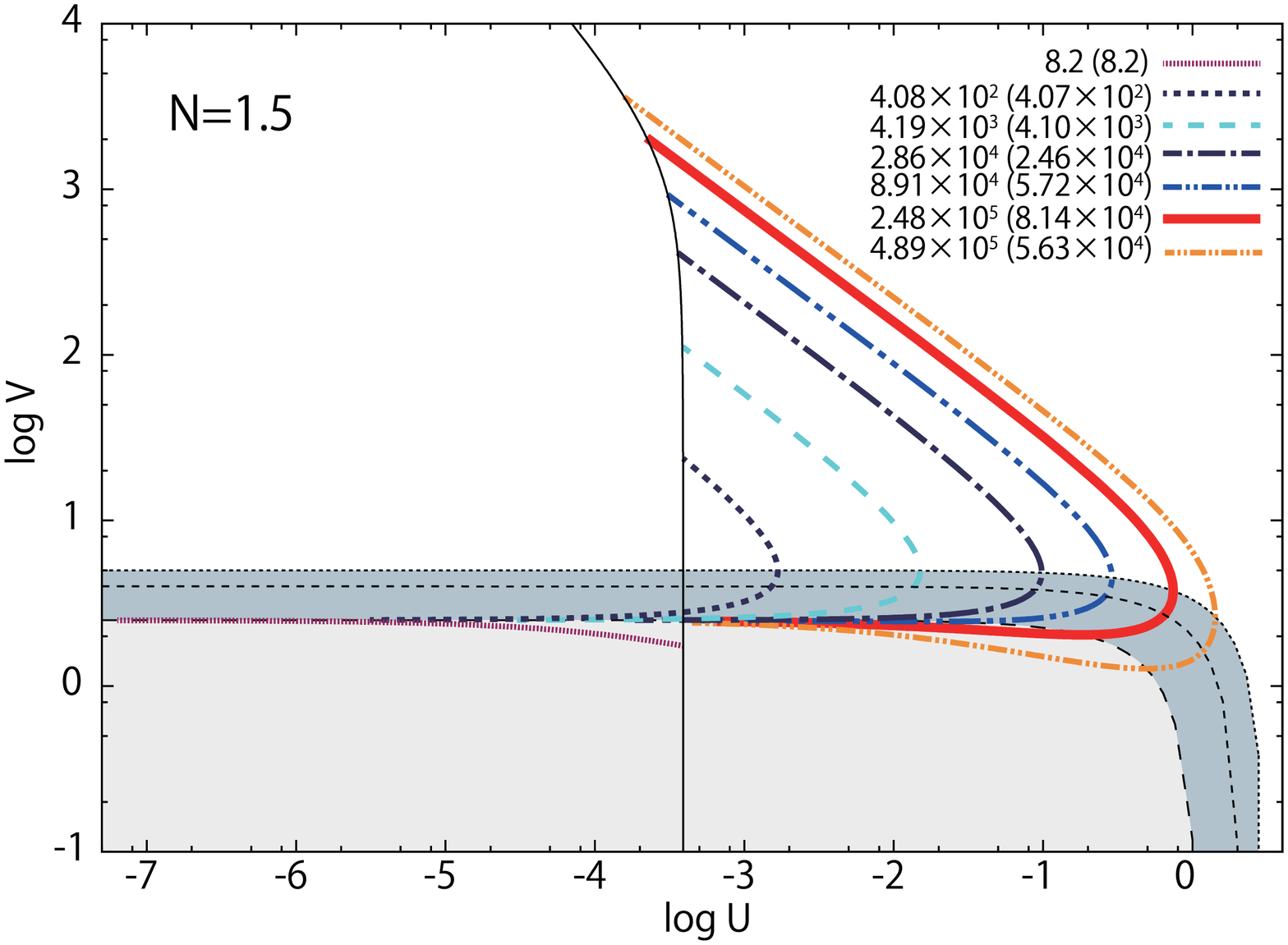}

	\includegraphics[width=0.45\textwidth]{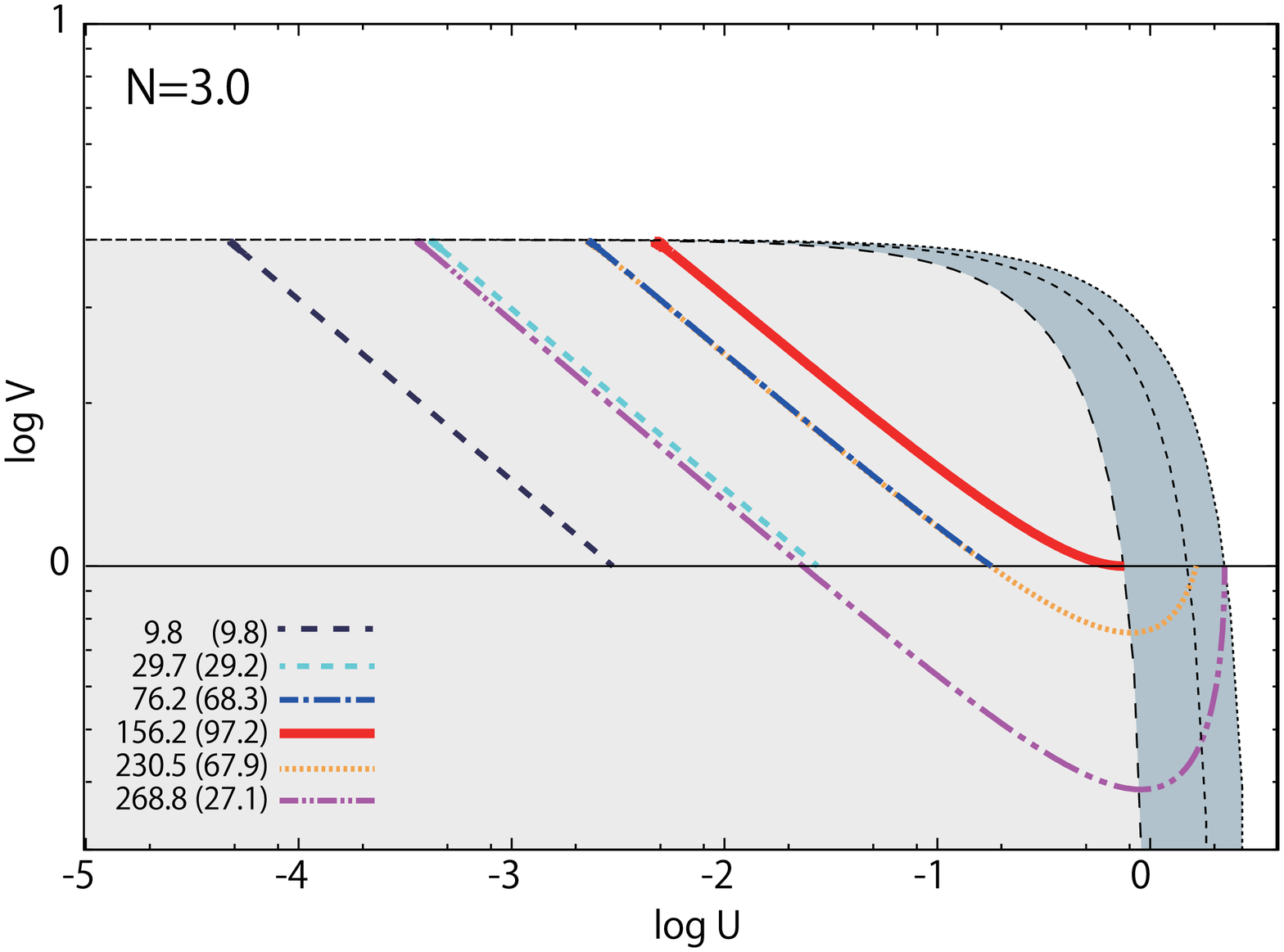}
	\hfil
	\includegraphics[width=0.45\textwidth]{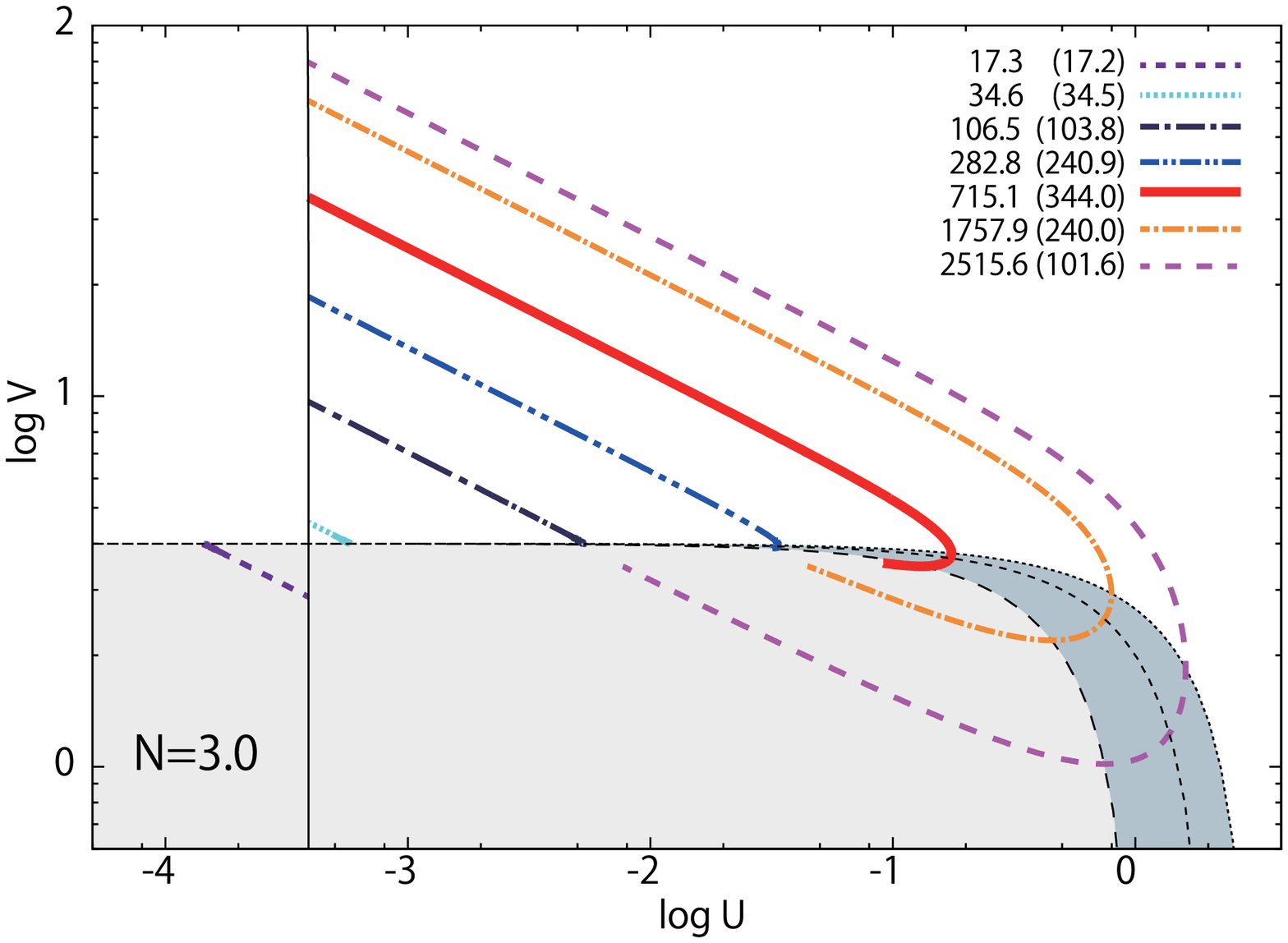}
	\caption{
	Structure lines of single polytrope models of various total masses with the Bondi radius (left panels) and the Hill radius (right panels) as the outer boundary, plotted on the characteristic plane of $\log U $-$ \log V$ diagram for polytropic indices of $N=1.5$ (top panels), $N=3$ (bottom panels);
	the planet total mass and the resultant core mass are given in the legend.
The critical models are denoted by red thick solid lines.
	In the top-left panel, the part of structure line in the core (rightmost vertical line) and those representing the jump condition between the top of core and the bottom of envelope are depicted, though they are omitted in the other panels for simplicity.
	Thin dotted, broken, and long-dashed lines denote the vertical, critical, and horizontal lines;
the region below the horizontal line is shaded by light grey and the region above the horizontal line and below the vertical line by thick grey.
	}
	\label{fig:uv_single_polytrope_stiff}
\label{fig:uv_p1.5+3.0_bondi}
\label{fig:uv_p1.5+3.0_hill}
\end{figure}
\begin{figure}
	\includegraphics[width=0.45\textwidth]{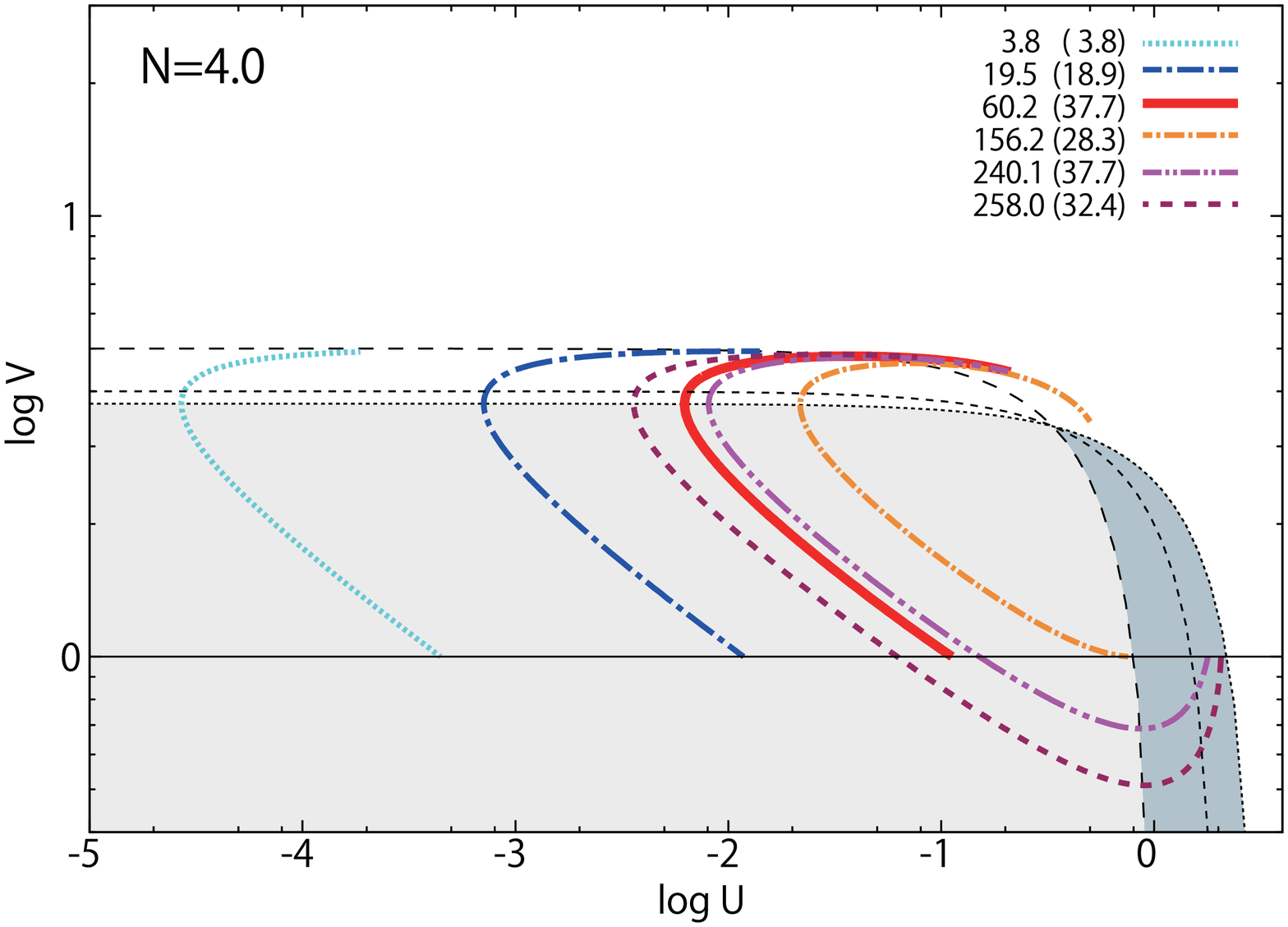}
	\hfil
	\includegraphics[width=0.45\textwidth]{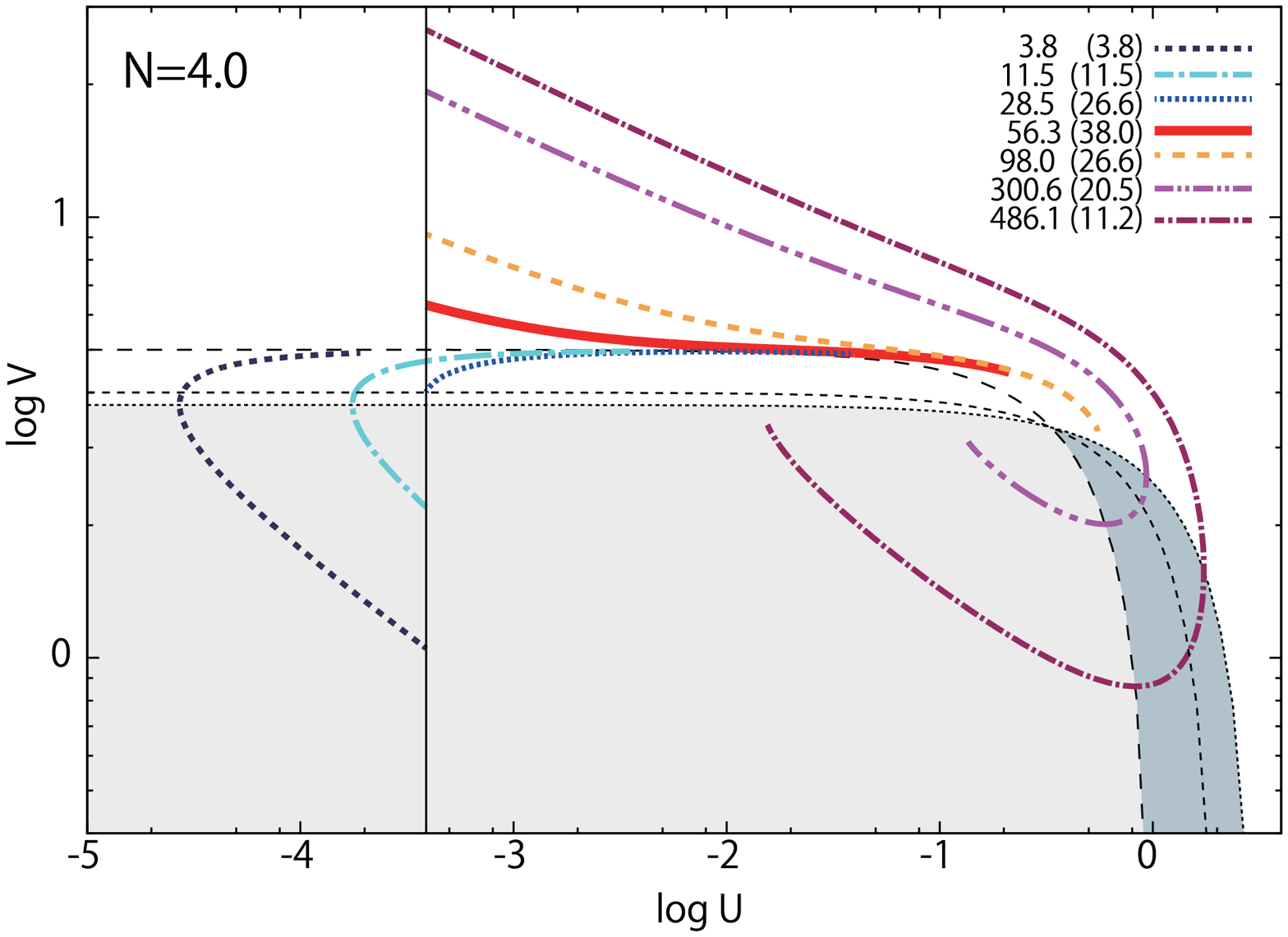}
	\caption{
	The same of Figure~\ref{fig:uv_single_polytrope_stiff}, but for polytropic index of $N=4$.
	}
	\label{fig:uv_single_polytrope_soft}
\label{fig:uv_p4.0_bondi+hill}
\end{figure}
We discuss how the transition occurs from the envelope structure, dominated by the core gravity, to that in which the self-gravity is important and the mechanisms that bring about the maximum core mass.
Figures \ref{fig:uv_single_polytrope_stiff} and \ref{fig:uv_single_polytrope_soft} illustrate the structure lines of models with various planet masses on the $\log U $-$ \log V$ diagram for the models of two stiff polytropes of $N=1.5$ and $N=3$ and a soft polytrope of $N = 4$ with both Bondi and Hill radius.
The detail behaviors and characteristics of these models are discussed in Appendix \ref{sec:nature_of_single_polytrope}.
Here we summarize the manifestations of the duplicity (or multiplicity) of envelope solutions and the conditions for the critical core mass.

On the characteristic plane, the solutions of envelope run between the inner edge and the outer boundary, following the equation of hydrostatic equilibrium (eq.~[\ref{eq:hydroeq_uv}]).
The location of outer boundary is determined as a function of the total mass when the physical conditions of the protoplanetary disk are specified for the Bondi radius (eq.~[\ref{eq:surf_uv_bondi}]), and in addition, the mass and separation of the host star for the Hill radius (eq.~[\ref{eq:surf_uv_hill}]).
Then, the location of inner edge, and hence the core mass, is determined by the condition of the radius ratio between the inner edge and surface of envelope (eq.~[\ref{eq:radius_condition}]).
The resultant inner edge of envelope satisfies the following relation;
\begin{eqnarray}
	\voe \uoe^{1/N} &=& (\rhocore / 3 \rhodisk)^{(N-3)/3N} (\qcore / \qchar)^{2/3}
\label{eq:env_inner_boundary}
\end{eqnarray}
derived from the jump conditions in eqs.~(\ref{eq:jump_uoe}) and (\ref{eq:jump_voe}).
For a given polytrope, the models with the same core mass locate their inner edges of envelope on the line of gradient
\begin{eqnarray}
	&& \delta \log \voe + (1/N) \delta \log \uoe = 0
\label{eq:inneredge_critmodel}
\end{eqnarray}
on this $\log U $-$ \log V$ diagram.
The loci of inner edges with constant core mass shift rightward as the core mass increases and also for larger polytropic index.

In the case of the stiff polytropes, Bondi models draw the structure line between the inner edge, just below the horizontal line, and the outer boundary on the line of $V = \gammad$ with decreasing $U/V$ outwards, as seen for top and bottom panels in the left column of Figure~\ref{fig:uv_single_polytrope_stiff}.
In this case, there are two sequences of the solutions, one which reaches to the outer boundary from above, and the other which intersects and meets it again from below.
The former sequence terminates and switches over to the latter sequence when the surface attains at the horizontal line, i.e.,
\begin{eqnarray}
\usurf =\usw&=& 1-\gammad / (N+1).
\label{eq:switch_point}
\end{eqnarray}
The solutions on the former sequence monotonically decrease $V$ outwards, which constitute a branch of models that the core mass increases with the total mass, including the envelope with the small mass, as discussed above.
The solutions on the latter sequence take a minimum value of $V$ when crossing the horizontal line, which represent the structure that the self-gravity of envelope is effective, and eventually lead to another branch of models with the core mass decreasing with the total mass.
For $N <3$, the model of the second branch with the same core mass as the models of $\usurf = \usw$ draws the structure lines slightly below since the structure line has less steep slope than the loci of inner edge with a constant core mass in eq.~(\ref{eq:env_inner_boundary}) below the horizontal line.
Then, the maximum core mass is reached for $\usurf$ slightly larger than $\usw$.
For $N = 3$, the both slopes agree and the structure lines of these two models with the same core mass overlap, so that the maximum core mass occurs at $\usurf = \usw$.

Accordingly, the criterion of maximum core mass approximates to $\usurf^{\rm crit} = \usw$.
This is converted into the condition on the ratio between the mean density, $\brho$, of planet and the surface (or environment) density, $\rhodisk$, as
\begin{eqnarray}
	\brho/\rhodisk \simeq 3/\left[ 1-\gammad/(N+1) \right].\label{eq:density_critmodel_bondi}
\end{eqnarray}
The Bondi model with the stiff polytrope is distinct from other models in that the ratio of the thermal energy to the gravitational energy is largest at the surface where most of mass is allocated, as discussed above, implying the configuration that the hot gas is confined by the external pressure rather than retained by the gravity.
The total mass, $\qtot^{\crit}$, of the critical models is given by
\begin{eqnarray}
	&&\qtot^{\crit}\simeq \gammad^{3/2}\left[ 1-\gammad/(N+1) \right]^{1/2} \qchar,
\label{eq:pmass_critmodel_bondi}
\end{eqnarray}
and nearly equal to the characteristic mass, $\qchar$, determined by the physical conditions in the surrounding gas alone aside from weak dependence on the polytropic index.
The core mass ensues from eq.~(\ref{eq:env_inner_boundary}) with the values of $\ucrit$ and $\vcrit$ in Table~\ref{tab:properties_single_model_BH};
$\vcrit \simeq N+1$ and $\ucrit$ decreases greatly for smaller $N$ since the horizontal line is more apart from the critical line, and hence, the structure line runs longer to satisfy the conditions of radius ratio.
Despite such a large difference in the interior, the fraction of core mass results very similar between $65 $-$ 62$ \% for $N = 1.5 $-$ 3$ because the structure is insensitive to polytropic index in the surface layer of small $V$.

In the case of stiff polytropes with the Hill boundary (top and bottom panels in the right column of Figure~\ref{fig:uv_single_polytrope_stiff}), two sequences are made by the models of $\vsurf > N+1$ that have the inner edge of envelope above and below the horizontal line, respectively, as discussed above for small mass envelope.
For the former, the structure line runs above the horizontal line all the way to the surface, and increases $V$ outwards to make the density drop steeper above the power-law part, as in the case for the envelope of the small mass, discussed above.
For the latter, the structure line first decreases $V$ outwards until it hits a minimum on the horizontal line, and then, turns to increase $V$, reaching the surface similarly to the former models;
a flatter density distribution develops with the decrease of $V$ outside the power-law part, which represents the solution that the self-gravity is effective.
As the core mass increases, the inner edges of envelope shifts toward larger $\uoe$ and approaches the horizontal line, and two sequences meet on the horizontal line.
These models form a linear series with the total mass as parameter, and the maximum core mass occurs in the model with the inner edge near and slightly below the horizontal line.

In the models, most of envelope mass is located in the flatter part near the intersection with the critical line, where $\vert d \log M_r / d \log P \vert = U/V$ is largest, whereas both the power-law and flatter part contribute to the radial extension, where the structure line runs along the critical line.
Since $\vsurf \gg N+1$, the pressure and density decreases greatly beyond the crossing of the critical line, and hence, the outer shells have little things to do with the inner structure.
Consequently, the behavior of models is determined by the inner structure, and eventually, by the location of the inner edge.
The values of $\ucrit$ and $\vcrit$ of critical models are determined by numerical computation and shown in Table.~\ref{tab:properties_single_model_BH}.
Although these models adopt the same centrally-condensed type structures as the above Bondi models, the structure line extends through the crossings of the horizontal and critical line.
Because of the contribution from these parts, the pass along the horizontal line is much shorter than in the Bondi models, resulting in $\ucrit$ larger by a factor of $\sim 800 $-$ 20$, while the relation of $\vcrit = N +1$ remains still good within $2 $-$ 12 \%$.
Given $\ucrit$ and $\vcrit$, the maximum core mass follows from eq.~(\ref{eq:env_inner_boundary}), and occupies the fraction of $0.33 $-$ 0.48$, smaller than for the Bondi models.
The total mass of the critical models is given by an order of the mass of Emden solution with the same polytropic equation of state and with the surface radius equal to the Hill radius (see \S \ref{sss:hill_stiff,result_of_polytrop} in Appendix).
The larger total and core masses are attributable to larger polytropic constant, $K$, for stiffer polytrope, which is taken to be the same as the physical condition in the planetary disk.

For the soft polytropes of $N > 3$, the envelope structure is characterized by a stronger inward increase in density and resultant concentration of envelope mass toward the innermost part, as discussed above for the envelope of small mass.
In the innermost shells, $U/V$ increases inwards, opposite to the models of the stiff polytropes, and for large total mass, the inner edge spirals around the singular point, as illustrated for $N = 4$ in Figure \ref{fig:uv_p4.0_bondi+hill} (bottom two panels).
In this case, two sequences are divided by the model that the inner edge of envelope just above the singular points, i.e., by the condition that $\uoe$ is smaller and larger than the critical value;
\begin{eqnarray}
	& & \uoe = \ucrit = (N - 3) / (N - 1).
\label{eq:ucrit_soft}
\end{eqnarray}
While $\uoe \ll \ucrit$, the power-law distribution expands to increase the density and both the envelope and the core mass augment with the total mass, as discussed for the envelope of small mass.
As $\uoe$ increases and the inner edge of envelope starts to spiral and decrease $\voe$, a flatter density distribution develops inside the power-law part, differently from the models with the stiff polytropes.
When the inner edge comes to right above the singular point of $\uoe = \ucrit$, the structure line has the same slope ($d \log V/ d \log U = -1/N$) as the loci of inner edge of envelope, specified by the jump condition for a constant core mass
{}\footnote{ In our numerical models, we should take into account the increase in the core density with $\uoe$ due to the penetration of gaseous component, which affects the slope of constant core mass, as discussed in Appendix~\ref{sec:nature_of_single_polytrope}
}.
This model marks the maximum core mass, indifferent of the outer boundary, since for still larger total mass, the inner edge spirals down whereas the loci of inner edge, specified by the jump condition, shifts upwards for larger core mass.
As the total mass increases beyond the critical model, the Bondi and Hill models exhibit different behaviors.
For the former, the inner edge turns back when the surface attains at $\usurf = \usw$, and then, the second maximum of core mass is reached at $\uoe = \ucrit$ on the way back to small $\uoe$.
For the Hill models, the inner edge keeps spiraling around the singular point and gives rise to a plateau of core mass when it reaches right below the singular point, where the structure line has the same slope of the loci of constant core mass.

The criterion for the critical model in eq.~(\ref{eq:ucrit_soft}) is transferred to the condition on the density at the inner edge of the envelope as
\begin{eqnarray}
	\rhooe &=& \rhooe^{\crit} = \rhocore \left[ (N-3) / 3 (N-1) \right].
\label{eq:density_ratio_soft}
\end{eqnarray}
The value of $\voe$ decreases as the inner edge spirals, and yet, $\vcrit$ is close to $N+1$ since the structure line crosses the horizontal line for $U \ll 1$.
The maximum core mass results from eq.~(\ref{eq:env_inner_boundary});
\begin{equation}
\qcore{}^{\rm crit} \simeq (3 \rhodisk / \rhocore)^{(N-3 ) / 2 N} (N + 1)^{3/2} [(N-3)/(N-1)]^{3/2N} \qchar ,
\label{eq:core_mass_crit_soft}
\end{equation}
and is a decrease function of polytropic index and much smaller than the models of the stiff polytropes.
The fraction of core mass is also larger;  $\qcore /\qtot = 0.63$ and 0.68 for $N = 4$ and $0.77$ for $N = 5$.

In summary, the envelope structure of protoplanets is described by the centrally-condensed type solutions.
These solutions are characterized by the value of $\voe \simeq N + 1$ at the bottom of envelope, and have two distinct configurations, one dominated by the gravity of core and the other in which the self-gravity of envelope is effective.
The maximum core mass is realized during transition between these two configurations, which take three different patterns depending on the polytropes and the boundary conditions.
Correspondingly, there are three different criteria for the critical models.
In the case of the stiff polytropes of $N \le 3$, the criteria are different according to the surface boundary.
For the Bondi radius, the criterion is given by the value of $\usurf$ at the surface, or the ratio between the mean density of planet and the density of environment in the protoplanetary disk; the envelope mass concentrates in the outermost part.
For the Hill radius, the criterion is given by the internal structure;
the critical core mass occurs for the model which has the inner edge on the horizontal line, while the envelope mass concentrates in the middle part.
In the case of the soft polytropes of $N >3$, the criteria is imposed on the value of $\uoe$, or the density ratio of $\rhooe / \rhocore$ at the inner edge of envelope, indifferent of the outer boundary;
the envelope mass also concentrates in the innermost shells, different from the Hill models of the stiff polytropes.

\section{Models with Composite Polytropes} \label{sec:realistic_model}

In the preceding section, we have discussed the envelope structure described by a single polytrope.
A realistic envelope of protoplanet may not have such a simple structure as the whole envelope is described only by a single polytrope, however.
The local polytropic index, defined in eq.~(\ref{eq:polytropic_eos}), may vary according to the local thermal condition in the envelope.   
In applying to the actual systems, we should consider the thermal property of the envelope and specify the relevant equation(s) of state in the form of polytrope(s).

\subsection{polytropic structure of envelope}
\label{sec:comp_polytrope}

As for the thermal structure of protoplanet envelope, it is argued that there is an isothermal layer in the outer part of envelope \citep{Mizuno_Nakazawa_Hayashi1978,Rafikov2006}.
With the accretion luminosity, $L_{\rm acc} = (G \qcore /\rcore) \qcdot$, of the core, where $\qcdot$ is the mass accretion rate of solid component, the radiative temperature gradient at the shell of mass $M_r$ in the envelope is estimated at
\begin{eqnarray}
	\vrad & = &0.05 \; {\kappa} \; M_r{}^{-1/3} \left( \frac{T^4/P \vert_{\rm disk}} {T^4 / P \vert_{M_r}} \right) \left( \frac{\qcore}{M_r} \right)^{2/3} \left( \frac{\qcdot}{10^{-6}\mperyr} \right),
	\label{eq:vrad_mag}
\end{eqnarray}
where $\kappa$ and $M_r$ are in units of $\hbox{cm}^{2} \hbox{ g}^{-1}$ and $\mear$, respectively.
Therefore, the outer envelope is likely to be nearly in an isothermal structure, unless the core accretion rate exceeds typical value of $\qcdot = 10^{-6}\mperyr$ \citep[e.g., see][and the references therein]{Ikoma_Nakazawa_Emori2000}.
In addition, \citet{Mizuno_Nakazawa_Hayashi1978} point out that an optically thin upper layer exists above the photosphere within the outer boundary.
The optical thickness of the Hill sphere is estimated at
\begin{equation}
	\tau_H = \kappa \rhodisk \rhill \simeq 9 \ \kappa \ (\qtot / \mear)^{1/3} (\dmean / 10^{-7} \hbox{ g cm}^{-3})^{1/3}.
	\label{eq:opt-depth_hill}
\end{equation}
If the dust is depleted in the protoplanetary disk, then, a large part of the Hill sphere can be optically thin until the protoplanet grows significantly massive.
Such a layer may also be regarded as isothermal through which the density increases with the weight of overlying layer.

In the inner layer, on the other hand, the temperature gradient increases with the density, and under radiative equilibrium, the thermal structure converges to the radiative zero-boundary solution \citep[e.g.,][]{Mizuno1980, Stevenson1982,Rafikov2006}.
This can be seen from the solution of radiative equilibrium;
\begin{eqnarray}
	&& \left( \frac{T}{\tdisk} \right)^{4+s+t}-1=\frac{4+s+t}{1+t}\left( \frac{\lacc/M_r}{\lchar/\qchar} \right) \left[ \left( \frac{P}{\pdisk} \right)^{1+t}-1 \right],
	\label{eq:tprelation_radiative}
	\label{eq:interior_condition_t}
\end{eqnarray}
derived by approximating the opacity to a power law function of $\kappa \propto T^{-s} \rho^t$ and neglecting the variations of mass and luminosity through the envelope \citep[e.g., ][]{Fujimoto1977}.
In the sufficiently deep interior of $P \gg \pdisk$, therefore, it converges to the radiative zero boundary solution of
\begin{eqnarray}
	&&\frac{a}{3}T^4=\frac{4+s+t}{4(1+t)}\left( \frac{\kappa \lacc}{4\pi G c \qtot} \right) P,
	\label{eq:interior_radiative_zero}
\end{eqnarray}
and the local polytrope decreases to approach a constant value of
\begin{eqnarray}
	&&N = \frac{1}{\vrad} - 1 =  (3+s)/(1+t). \label{eq:pindex_deep_t}
	\label{eq:poly_index_zero}
\end{eqnarray}
As a corollary, the structure of the inner part may well approximate to the polytrope with the index $N=3$ when the opacity is regarded as constant, and to a soft polytrope of $N = 3.25$ for the Kramers opacity ($s=3.5$, $t=1$).
This holds good as long as the radiative region extends over several pressure scale heights.

In the interior of protoplanet envelope, however, convection may appear when the temperature is high enough for the dissociation of ${\rm H}_2$ molecules and for the ionization of hydrogen, which decreases the adiabatic temperature gradient, $\vad = d \log T / d \log P \vert_{\rm ad}$, below the radiative temperature gradient.
In addition, the convection may persist inside these regions because of concomitant increase of opacity due to H$^-$ and bound-free and free-free transitions of hydrogen atoms
\citep[e.g.,][]{Mizuno_Nakazawa_Hayashi1978}.
The convection affects the local polytropic index in two opposite ways.
In the regions of dissociation or the ionization, $\vad$ decreases to give the local polytropic index even larger than $N_{\ad} > 3$, which also brings about a similar effect to the isothermal structure, i.e., the variation of the density with little change in the temperature.
In the convective zone below these layers, the local polytropic index decreases downs to $N=1.5$ and 2.5, which corresponds to the adiabatic temperature gradients of $\vad = 2/5$ and $2/7$, respectively.

Consequently, we may approximate the structure of the envelope by two composite polytropes, an isothermal outer layer and an inner layer of finite polytropic index, $\nbe$, in the similar way to \citet{Mizuno_Nakazawa_Hayashi1978}.
It has been demonstrated that the existence of an outer isothermal layer decreases the critical core mass \citep[e.g.,][]{Mizuno_Nakazawa_Hayashi1978}.
In an outer isothermal layer, the entropy decreases inwards with the increase in the density, which embodies cooling in the envelope.
In the following, we examine the effects of the overlying isothermal layer on the structure of the envelope to investigate the dependence of the critical core mass on its width, or the degree of cooling in the envelope.

\begin{figure}
	\resizebox{\textwidth}{!}{\plottwo{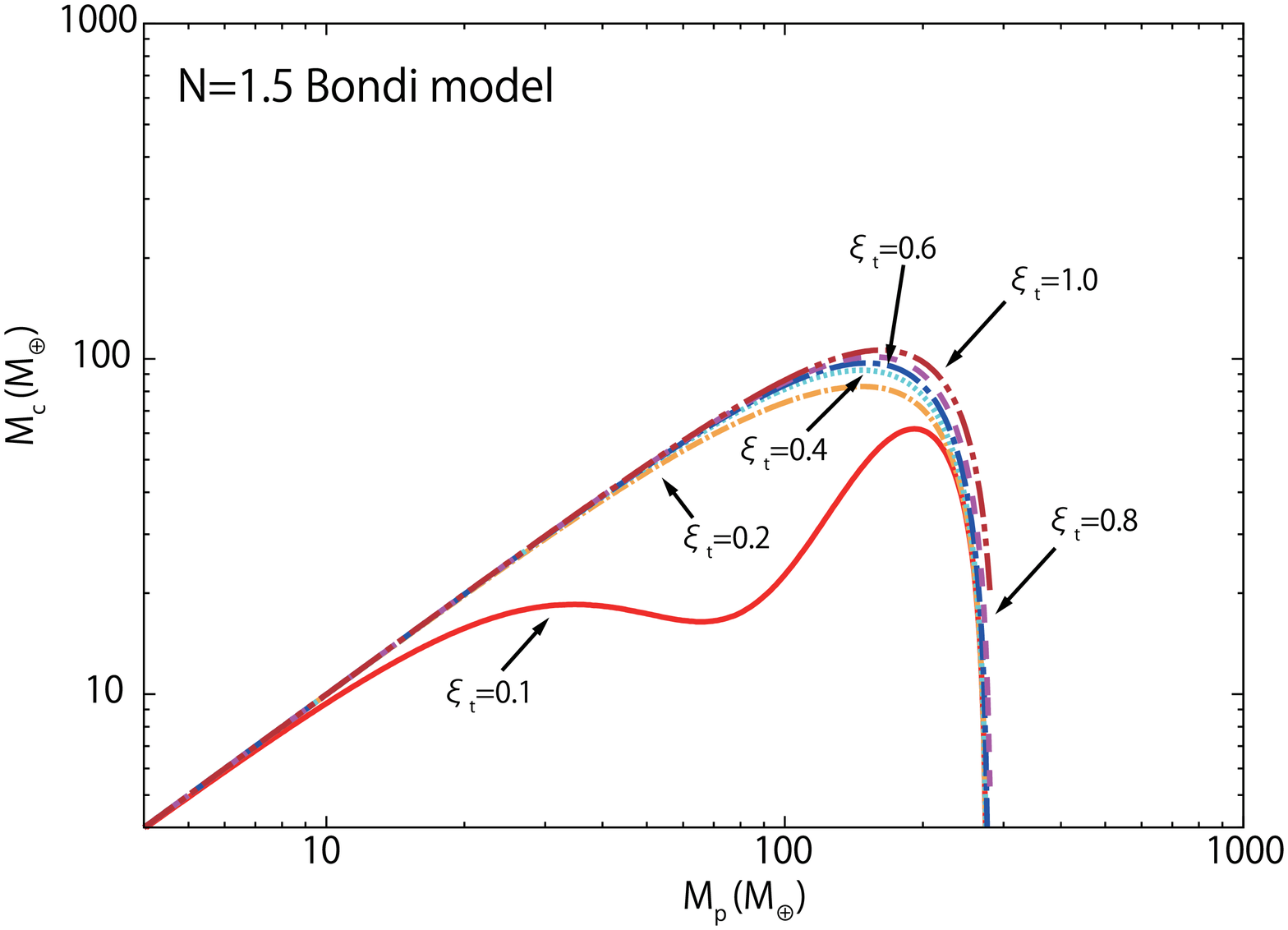}{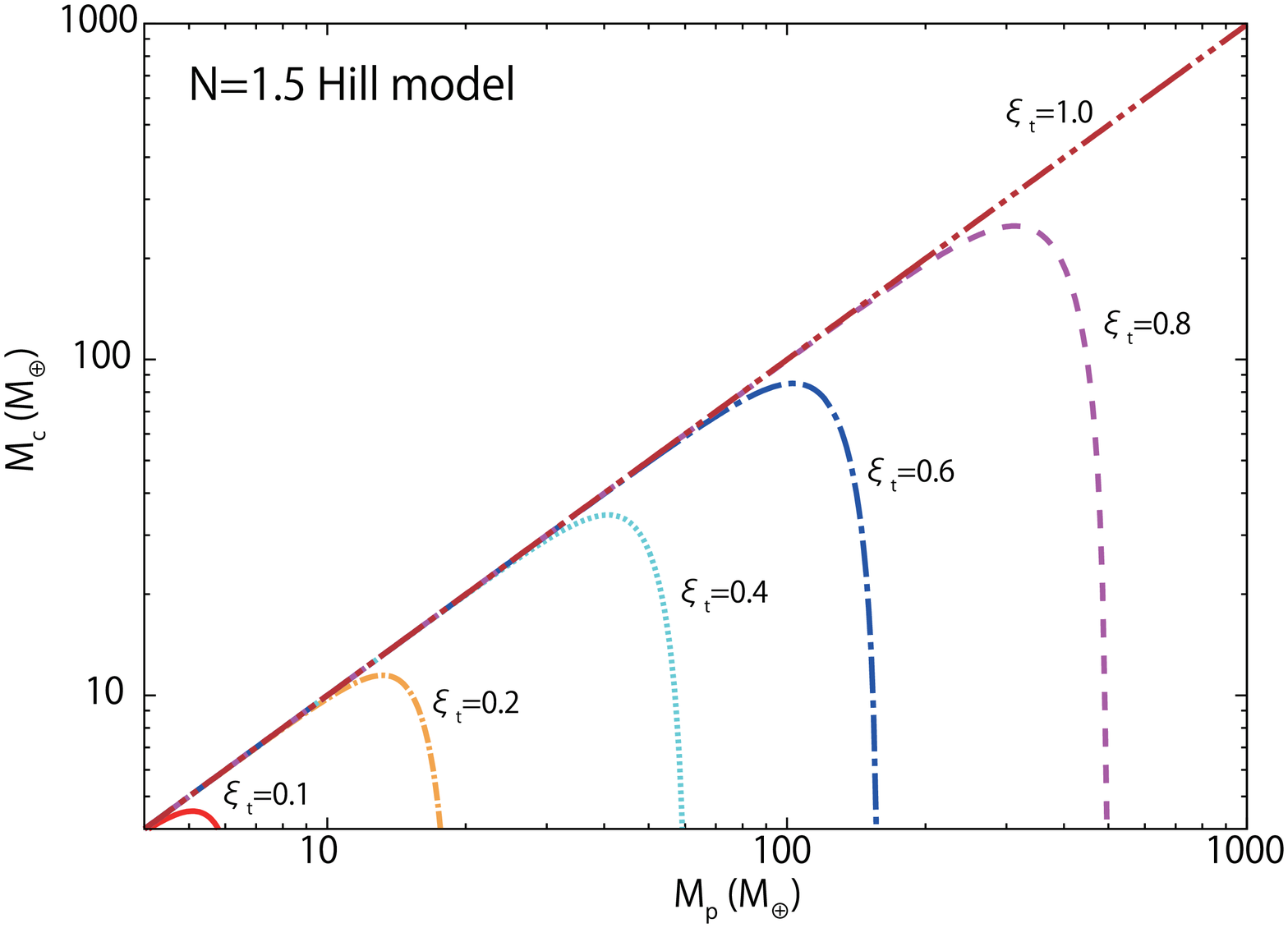}}
	\resizebox{\textwidth}{!}{\plottwo{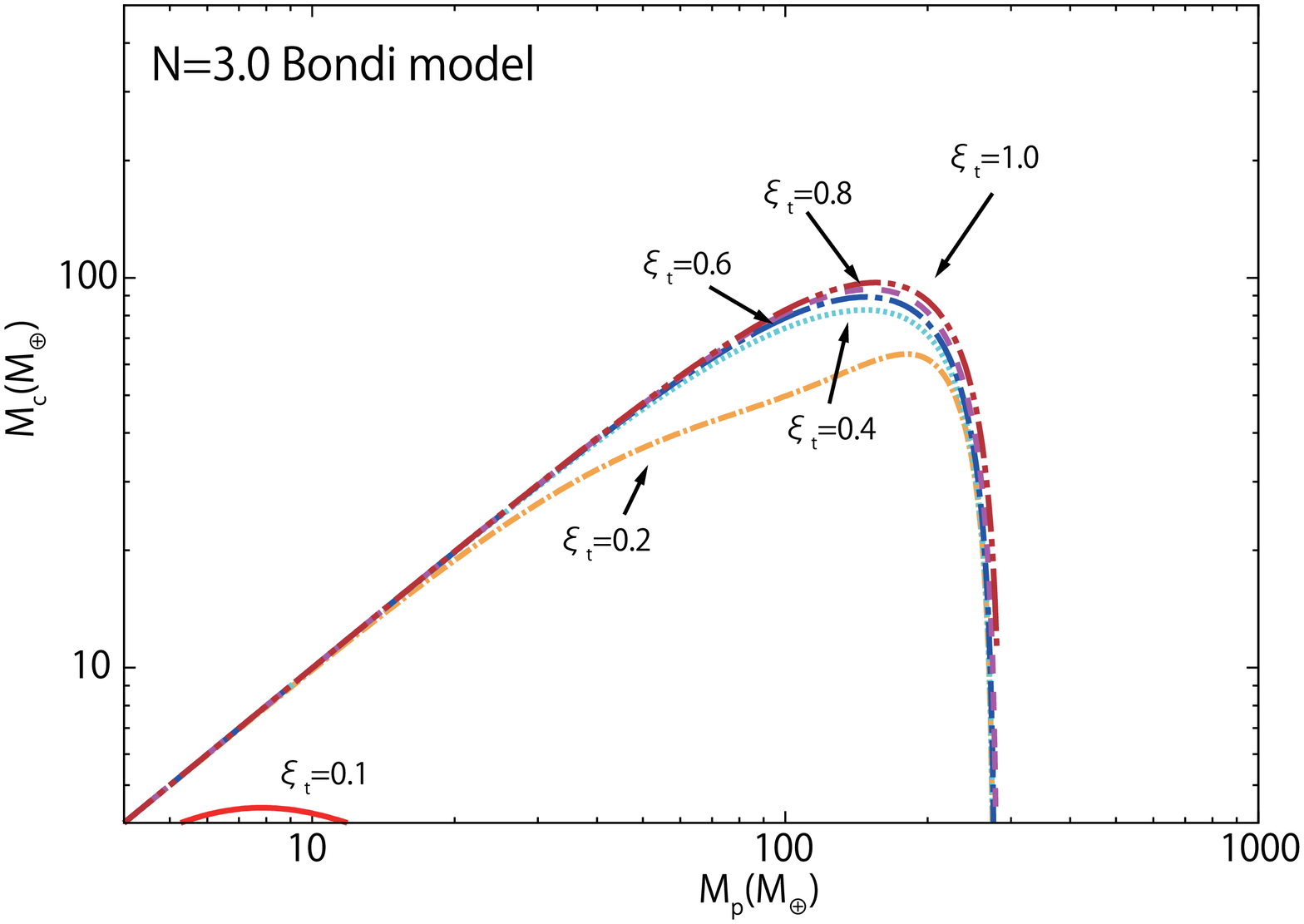}{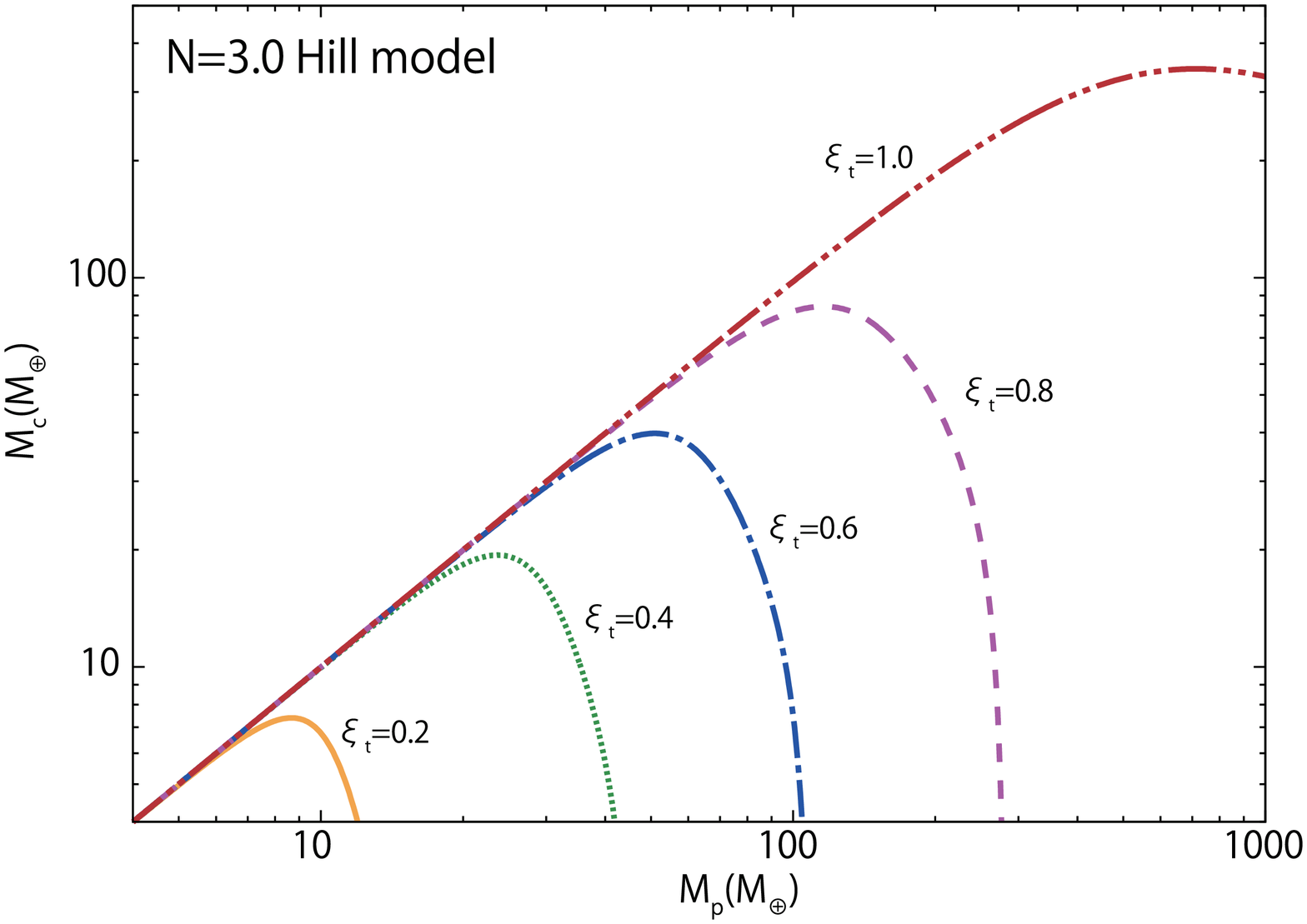}}
	\resizebox{\textwidth}{!}{\plottwo{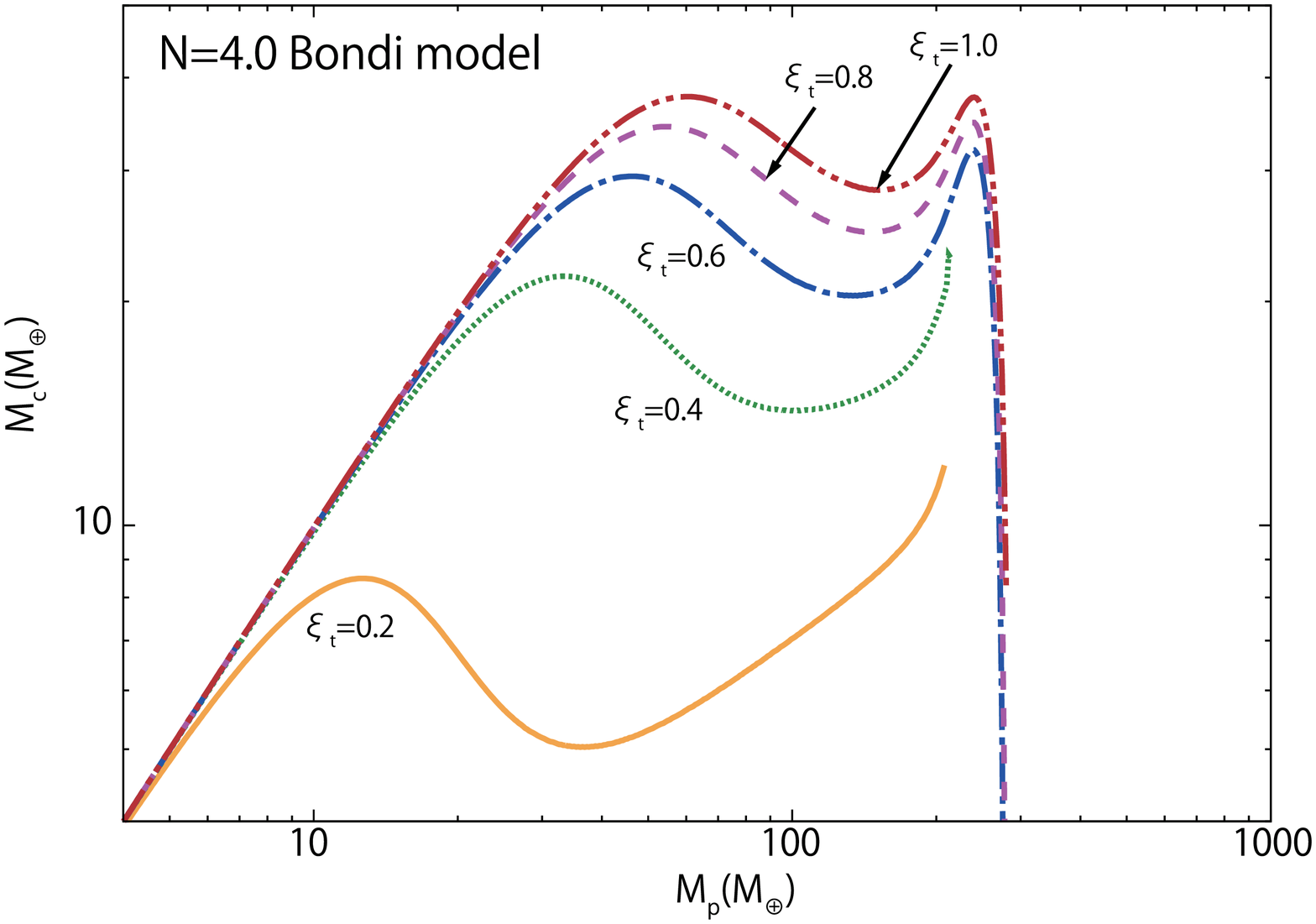}{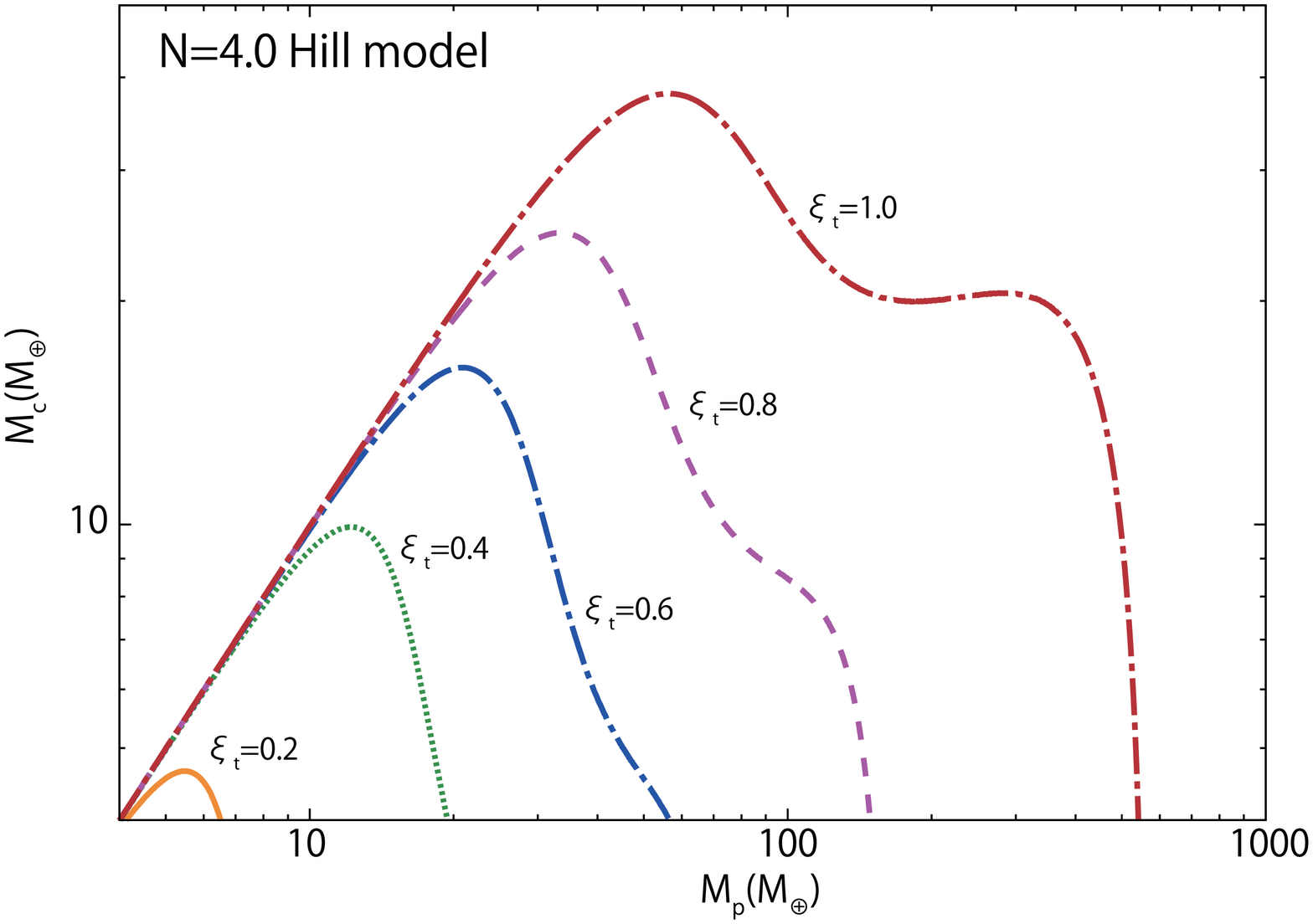}}
\caption{
The core mass, $\qcore$, as a function of the total mass, $\qtot$, for the composite polytrope model with an outer isothermal layer.
Left and right panels show the models with Bondi and Hill radius, and upper, middle and lower panels do the models with the inner polytrope of index $N = 1.5$, 3.0, and 4.0, respectively.
Model parameter, $\xit = r_{t} / \rtot$, is attached to each curve, which denotes the ratio of the transition radius, $r_t$, between the interior of polytrope of finite index and the outer isothermal layer to the planet radius.
}
\label{fig:qtot_vs_qcore_with_isothermal_layer}
\end{figure}

\begin{figure}
	\resizebox{\textwidth}{!}{\plottwo{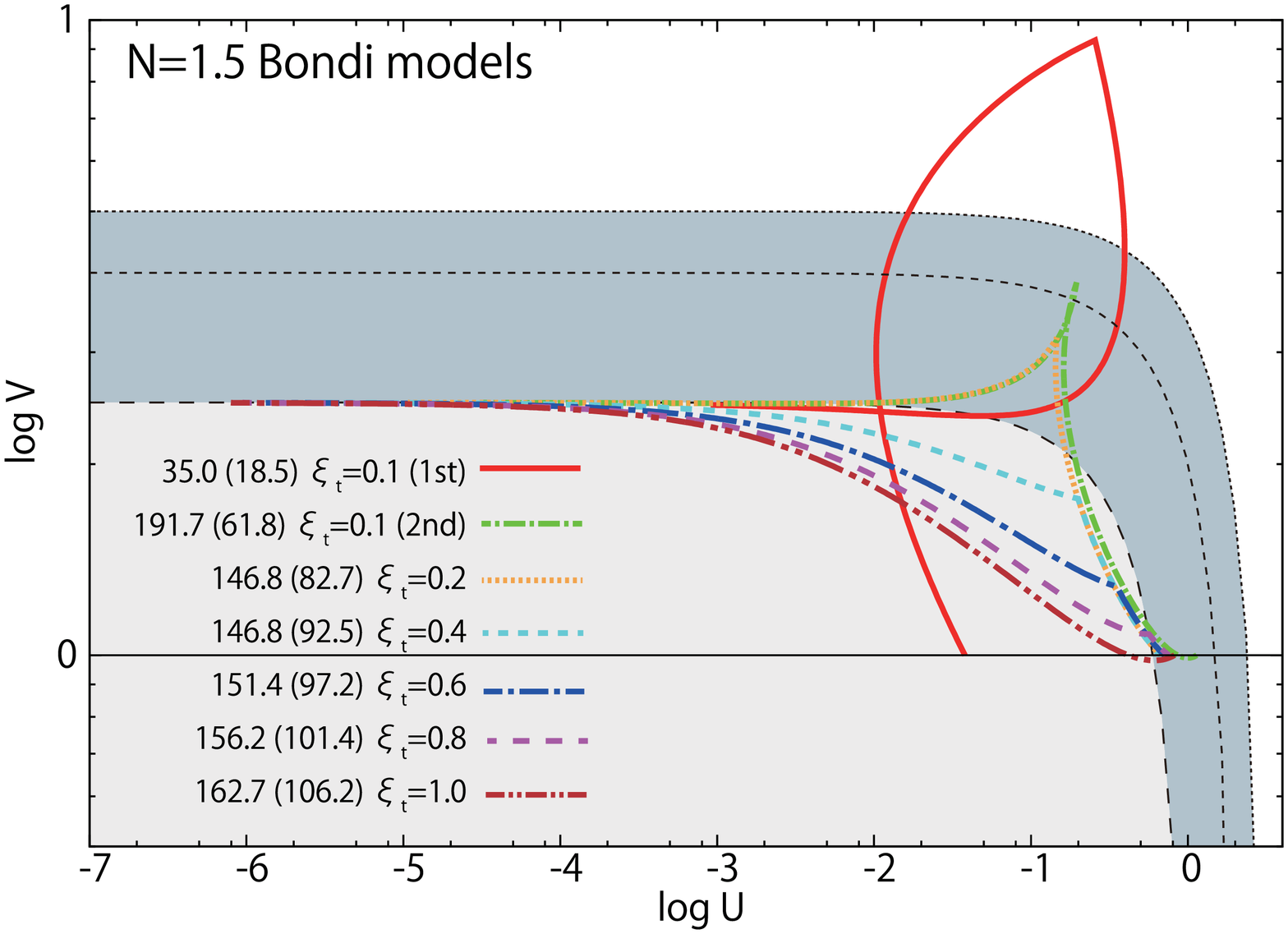}{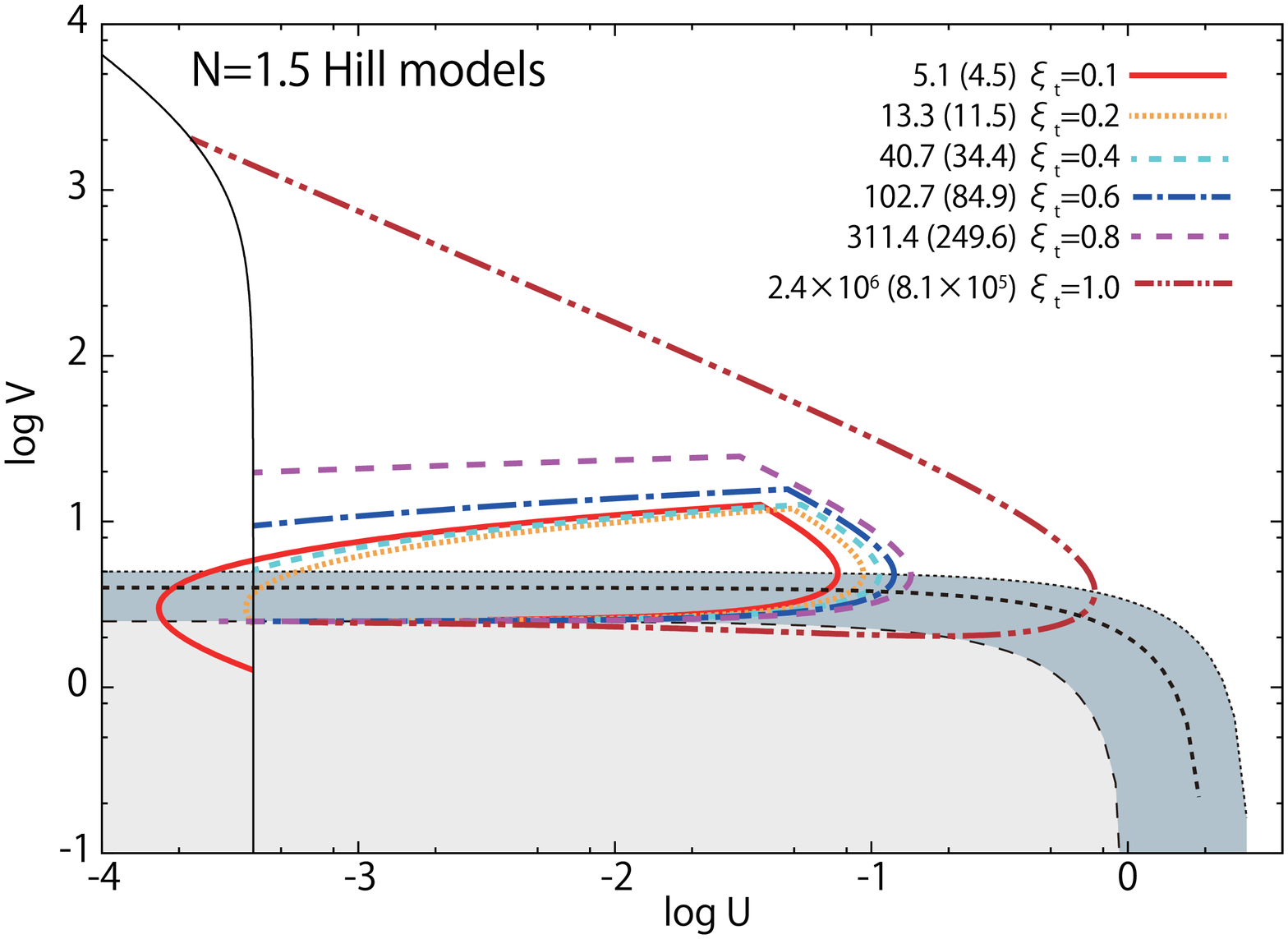}}
	\resizebox{\textwidth}{!}{\plottwo{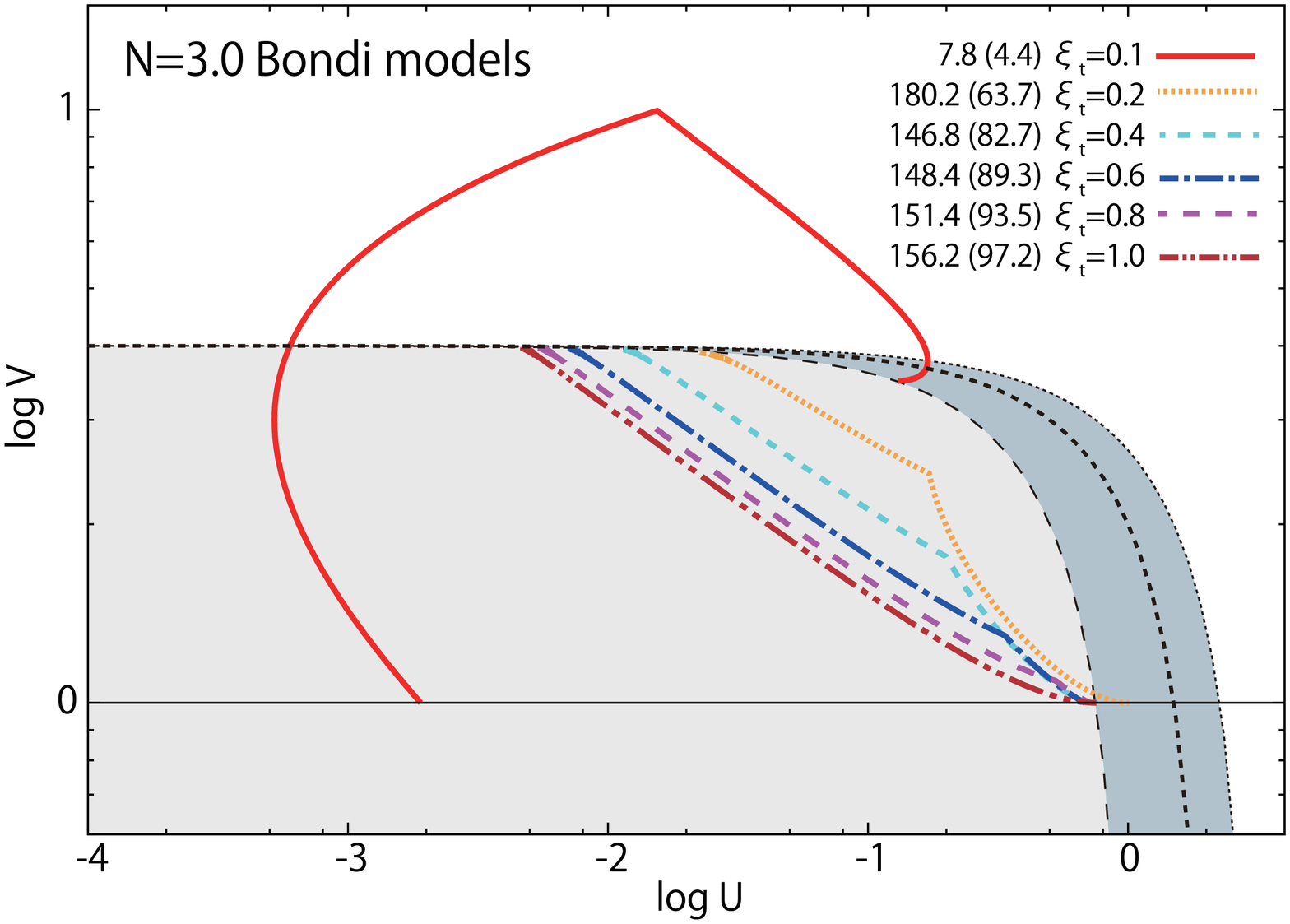}{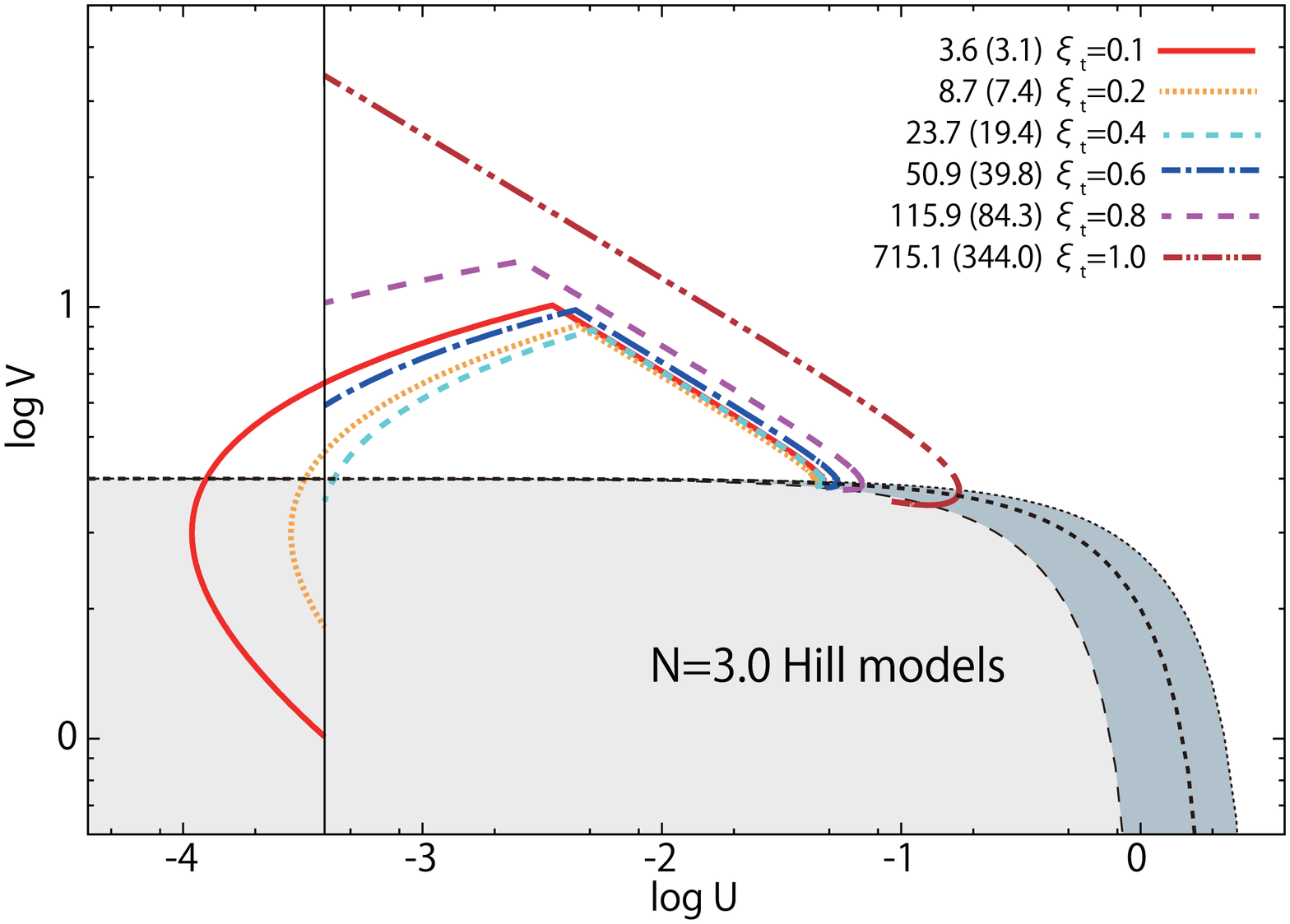}}
	\resizebox{\textwidth}{!}{\plottwo{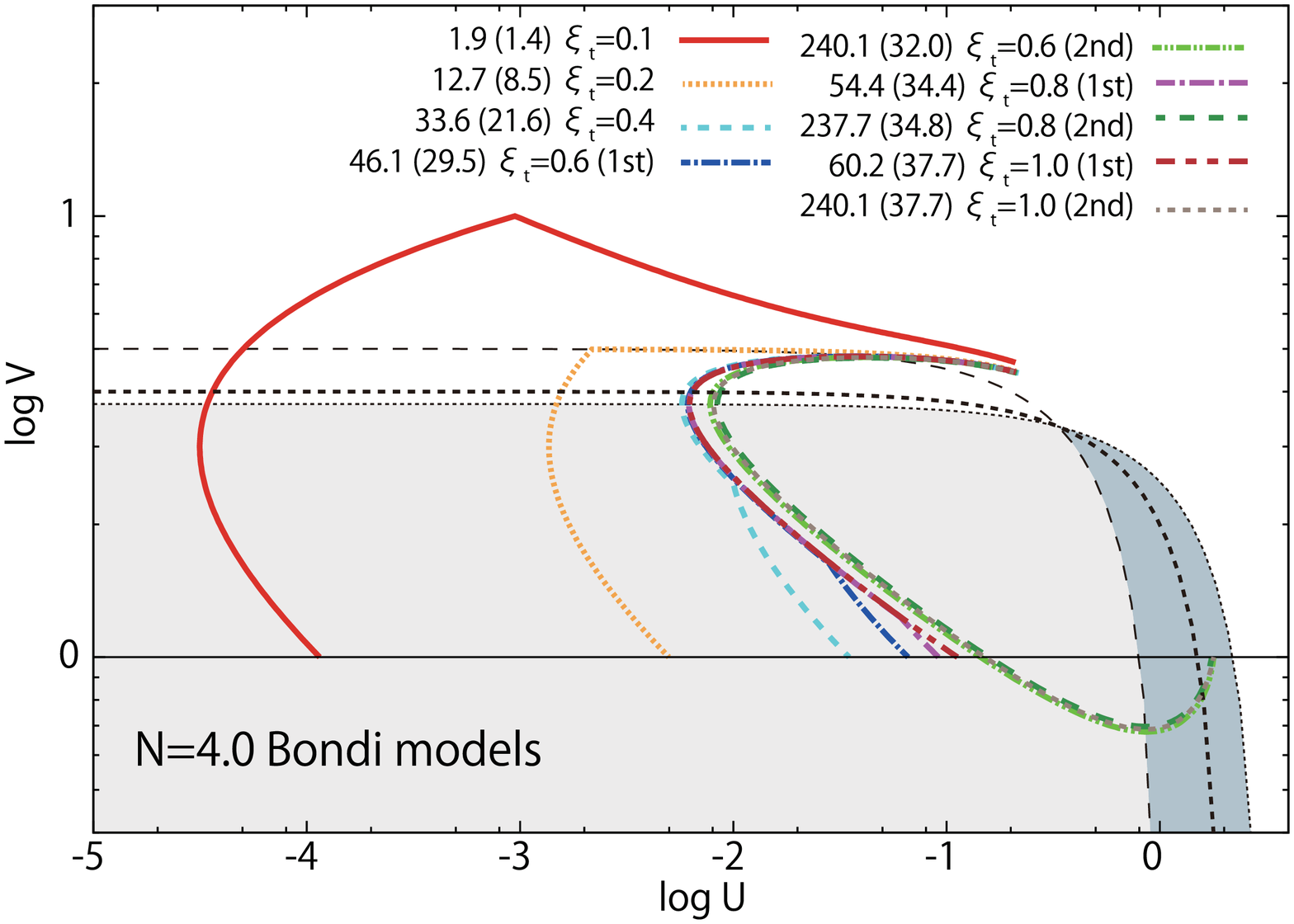}{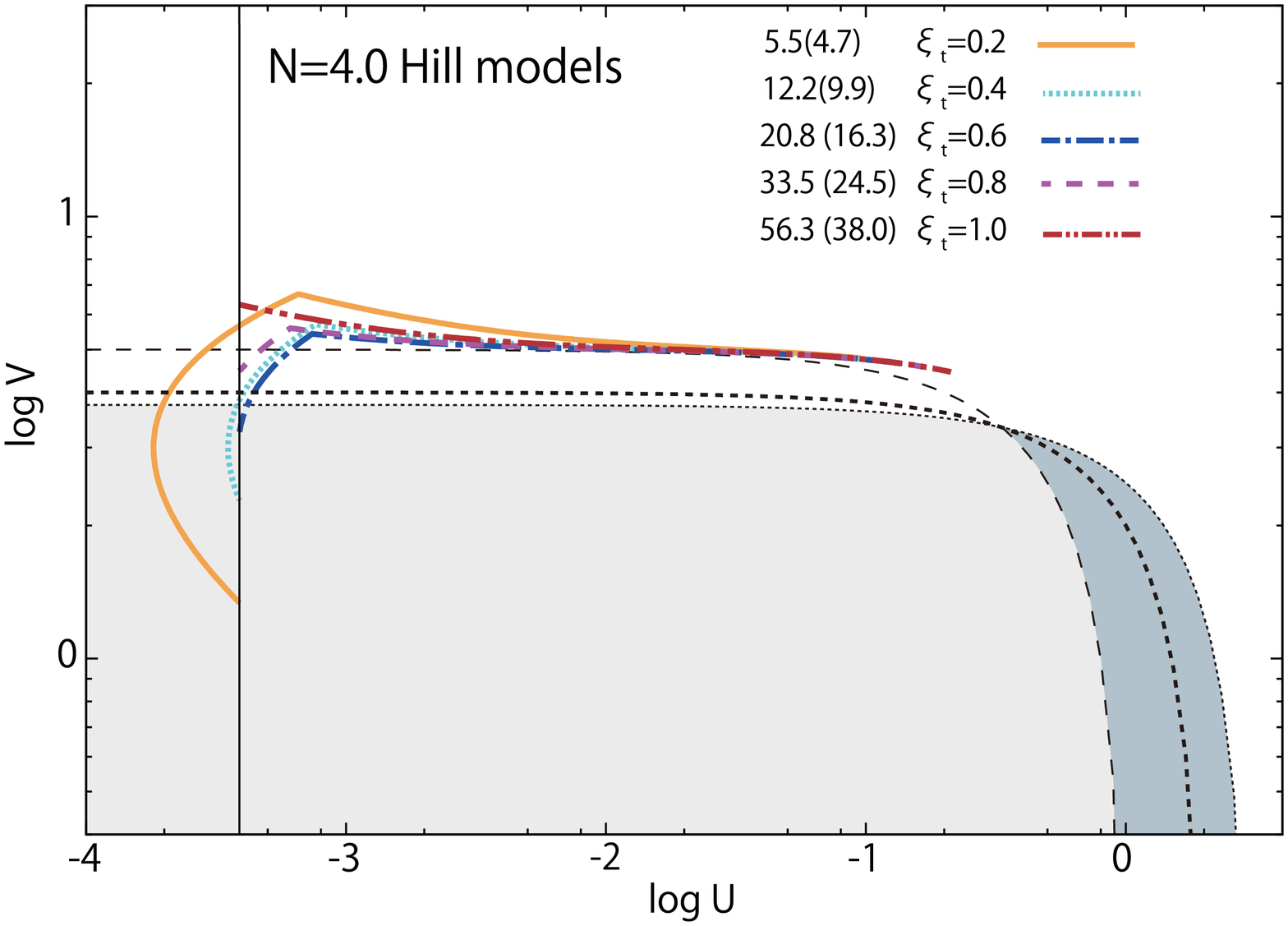}}
\caption{Behaviors on the characteristic plane for the critical models with an isothermal outer layer of different thickness. }
\label{fig:uv_with_isothermal_layer}
\end{figure}


\begin{table*}
	\begin{center}
		\caption{Properties of critical models with an isothermal outer layer \label{tab:isothermal_model}}
		\begin{tabular}{clccccccc}
			\tableline\tableline
			$N$ & $\xit$ & $\log(\ucrit)$ & $\log(\vcrit)$ & $\qtot^{\crit}(\mear)$ &
			\multicolumn{1}{c}{$\qcore^{\crit}(\mear)$\tablenotemark{a}}& $\log(\rhot/\rhodisk) $ & $T_{1,e}/\mu$ $(10^{4}K)$\\
			\tableline
\multicolumn{8}{c}{Bondi boundary condition} \\
			\tableline
			1.5 & 0.1 (1st)& -3.04 & 0.39 & 36.0  & 18.5 (0.53)& 3.72 & $ 2.14$ \\
			1.5 & 0.1 (2nd)& -4.51 & 0.40 & 191.7 & 61.8 (0.32)& 1.82 & $ 4.82$ \\
			1.5 & 0.2      & -4.97 & 0.40 & 146.8 & 82.7 (0.56)& 1.20 & $ 5.87$ \\
			1.5 & 0.4      & -5.63 & 0.40 & 146.8 & 92.5 (0.63)& 0.50 & $ 6.21$ \\
			1.5 & 0.6      & -5.88 & 0.40 & 151.4 & 97.2 (0.64)& 0.23 & $ 6.42$ \\
			1.5 & 0.8      & -6.03 & 0.40 & 156.2 & 101.4(0.65)& 0.10 & $ 6.79$ \\
\hdashline
			3.0 & 0.1      & -0.88 & 0.54 & 7.84 & 4.37  (0.56)& 3.88 & $ 0.43$ \\
			3.0 & 0.2      & -1.64 & 0.59 & 180.2& 63.74 (0.35)& 0.99 & $ 0.77$ \\
			3.0 & 0.4      & -1.94 & 0.60 & 146.9& 82.70 (0.56)& 0.52 & $ 1.46$ \\
			3.0 & 0.6      & -2.15 & 0.60 & 148.4& 89.29 (0.60)& 0.25 & $ 2.38$ \\
			3.0 & 0.8      & -2.26 & 0.60 & 151.4& 93.48 (0.62)& 0.09 & $ 4.00$ \\
\hline
			4.0 & 0.1      & -0.68 & 0.67 & 1.93  & 1.37 (0.71) & 3.92 & $ 0.24$ \\
			4.0 & 0.2      & -0.67 & 0.65 & 12.7  & 8.5  (0.67) & 1.73 & $ 0.83$ \\
			4.0 & 0.4      & -0.66 & 0.64 & 33.6  & 21.6 (0.64) & 0.64 & $ 1.60$ \\
			4.0 & 0.6 (1st)& -0.68 & 0.65 & 46.1  & 29.5 (0.64) & 0.28 & $ 1.91$ \\
			4.0 & 0.6 (2nd)& -0.72 & 0.65 & 240.1 & 32.0 (0.13) & 0.19 & $ 1.96$ \\
			4.0 & 0.8 (1st)& -0.68 & 0.65 & 54.4  & 34.4 (0.63) & 0.11 & $ 2.14$ \\
			4.0 & 0.8 (2nd)& -0.67 & 0.64 & 237.7 & 34.8 (0.15) & 0.08 & $ 2.18$ \\
			\tableline
\multicolumn{8}{c}{Hill boundary condition} \\
			\tableline
			1.5 & 0.1 & -2.50 & 0.41 & 5.08  & 4.52  (0.89) & 4.84 & $ 0.81$  \\
			1.5 & 0.2 & -2.86 & 0.40 & 13.3  & 11.5  (0.86) & 4.20 & $ 1.53$  \\
			1.5 & 0.4 & -3.25 & 0.40 & 40.8  & 34.4  (0.85) & 3.20 & $ 3.20$  \\
			1.5 & 0.6 & -3.44 & 0.40 & 102.7 & 84.9  (0.83) & 2.67 & $ 5.85$  \\
			1.5 & 0.8 & -3.54 & 0.40 & 311.4 & 249.6 (0.80) & 2.19 & $ 12.3$  \\
\hdashline
			3.0 & 0.1 & -1.32 & 0.60 & 3.60  & 3.15 (0.87)& 3.90 & $ 0.63$ \\
			3.0 & 0.2 & -1.36 & 0.59 & 8.67  & 7.40 (0.85)& 3.15 & $ 3.20$ \\
			3.0 & 0.4 & -1.35 & 0.58 & 23.7  & 19.4 (0.82)& 2.25 & $ 3.76$ \\
			3.0 & 0.6 & -1.32 & 0.58 & 50.9  & 39.8 (0.78)& 1.64 & $ 3.88$ \\
			3.0 & 0.8 & -1.24 & 0.58 & 115.9 & 84.4 (0.73)& 1.11 & $ 3.95$ \\
\hline
			4.0 & 0.2 & -1.02 & 0.68 & 5.48  & 4.66 (0.85) & 2.29 & $ 0.48$ \\
			4.0 & 0.4 & -0.93 & 0.67 & 12.2  & 9.93 (0.81) & 1.49 & $ 0.81$ \\
			4.0 & 0.6 & -0.87 & 0.67 & 20.9  & 16.3 (0.78) & 0.94 & $ 1.16$ \\
			4.0 & 0.8 & -0.78 & 0.66 & 33.5  & 24.7 (0.74) & 0.43 & $ 1.60$ \\
			\tableline
			\tableline
		\end{tabular}
		\tablenotetext{a}{the value in brackets denotes a ratio of $\qcore$ and $\qtot$}
	\end{center}
\end{table*}

\subsection{Roles of the overlying isothermal layer} 
\label{sec:isothermal_layer}

We will study the properties of models, composed of two polytropes, an outer isothermal layer of $N = \infty$ and an inner shells with a polytrope of finite index, $\nbe$.
Here we regard the transition radius, $r_t$, between two layers as a parameter for simplicity, and specify it by the radius ratio, $ \xit$, to planetary radius ($ = r_{t} / \rtot$).
We compute a series of models by varying the transition radius down to $\xit = 0.1$ at interval of 0.1 dex for both stiff and soft inner polytropes of $\nbe=1.5$, 3 and 4, and for Bondi and Hill boundary radius.
Figure~\ref{fig:qtot_vs_qcore_with_isothermal_layer} shows the resultant relationship between the core and total mass, and the characteristics of the models with the maximum core mass are summarized in Tables~\ref{tab:isothermal_model}.
In general, the critical core mass decreases as the transition radius ratio grows small or the fraction of overlying isothermal layer grows large.
For the smallest transition radius of $\xit = 0.1$ computed, the critical core mass decreases below $\qcore^{\rm crit} = \hbox{ a few - several }\mear$, to be as small as, or smaller than, the smallest critical core mass obtained for the realistic models, except for Bondi models of $\nbe = 1.5$.

In the case of the stiff inner polytropes with the Bondi radius, the reduction remains small for $\xit \gtrsim 0.2$ and grows significant only below $\xit = 0.1$.
In the stiff inner polytrope with the Hill radius, a small fraction of outer isothermal layer has a conspicuous effect and the reduction of critical core mass is as large as by factors of $\sim 300$ and 4 times between $\xit = 0.8$ and 1.0 for $\nbe = 1.5$ and 3, respectively.
As $\xit$ decreases further and the fraction of isothermal layer grows thick, the critical core mass diminishes steadily, to be smaller than for the Bondi models with the same $\xit$, though the reduction rate somewhat decreases for smaller $\xit$
In the case of the soft inner polytrope of $ \nbe = 4$, the effect of isothermal layer is more or less constant, and yet, a different tendency is discernible between the Bondi and Hill models similar to the models of $\nbe \le 3$.

Figure~\ref{fig:uv_with_isothermal_layer} shows the behaviors of critical models on the $U $-$ V$ plain.
For all of the models, their inner structure is described by the centrally-condensed type solutions with their polytropic index, $\nbe$, as in the case for their single polytrope, indifferent to thickness of the isothermal layer.
In particular, it holds to a good approximation that $\voe = \nbe + 1$ for $N \le 3$ and also for $\nbe > 3$.

The overlying isothermal layer plays two roles in determining the critical core mass.
One is to decrease the polytropic constant, or the cooling of gaseous envelope, as discussed by \cite{Mizuno_Nakazawa_Hayashi1978}.
The effect of the cooling by the isothermal layer is related to the increase in density through the outer isothermal layer as
\begin{eqnarray}
	\log(\kt/\kd)&=&-\left( 1/\nbe \right)\log\left( \rhot/\rhodisk \right)\nonumber\\
	&=& -\left( 1/\nbe \right) \int_{\xit}^{1}Vd\log \xi.
\label{eq:rhot}
\end{eqnarray}
The cooling effects is larger for larger value of $\vsurf$, and hence, for Hill models than for Bondi models.
A smaller $\kt$ implies that a larger density is necessary to produce the required pressure for hydrostatic equilibrium, and hence, enhances the contribution of the gaseous component to the gravity.
Along with it, mass of the core decreases as
\begin{eqnarray}
	&&\qcore = \left( \frac{3\rhodisk}{\rhocore} \right)^{(N-3)/2N} \left( \voe \uoe^{1/N} \right)^{3/2} \left( \frac{\kt}{\kd} \right)^{3/2} \qchar, \label{eq:ccm_composite}
\end{eqnarray}
for given values of $\uoe$ and $\voe$.
For the critical core mass, we should take into account the effect on the value of $\ucrit$, in particular for small $\nbe$, as discussed below.

The other effect is that the outer isothermal layer enables the structure line to cross the horizontal and critical line as a result of the inward decrease in its thermal energy relative to the gravitational energy.
In the isothermal layer, the value of $V$ increases inwards for $U < 1$ and the value of $V$ at the transition shell is given by
\begin{eqnarray}
	&&\log(\vt/\vsurf)=\int_{\xit}^{1}(1-U)d\log \xi.\label{eq:vt}
\label{eq:effect_v_isothermal}
\end{eqnarray}
For Bondi models, the structure line of the isothermal part moves above the critical line if the isothermal layer grows thicker than $\xit \simeq 4^{- 1 / \langle 1 - U \rangle}$, which may change the nature of solution.
For Hill models, $\vsurf$ can enter below the critical line and horizontal line, starting from the structure with the inner edge above the horizontal line with the$\vt$ above the critical line, and hence, without changing the nature of internal structure.

As long as the transition shell remains below the critical line, Bondi models retain the same characteristics of envelope structure as single polytrope models, as seen in Fig.~\ref{fig:uv_with_isothermal_layer} (left top and middle panels).
The occurrence of the maximum core mass is regulated by the outer boundary in the same way as discussed above.
The critical models decrease only slightly with the reduction of $\xit$ since the effect of a decrease in $\kt$ is largely offset by increase in $\ucrit$;
the structure line draws closer to the critical line in the isothermal layer, as compared with single polytrope models with the same polytrope as the inner layer, and hence, needs to run shorter distance along the horizontal line to satisfy the condition of radius distance.
As the isothermal layer grows sufficiently thick, on the other hand, structure lines of the isothermal part may extend beyond the critical line, as seen from models of $\xit = 0.1$.
Then models can adopt the envelope structures with their inner edge above the horizontal line, and behaves similar to the Hill models.
Consequently, the critical model comes to be determined by its internal structure.
In this case, the structure line in the isothermal part runs nearly at right angle with the critical line and the effect on $\uoe$ remains relatively small, and hence, the critical core mass decreases largely as the result of decrease in $\kt$.

For Hill models of $N \le 3$, the upper isothermal layer has a significant quantitative effect, even for a small fraction of overlying isothermal layer.
The large decrease of $\vsurf$ between the models of $\xit = 0.8$ and 1.0 much exceeds the difference between $\vsurf$ and $\vt$ in eq.~(\ref{eq:effect_v_isothermal}) which proportional to $\xit$ for $U \ll 1$.
Rather it is mostly attributable to the difference in the slope of structure lines between the isothermal and finite index polytropes;
the isothermal structure line has a positive slope of $d $log V / d $log U \simeq 1/V$ for $V \gg N + 1$, while the polytrope of finite index has a negative slope of $d $log V / d $log U \simeq -1/N$, the steeper for the stiffer polytrope, because of the outward decrease in the temperature. 
For sufficiently smaller $\xit$, the ratio of $ \vsurf / \vt $ decrease and enables $\vsurf$ to enter into below the horizontal line, differently from the single polytrope models.
Accordingly, the total planet mass of critical models, and hence, the critical core mass, can be smaller than
$\qtot^{\crit} < (N+1) (\rhodisk / \dmean)^{1/2} \qchar$, which is below $(N+1) 3.5 \mear$ for our parameters.
Even in this case, the transition shell remains above the horizontal line so that the nature of solution is kept unchanged.
Accordingly, the decrease in the critical core mass is attributed mainly to the decrease in the $\kt$ in eq.~(\ref{eq:ccm_composite});
variations of $\uoe$ remains relatively small, though larger for stiffer polytropes, since the decrease of $\kt/\kd$ occurs in the outer layer, and hence, the structure line runs more distant from the critical line and only weakly affects the value of $\uoe$, which is determined from the condition of the radius ratio. 

For the polytrope of $N > 3$, the behavior of models on the $U$~-~$V$ diagram is almost same as for the single polytrope models, except for the steep inward increases of $V$, or of density, relative to temperature in the upper isothermal part.
Consequently, the total mass and core mass decrease.
As in the case of single polytrope models, the interior structure of envelope little differs between Bondi and Hill models.
Since the duplicity (or multiplicity) of configuration stems from the inner structure, the overlying isothermal layer little affects the structural characteristics;
as in the case of stiff polytropes, the thickest upper isothermal layer allows Bondi model to enter into the region above $V > N+1$.
Thus, resultant variations of $\ucrit$ are rather small for different thickness of the isothermal layer, remaining almost unchanged for Bondi models while decreasing by a factor of $\sim 2$ for Hill Models.

In summary, the overlying isothermal layer in generally decrease the critical core mass through the decrease in the thermal energy, or the decrease in the polytropic index, $\kt$, which represents the effect of cooling of envelope.
Despite the modification of surface structure, the structural characteristics are kept basically unchanged, and the criteria are determined by the polytropic index, $\nbe$, in the bottom of envelope except for the Bondi models of the stiff inner polytropes of $\nbe \le 3$ with a thick isothermal layer.
In particular, for the Hill boundary of $\nbe \le 3$, the reduction of the critical core mass is conspicuous because of the largest $\vsurf$ in a single polytrope models;
for sufficient thick isothermal layer, $\vsurf$ enters below the horizontal line and enables a small critical core mass of order of the earth mass.
On the other hand, the surface isothermal layer has rather small effect on the Bondi models of $\nbe \le 3$ except for sufficiently thick isothermal layer.
In the latter case, the Bondi models behave like the Hill models, yielding a large decrease in the critical core mass.
The present results suggest that the cooling of envelope is essential in determining the critical core mass, and the critical core mass reduces to be smaller as the cooling is more effective in the envelope.

\section{Discussion} \label{sec:Discussion}
\subsection{Parameter dependence and comparisons with former works}

A salient feature of the centrally-condensed type solutions is the relationship of $\voe = \nbe + 1$ at the inner edge of envelope, which characterizes the density distribution in power-law around the core.
This holds to a good approximation not only for the inner layer of stiff polytropes but also for the soft polytropes, although we should into account the contribution of $\uoe$ for the latter.
Since it represents the balance between the thermal energy and the gravitational energy, this connect the temperature (divided by mean molecular weight) at the inner edge of envelope to the gravitational energy of core by
\begin{eqnarray}
	\qcore &=& \left( \frac{3}{4 \pi \rhocore} \right)^{1/2} \left( \frac{ \nbe + 1} {G} \ \frac{ \bolz }{ \matm } \right)^{3/2} \left( \frac{ \toe }{ \mu_{\be} } \right)^{3/2}.
	\label{eq:core_mass_t1e_crit}
\end{eqnarray}
In other words, the temperature at the bottom of envelope is given as a function of core mass and turns out to be an order of $\toe / \mu_{\be} \simeq 10000$ K for $\qcore \simeq 10 \mear$.
In addition, once $\ucrit$ is specified, the critical core mass follows from eq.~(\ref{eq:ccm_composite}), and is written with the polytropic constant in the inner shells, as
\begin{eqnarray}
	\qcore^{\crit} &=& \left( \frac{1}{4 \pi}\right)^{1/2} \left( \frac{\rhocore}{3} \right)^{-1/ 2} \left[ \frac{ \nbe + 1}{G} \, (\ucrit \rhocore / 3) ^{{1} / { \nbe}} \, {\kt} \right] ^{3/2}.
	\label{eq:core_mass_pc1e_crit}
\end{eqnarray}
The values of $\ucrit$ depend on the stiffness of polytropes and on the outer boundaries, and also, on the thickness of thermal structure, or the degree of cooling, in the upper layer.
The latter variation decreases with the polytropic index and the dependence grows weak or vanishes for $\nbe \gtrsim 3$ except for the Bondi models of stiff polytropes.
This relation indicates that given the condition in the protoplanetary disk, the cooling in the envelope is essential to determine the critical core mass.
The critical core mass is smaller for smaller polytropic constant, $\kt$, in the bottom of envelope, although it is partly compensated by the variation of $\ucrit$ due to an increase in the density, $\rhooe$.
The composite models with an isothermal and/or optically-thin upper layers or with the layer of larger local polytropic index than $N > 3$ decrease $\toe / \mu_{\be}$ and $\kt$ as compared with the single polytrope models of the same index, $\nbe$, giving rise to smaller critical core mass.

Parameter dependences of the critical core mass have been discussed by previous works not only with the numerical computations but also with recourse to various analytical approximations.
In particular, \citet{Stevenson1982} derive the critical core mass for a radiative envelope by applying the radiative zero-boundary solution with a constant opacity.
\citet{Wuchterl1993}, on the other hand, considers the critical core mass for a wholly convective envelope.
We compare our and their results, derived from our polytrope models, with the former works and elucidate relevance of our criteria for the critical core masses.

For radiative envelopes, the radiative zero-boundary solution applies for the bottom of the envelope where $d M_r /d \log P \vert_{\be} = (U/V)_{\be} \ll1$.
By using the polytropic index and constant in eqs.~(\ref{eq:poly_index_zero}) and (\ref{eq:interior_radiative_zero}), we may express the critical core mass in terms of model parameters as follows;
\begin{eqnarray}
	\qcore^{\crit}  = \left[ \left( \frac{1}{4 \pi} \right)^{{(2-s)}/{3}} \left( \frac{\rhocore}{3} \right)^{{(1+s+3t)}/ 3} (N+1) (\vcrit \ucrit{}^{1/N}){}^{3-s} \right. \phantom{{}^{{3}/{(7-2s)}}}& & \nonumber \\
	\left. \times \ \frac{3}{16 \pi G^{3-s} a c} \left( \frac{\bolz}{\mu_{1e} \matm} \right)^{4-s} (\kappa_{0} \dot{M}_{\solid}) \right]^{{3}/{(7-2s)}}.  & &
\label{eq:radiative_zero_qcore_crit}
\end{eqnarray}
Accordingly, the critical core mass is determined only by the properties of constituent materials (the density of solid component, and the opacity and mean molecular weight of gaseous components) and the core accretion rate, $\dot{M}_{\solid}$, irrespective of the thermal state in the upper layer and the boundary condition.
This gives the same parameter dependences as the formula derived by \cite{Stevenson1982}, for a constant opacity ($s = t = 0$).
Since the thermal structure rapidly converges to it in the interior, the radiative-zero solution is applicable as long as the radiative zone extends over sufficient pressure scale height \citep[e.g.,][]{Fujimoto1977}.
In addition, for $\nbe \gtrsim 3$, the value of $\ucrit$ is little dependent on the thickness of the outer isothermal layer.
This may explain that the critical core mass is insensitive to the outer boundary, as argued in previous works \citep{Mizuno1980,Stevenson1982,Rafikov2006}.

For a wholly convective envelope, the critical core mass is obtained by eq.~(\ref{eq:core_mass_pc1e_crit}) with $\kt$ given for the entropy in its photosphere.
As noted by \citet{Wuchterl1993}, however, we should take into account variations of adiabatic exponent due to molecular dissociation and ionization.
These phase transitions may reduce the polytropic constant, $K_{\be}$, or temperature, $\toe/\mu_{\be}$, below the values inferred for the polytrope in the photosphere.
Furthermore, the adiabatic temperature gradient may decrease below $\vad < 1/4$ to raise the polytropic index above $N_{\ad} > 3$.  
The layer of $N > 3$ plays the similar role to an isothermal layer and affects structure in the inner layer.
Both of them work to reduce the mass of the critical models.
Numerical models in his Table~1 give the total and core masses significantly smaller than the characteristic mass of $\qchar = 73.6 (\tdisk / 100 \hbox{ K})^{3/2} (\rhodisk / 10^{10} \hbox{ g cm}^{-3})^{1/2} \mear$.
In addition, some of his models have a radiative zone in the envelope, which also contributes to effective cooling and decreases the polytropic constant.

\citet{Perri_Cameron1974}, on the other hand, study the convective envelope assuming an isentropic structure in whole interior with entropy in the surrounding gas.
In actuality, their model has the region of $N_{\ad} > 3$ in the middle of the envelope that stretches over more than a factor of 10 in radius, as shown in their Fig.~2.
Between $T \simeq 2000 $-$ 10000$ K, the adiabatic exponent decreases below $\gamma < 4/3$ so that the adiabatic polytropic index reaches to $N_{\ad} \simeq 9$, while it increases and approaches to $N_{\ad} \simeq 1.5$ in the further interior..
They also find the critical core mass of mass $\sim 115 \mear$, smaller than $\qchar = 257 \mear$ for $\tdisk = 125$ K and $\pdisk = 0.08 \hbox{ dyn cm}^{-2}$).
They set the outer boundary at $\rtot = A (\qtot / \mear)^{1/2} R_{\oplus}$, which gives $\usurf = 0.0117 (A /500)^3 (\qtot /115 \mear)^{1/2}$ and $\vsurf = 2.60 (\tdisk /\hbox{125 K})^{-1} (A /500) (\qtot /115 \mear)^{1/2}$.
Since the surface lies below the critical line, their model first increases $V/U$ inwards in the outermost layer of $N_{\ad} < 3$, and then, moves upward to cross the critical line in the middle layer of $N_{\ad} > 3$.
Afterward, it turns to decrease $V/U$ inwards, and finally connects with the centrally-condensed type solution of $\nbe < 3$ in the inner part of envelope.
The intervening layer of $N_{\ad} > 3$, therefore, plays the similar role as discussed above for the outer isothermal layer to reduce the total and core masses of critical core masses.

The problem of radiative and convective envelopes is revisited by \citet{Rafikov2006}, although he has overexploited the approximation of power-law distribution, similarly to other previous authors, which is valid only for the envelope of mass much smaller than the core (see \S~\ref{ss:small_envelope,result_of_polytrope}).
The upper radiative and convective zones exert similar effects on the structure of protoplanets but differ in extent, as discussed above.
In the radiative zone, the temperature gradient is smaller than in the adiabatic one and entropy decreases inwards.
Accordingly, the envelope with the upper radiative zone gives rise to a smaller value of $\kbe$ in the bottom of envelope than the envelope with the upper convective zone.
In this case, the thermal structure tends to be determined by the luminosity and opacity under the thermal equilibrium, as discussed in \S~\ref{sec:comp_polytrope};
models with smaller opacity and/or smaller luminosity are attendant with thicker isothermal and/or optically-thin layers, and bring about smaller $\kbe$.
In convective zones, the value of $\kbe$ decreases below its surface value because of dissociation of hydrogen molecules and/or ionization of hydrogen atoms even under an isentropic structure, though the degree is smaller as compared with the radiative zone.

On the other hand, \citet{Pollack_etal1996} and \cite{Ikoma_Nakazawa_Emori2000} find that the runaway gas accretion occurs without the growth of a solid core.
This is also related to a decrease of entropy in the bottom of envelope.
In response to a decrease of entropy, the envelope is contract.
While the envelope structure is governed by the gravity of core, the rise of temperature due to the compression compensates the decrease due to energy loss so that $\toe$ is maintained to meet the conditions of centrally-condensed type solutions in eq.~(\ref{eq:core_mass_t1e_crit}).
The contraction of envelope increases the envelope mass and the gravitational energy of gas component relative to the solid component.
Namely, $\kbe$ decreases and $\rho_{\rm 1e}$ increases due to compression, which reduces the critical core mass, as seen in eq.~(\ref{eq:core_mass_pc1e_crit}).
When $\qcore^{\crit}$ grows smaller than the core mass and the self-gravity of gaseous component grows important, on the contrary, the effect of compression overweighs the effect of energy loss so that a decrease of entropy results in an increase of temperature in the bottom of envelope, which further promotes the cooling and enhances the entropy decrease, leading to the onset of the runaway gas accretion.
This implies that the envelope changes the response to energy loss with the hydrostatic readjustment included, i.e., the gravo-thermal property, at the critical model.
While $M_{\rm env} \ll \qcore$ and the structure is governed by the gravity of core, the gravo-thermal heat capacity of envelope with the hydrostatic readjustment included is positive, and hence, the structure is thermally stable against the energy loss.
When the self-gravity of envelope grows important, the structure is thermally unstable against the energy loss.

Grain opacity has nothing directly to do with the thermal structure in the bottom of envelope since dust grains essentially all evaporate at high temperatures above $T > 1800$ K.
It may, however, affect the latter through an entropy decrease in the upper layers;
the smaller the grain opacity, the smaller the entropy results in the bottom of envelope.
\cite{Movshovitz_Bodenheimer_Podolak_Lissauer2010} consider grain opacity taking into account the size distribution and sedimentation, and shows that the runaway gas accretion is caused earlier than in models that assume constant grain opacity.
Such a tendency of decreasing core mass is also confirmed by \citet{Hori_Ikoma2010} who consider an alleged metal-free envelope.

\subsection{the Validity of Outer Boundary Conditions}

In this paper, we have adopted both the Bondi and Hill conditions for the purpose of comparison with the former works.
Now we may discuss the relevance of these outer boundary conditions on the basis of the present results.

The Bondi radius was introduced in relation to gas accretion onto a protoplanet on a dynamical timescale \citep{Cameron_Decampli_Bodenheimer1982}.
The Bondi radius is derived from the equality of the dynamical timescale of sound propagation and the free-fall timescale.
Accordingly, it may have physical meaning only when the accretion flow approaches sonic velocity, and intrinsically, has nothing to do with the structure in hydrostatic equilibrium.
In fact, we have demonstrated in this paper that the Bondi models extract the inner part of the centrally-condensed solution by the condition of $\vsurf = \gammad$, which otherwise extends outwards up to a large $\vsurf$.
If the Bondi radius is within the Hill radius, therefore, the protoplanet attracts surrounding gas so that the gaseous component may pile up beyond the Bondi radius in hydrostatic and thermal equilibria as long as the evolutionary timescale is longer than the dynamical timescale of the protoplanetary disk and the gas accretion rate remains slower than the Bondi accretion rate.
In actuality, we see for the critical models that there exists the layer of $V < 1$ inside the Bondi radius and $V$ is increasing outwards, rather than decreasing, at the Bondi radius, as seen in Fig.~\ref{fig:uv_single_polytrope_stiff}.

The Bondi accretion rate, $\dot{M}_{\rm Bondi}$, is given by
\begin{equation}
	\dot{M}_{\rm Bondi} = 4 \pi \rbondi^2 \rhodisk \sonic = (\qtot / t_{\rm free \ fall}) (\qtot /\qchar)/\gammad^{3/2},
	\label{eq:bondi_acc_rate}
\end{equation}
where $t_{\rm free \ fall}$ is the free fall timescale in surrounding disk $(= 1/ \sqrt{4 \pi G \rhodisk})$.
Corresponding to the model parameters in Table~1, we have $t_{\rm free \ fall} = 4.7 \ \hbox{ yr}$ and $\dot{M}_{\rm Bondi} = 1.2 \times 10^{-3} (\qtot /\mear)^2 \ \mear \hbox{ yr}^{-1}$.
The mass accumulation rate of the gaseous component remains smaller than of an order of the core accretion rate, $\qcdot$ until the envelope mass grows as large as $\qcore \ll \qchar$ (see \S~\ref{ss:small_envelope,result_of_polytrope}).
It is hence much smaller than the Bondi gas accretion rate since the latter rate is typically $\qcdot \simeq 10^{-6} \ \mear \hbox{ yr}^{-1}$.

In contrast, the Hill condition defines the contour which mechanically demarcates the protoplanet from the surrounding gas.
In our computations as well as in most of the former works, gravity is usually set to be finite at the Hill radius, neglecting tidal force by the host star.
For example, \cite{Lissauer_Dangelo_Bodenheimer2009} argues based on their 3D hydrodynamical simulation that only gas within $\sim 0.25 \rhill$ remains bound to the planet.
It is to be noted, however, that the inflow or outflow of gas from the Hill sphere has little thing to do with the stratification in hydrostatic equilibrium as long as the flow velocity remains below the local sound velocity.
In any case, that the present approximations yield only small differences in the mass of protoplanets as long as $\usurf \ll 1$ ( $\Delta \log \qtot \simeq \usurf \Delta \log R$).
Accordingly, the Hill boundary is relevant in discussing the critical core mass.
In considering the critical core masses of existent models with the Bondi surface radius, therefore, the critical mass should be smaller if we into account overlying layers within the Hill radius.
It should be noted that the upper layer are likely to be isothermal and/or optically thin.
If the latter layers are sufficiently thick, then, the Bondi models behave like the Hill models and it is possible that $\usurf \ll 1$.
In this case, the resultant critical core mass is not so much different if computed with the Hill outer boundary.

\subsection{Implication for Planet Formation}
\label{ss:discussion_discussion}

Present results may give an insight into the formation of gas giant planets in the protoplanetary disks.
As inferred from extrasolar planets, the physical conditions in the protoplanetary disks may vary from system to system;
the temperature and density of protoplanetary disks may have different dependences on the distance from the host star with the MMSN model.
Since we formulate the criteria of critical models by using non-dimensional forms with homology invariants, our results may be applicable to a wide range of parameters.

In \S \ref{sec:single_polytropic_model} and \S \ref{sec:realistic_model}, we treat the polytropic index as a free parameter.
However, it should be determined by the physical conditions of the interior.
One of the robust results is that the relationship $\voe = \nbe+1$ holds regardless of other conditions.
>From this, we can estimate the temperature at the inner edge of envelope as
\begin{eqnarray}
	&&\toe/\mu_{\be} = 3.0 \times 10^3 \left( \qcore/1\mear \right)^{2/3}\left( \rhocore / 5.5 \ \gcm \right)^{1/3}(\nbe+1 / 2.5)^{-1} \hbox{ K}.\label{eq:temp_bottom}
\end{eqnarray}
For $\qcore \gtrsim 1 \mear$, therefore, convection develops in the envelope because of the dissociation of hydrogen molecules and the concomitant increase in opacity and the adiabatic polytropic index may reaches $N_{\ad} > 3.0$ at the largest \citep[e.g.,][]{Mizuno_Nakazawa_Hayashi1978,Tajima_Nakagawa1997,Movshovitz_Bodenheimer_Podolak_Lissauer2010}.
For $\qcore \simeq 10 \mear$, then, $\toe \simeq 10^4$ K so that hydrogen molecules are likely to dissolve almost completely and the adiabatic polytropic index approaches to $N_{\ad} = 1.5$ and so does $\nbe$.

\begin{figure}
	\plotone{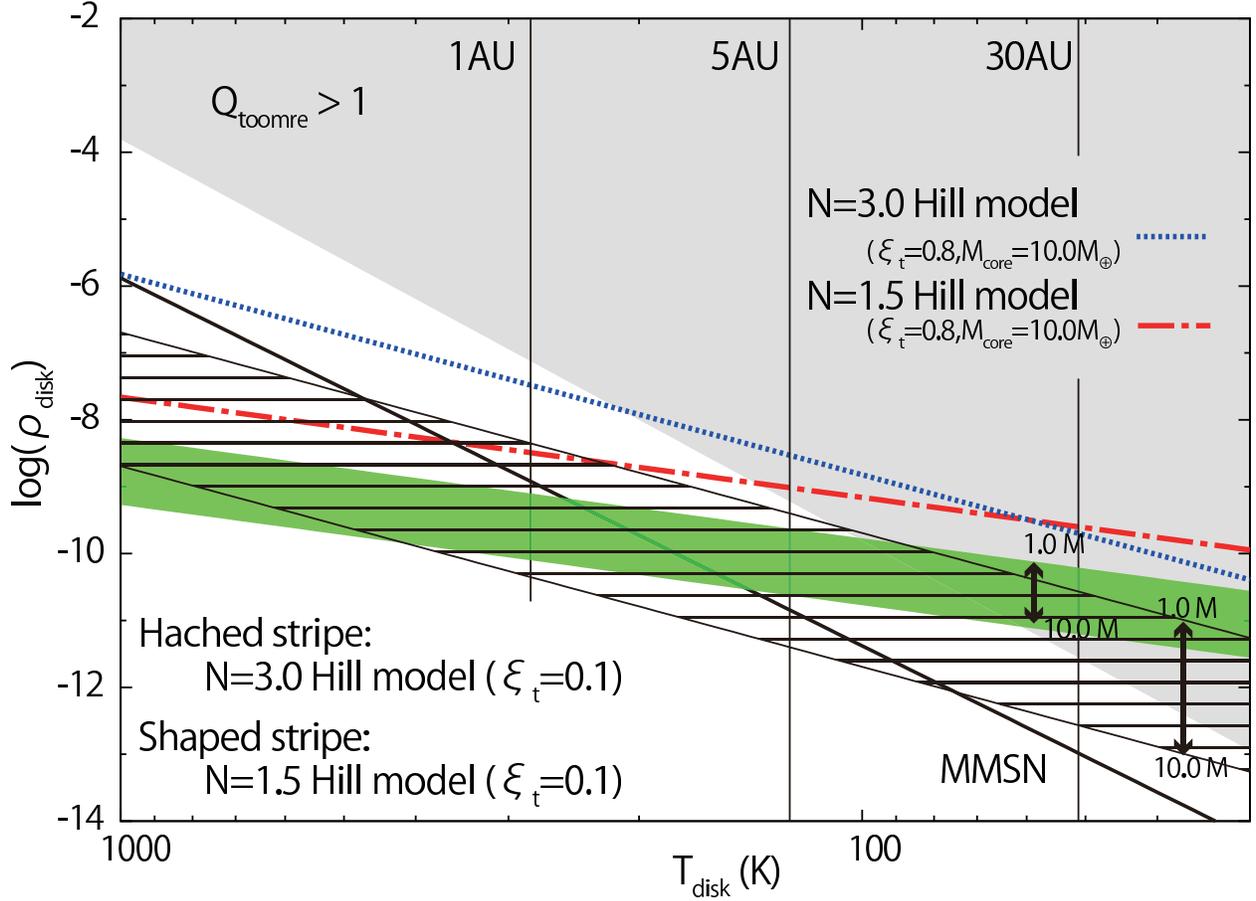}
	\caption{Critical core masses, $\qcore^{\crit}$, as a function of the density and temperature in the protoplanetary disk.
	Shaded and hatched stripes show the regions of $\qcore^{\crit} = 1 $-$ 10 \mear$ for the thickness of isothermal layer of $\xit = 0.1$ for Hill models with $\nbe = 1.5$ and 3, respectively.
	Also plotted by lines are estimates of $\qcore^{\crit} = 10 \mear$ for Hill models with $\xit = 0.8$ and $N = 1.5$ and 3.
	Thick solid line and gray shaded area denote the density and temperature of the MMSN model and the Toomre's instability parameter \citep{Toomre1964} $\qtoomre > 1$, respectively, for a host star of solar mass and luminosity;
	the separation from the host star is given on the top. }
	\label{fig:rhod_variation_vst}
\end{figure}

Furthermore, if we can specify the polytropic constant, $\kbe$, in the bottom, the critical core mass follows from eq.~(\ref{eq:core_mass_pc1e_crit}).
For a given physical condition in the protoplanetary disk, the physical thermal state in the bottom of envelope is determined through thermal processes in the envelope, which depends on the opacity and composition of gas and the core accretion luminosity.
This ``cooling'' effect of overlying layers may be expressed by the ratio, ${\kt}/{\kd}$, of the polytropic constant to its surface value in eq.~(\ref{eq:ccm_composite}), as discussed in \S~\ref{sec:isothermal_layer}.
Then the critical core mass is estimated as a function of its environment density and temperature.

We illustrate the critical core mass for models of fixed $\xit$ with ${\kt}/{\kd}$ taken from Table \ref{tab:isothermal_model} on the diagram of disk density and temperature of Figure~\ref{fig:rhod_variation_vst}.
Ranges of disk density and temperature that give the critical core mass between $\qcore^{\crit} = 1 $-$ 10 \mear$ are plotted for $\nbe=1.5$ and 3 both with $\xi = 0.1$.
Here we consider Hill models only, as discussed above.
Since $\qcore^{\crit} \propto (\tdisk / \rhodisk{}^{1/\nbe})^{3/2}$, as seen from eq.~(\ref{eq:ccm_composite}), the critical core mass is larger for smaller density and for higher temperature in the planetary disk;
the range of $\qcore^{\crit} = 1$ and $10 \mear$ is wider in $\rhodisk$ ($\propto \qcore ^{\crit}{}^{2N/3}$) and the slope is steeper ($\rhodisk \propto \tdisk{}^{\nbe}$) for larger polytropic index.
If the isothermal layer is thinner, the critical core mass is larger for a given density and temperature in disks, as seen from the comparison with the model of $\xit = 0.8$.

The temperature in disks is related to distance, $\sma$, from the host star if we assume radiative equilibrium with irradiation from the host star onto dust particles in disks, as shown on the top for $L_{\ast} = 1 L_{\odot}$.
For the density and temperature of MMSN model ($M_{\ast}=1\msun,L=1\lsun$) (thick solid line), our models of $\xit = 0.1$ give the critical core mass of $\qcore^{\crit} \simeq 10 \mear$ around $\sma \simeq 5 $ AU, in accordance with existent numerical models.
The MMSN has much steeper slope ($\rhodisk \propto \sma{}^{-11/4}$) than loci of constant critical core mass with a fixed $\xit$ both for $\nbe = 1.5$ and 3 ($\rhodisk \propto \propto \sma{}^{-\nbe/2}$).
This may suggest a smaller (larger) critical core mass for the inner (outer) part of disk, although we should take into account the variation of $\xit$ along the MMSN.

We may assume that the thickness of optically-thin and isothermal layer is correlated with the optical thickness of Hill sphere and the thickness of isothermal layer with the temperature gradient at the photosphere of protoplanets, respectively.
The optical thickness of Hill sphere is estimated by $\tauh = \kappa \rhodisk \rhill $, and the temperature gradient at the surface by $\vrad{}_{,\surf} \propto \kappa \lacc \rhodisk / \tdisk^3 $.
Along the loci of constant critical core mass with a fixed $\xit$, therefore, we have $\tauh \propto \kappa \sma ^{(2-\nbe)/2}$ and $\vrad{}_{,\surf} \propto \kappa \lacc \sma{}^{(3 - \nbe)/2}$.
The former changes the dependence on $\sma$ with the polytropic index as it varies between $\nbe = 1.5$ and 3.
For the latter, we should take into account the variation of luminosity, or core accretion rates, with the distance.
Along the MMSN, on the other hand, $\vrad{}_{,\surf} \propto \kappa \lacc \sma{}^{-5/4}$ and $\tauh \propto \kappa \sma{}^{-7/4}$, and hence, both $\vrad{}_{,\surf}$ and $\tauh$ decrease as a function of $\sma$.

In the inner disk, $\vrad{}_{,\surf}$ grows smaller for a constant opacity and luminosity (or core accretion rate) along the constant critical core mass with a fixed $\xit$, though the core accretion rate increases with the disk density.
On the other hand, $\tauh$ decreases for $\nbe = 1.5$ and increases for $\nbe = 3$.
As for the MMSN model, both are likely to increase with the density of disks, and hence, the isothermal and optically thin layers decreases to increase $\xit$.
Considering models of $\xit = 0.8$ and the smaller slope, it is possible to have smaller critical core mass for the inner disk of small $\sma$.

For the outer part of disk, the loci of constant critical core mass, shown in the figure, enter the region of disk instability, i.e., $\qtoomre > 1$, as shown by gray shade in the figure.
In order for the core accretion model to work, therefore, it demands thicker isothermal and/or optically thin layers than for the inner disk.
In the viewpoint of critical core masses, formation of gas giant planets is possible in the outer disk since $\tauh$ and $\vrad{}_{,\surf}$ decreases for further outward in the MMSN model, although the growth of core takes longer time because of lower density in the disk.

\subsection{Stability and the Trigger of Runaway Gas Accretion}

We have shown that protoplanets, embedded in a given thermal condition of a protoplanetary disk, constitute linear series with the total mass as the parameter.
It is known that such a linear series of equilibria may change the stability at the point where the series turns back through the parameter space already traversed.
In our case, this corresponds to the critical model which divides models of the core mass increasing and decreasing with the total mass.

For models on the sequence of increasing core mass, an addition (removal) of gaseous mass may demand larger (smaller) gravity of the core to accommodate it within the surface boundary under thermal equilibrium, represented by the polytrope.
If gas that follows the same polytrope is added with the fixed core mass, therefore, the envelope expands beyond its surface boundary to thrust out the added gas, while it reacts in an opposite way against a removal of gas mass, recovering the original configuration in equilibrium.
For models on the sequence of decreasing core mass, on the contrary, the gravity of core has to reduce (or enlarge) in order to sustain an envelope of larger (or smaller) mass in hydrostatic and thermal equilibrium with the same polytrope.
Without decreasing the core mass, therefore, an addition of gas mass causes the envelope to contract because of stronger gravity with increased total mass than in thermal equilibrium.
This promotes further contraction of the envelope and inflow of gas from the surface, which entails instability.

After the onset of runaway accretion, the gaseous envelope of gas planets is governed by its self-gravity.
After the onset of runaway gas accretion, its thermal structure, i.e., its polytropic index and $\xit$, etc., should be adjusted to keep hydrostatic equilibrium, while the thermal equilibrium breaks and the thermal imbalance leads to so-called the gravitational contraction.
for a red giant, it is stable because of strong temperature dependence of nuclear burning while the gas giant is unstable.
Gravitational contraction of the envelope proceeds with the runaway gas accretion of gas from its surface in protoplanetary disks so that its structure line comes to cross the characteristic lines at larger $U$ and smaller $V$.
Along with the increase of envelope mass, its structure line departs from the critical line below the horizontal line;
$\uoe$ grows larger and $\voe$ grows smaller owing to the increase of $\rhooe$ and rise of $\toe$.
At the same time, the degree of mass concentration shifts into its interior with the increase of $\uoe$.
As the envelop mass increases, the structure line comes to cross the critical line at very small $V$, and resembles the dwarf structure, such as described by the Emden-type solution, except for the core in the very center of mass much smaller than the envelop.
These processes take places by the Kelvin-Helmholtz timescale in the interior, and the latter timescale becomes shorter with increase of envelope mass, to be even smaller than gas supply timescale from the planetary disk.
If there remains gas in the surrounding planetary disk, therefore, the accretion flow eventually exceeds gas supply speed and the protoplanet becomes decoupled from the protoplanetary disk with the accretion shock on its surface.
In summary, the runaway gas accretion is the process of transition from the red-giant like configuration to the dwarf structure.

\section{Conclusions} \label{sec:Conclusion}
We have analyzed structural characteristics of protoplanets, embedded and accreting gas in protoplanetary disks, by applying the theory of stellar structure with characteristic variables, $U$ and $V$.
Our results are summarized as follows;

\begin{enumerate}
	\item
	The envelope structure of protoplanets is described in terms of the so-called centrally-condensed type solutions in common to envelope of red giants.
This is a requisite to maintain a tenuous, extended envelope around a high-density core.
	The occurrence of the maximum core mass arises from the properties of the two-component gravitating systems in hydrostatic and thermal equilibrium, which allows two modes of structures:
	While core gravity is dominated, the system can accommodate more envelope mass by compressing it with larger gravity.
	When self-gravity becomes important, the envelope can retain more mass only by shifting it outward to reduce its self-gravity, which also demands a decrease in core gravity.
The maximum core mass is realized during the transition from the envelope structure, governed by an external field of core, to that governed by a self-gravity of envelope.
	\item
	The centrally-condensed type solutions depend on the stiffness of the equation of state in the envelope, or the polytropic index, $N$.
In addition, the protoplanet envelope has distinct character according to whether the Bondi or Hill radius is assumed.
	Accordingly, the manifestation of critical core mass takes different forms for the following three groups;
(I) the stiff polytrope models of $N \le 3$ with the Bondi radius as the outer boundary, (II) the stiff polytrope models with the Hill radius as the outer boundary, and (III) the soft polytrope models of $N > 3$, regardless of outer boundary.
The models of three cases have different structure, and hence, there is three different conditions for the critical core mass:
for case (I), the criterion is imposed on the surface value of $\usurf$, or the ratio between the mean density of planet and the surface density: for case (II), as to the location of the inner edge of envelope on the horizontal line: and for case (III), on the value of $\uoe$ at the inner edge of envelope, or the ratio of density at the bottom of envelope and the mean density of core.
	\item
	Under the realistic conditions, there develops outer layer of small temperature gradient, such as can be approximated to isothermal layers.
	We also consider the effects on the envelope structures to show that the structural characteristics are kept basically unchanged, and the criteria are determined by the thermal property, or the polytropic index, $\nbe$, in the bottom of envelope.
The reduction of thermal energy through the isothermal layer envelope leads to small critical core mass in general.
The effect is largest for the models of the stiff polytropes of $\nbe \le 3$ with the Hill radius as boundary, and enables the critical core mass as small as an order of earth mass.
The models with the Bondi radius changes the structure characteristics for sufficiently thick isothermal layer, and behave like the Hill models to have as small critical core as theirs.
	\item
	Since our criteria are given by using the homology invariants, and hence, applicable to a wide range of parameters.
	One of the salient features is that the relationship of $\voe = \nbe+1$ hold to a good approximation, common to all three cases.
	The value of $\uoe$, on the other hand, varies with the polytropic index and the outer boundary condition.
	We derive the values of $\ucrit$ and $\vcrit$ in the critical models for some typical circumstances by using the polytropic equation of state, which provides a formula for the maximum core mass.
	Base on the results, we discuss comparisons with the former works and the parameter dependences of gas giant planets under various conditions.
In particular, we discuss the relevance of outer boundary condition, to point out the Bondi radius is irrelevant for the static structure.
	\item
	Once the core reaches the critical core masses, runaway gas accretion is initiated, which leads to the gravitational contraction of envelope.
	This proceeds in the Helmholtz - Kelvin timescale in the envelope, which grows shorter as the envelope mass increases to accelerate the mass inflow rate from the surface.
	From the structural viewpoint, the runaway gas accretion is the transition from the red-giant like structure to the dwarf-like structure, attendant with predominance of envelope mass over the core mass.

\end{enumerate}

\acknowledgments
We want to thank Yasunori Hori for his helpful discussions and comments; Masahiro Ikoma for the discussions and useful advaices.
K.D.K. is supported by Center for Planetary Science running under the auspices of the MEXT Global COE Program entitled "Foundation of International Center for Planetary Science".
This work has been partially supported by Grant-in-Aid for Scientific Research (18104003,23224004), from the Japan Society of the Promotion of Science.

\appendix
\section{structural characteristics of single polytrope models}
\label{sec:nature_of_single_polytrope}

In this section, we explore the behaviors of models on the diagram of homology invariants, $\log U$ and $\log V$.
   The single polytrope models are divided into the four groups that have distinct characteristics according to the stiffness of polytrope and to the outer boundary conditions.
   Accordingly, we discuss the models with the different outer boundary conditions and with the stiff and the soft polytrope, $N \le 3$ and $N>3$, separately, to elucidate the nature of these solutions in the details and reveal the conditions giving rise to the critical more masses.

\subsection{Bondi Models with Stiff Polytropes}
\label{sss:bondi_stiff,result_of_polytrop}

The envelope structures in the single polytrope models with the Bondi boundary condition are shown for the polytropes of $N=1.5$ and $N=3$ on the characteristic diagram of $\log U$ - $\log V$ in top-left and bottom-left panels of Figure~\ref{fig:uv_p1.5+3.0_bondi}, respectively.
   Their structure lines run straightforwardly between the inner edge of envelope, just below the horizontal line, and the outer boundary on the line of $\vsurf = \gammad$.
    The inner edge satisfies the relationship in eq,~(\ref{eq:env_inner_boundary}), which follow from the jump condition, and for given core mass, lies on a straight line in this $\log U $-$ \log V$ diagram.
   The structure line reaches it as it converges to $\voe = N+1$ along the horizontal line with decreasing radius, or vanishing $\uoe$, toward the center.
This is the well-known property of the centrally condensed-type solution for $N \le 3$ \citep[e.g.][]{Hayashi_Hoshi_Sugimoto1962}, and holds regardless of the core and envelope mass as long as $\rcore \ll \rtot$, and hence, for $\qcore \gg \qcoremin$.

The different asymptotic value of $\voe$ arises from the stiffness of equation of state, as stated in \ref{sec:small_envelope}.
The latter also takes part in determining the slope and extension of structure lines.
For $U \ll 1 - V/(N+1)$, the slope is approximated to
\begin{eqnarray}
	\dpar{\log V}{\log U} \simeq -\frac{1}{3} \left[ 1-\frac{(3-N)V}{3(N+1)-NV} \right],\label{eq:slope_struct}
\end{eqnarray}
and less steep for a smaller polytropic index.
Furthermore, for a smaller polytropic index, the structure line has to run over a longer path in $\log (V/U)$ along the horizontal line in order to satisfy the radius ratio condition in eq.~(\ref{eq:radius_condition}), because of a greater distance of the horizontal line from the critical line, as seen for $N = 1.5$ in this figure.
For $N \simeq 3$, on the other hand, the structure line need not run along the horizontal line because of the very close proximity to the critical line.
This explains the difference of values of $\uoe$ between models of $N = 1.5$ and $N = 3$.

As the core mass increases, the structure line shifts rightwards for larger $\uoe$, as seen from eq.~(\ref{eq:env_inner_boundary}), and reaches the outer boundary at a larger $\usurf$ since structure lines with the same polytropic index cannot intersect with each other.
Accordingly, the model with a larger core mass has a larger total (envelope plus core) mass [see eq.~(\ref{eq:surf_uv_bondi})].
When $\usurf$ attains at the intersection with the horizontal line, the structure line takes a local minimum of $V$ and touches tangentially the line of outer boundary.
For the models with a larger core mass, therefore, the structure line can no longer contact with the line to satisfy the outer boundary condition.
Consequently, this sequence of models terminates when $\usurf=\usw$, defined by eq.~(\ref{eq:switch_point}) and
   forms a sequence of models that the core mass increases with the total mass.

There follows another sequence of models that can satisfy the outer boundary condition if we allow the structure line of models to extend beyond the line of $V = \gammad$.
After entering below the horizontal line, the structure line continues to decrease the value of $V$, and after crossing the horizontal line, it turns to increase the value of $V$.
Eventually, the model reaches the outer boundary again at $\usurf$ larger than $\usw$ to have the total mass larger than the above sequence.
Accordingly, this new sequence of models are distinguished from models of the above sequence by the number of intersections with the outer boundary line, or the existence of the local minimum of $V$ on the structure line.
Models of the both sequences share a part of structure line of $V > \gammad$.
Since the inner edge of structure line ensues from the condition of the radius ratio in eq.~(\ref{eq:radius_condition}) for a given value of $\usurf$ or a given total mass, values of $\uoe$ are larger for models with the $V$ minimum than for their counterparts that share the part of structure lines because of larger contribution to the integral near the surface.
   For a given core mass, on the other hand, the loci of inner edge of envelope lie on the line, specified by the jump condition of eq.~(\ref{eq:env_inner_boundary}) in the characteristic plane, which has a slope of $\delta \log \voe / \delta \log \uoe $ in eq.~(\ref{eq:inneredge_critmodel}).
   For $N < 3$, this slope is steeper than the slope of structure line in eq.~(\ref{eq:slope_struct}), and hence, models of two intersections have a larger core masses than their counterparts of the single contact.
   For $\usurf$ that sufficiently surpasses $\usw$, however, the core mass starts to diminish with the total mass because contribution of the outer part of structure line beyond the intersection with the horizontal branch to the radius integral grows small, relative to an increase in surface radius ($\delta \log \rbondi = \delta \log \qtot = \delta \log \usurf /2$) since the structure line makes a larger excursion into smaller $V$ below the outer boundary line and departs more from the critical line;
   note that the minimum of $V$ grows small as $\delta \log V_{\rm min} \sim - (\log \usurf - \log \usw) \delta \log \usurf / 2$.
   Since the counterparts have smaller core mass for smaller $\usurf$, the core mass also turns to decrease with $\usurf$ for models of two intersections.
Consequently, models of two intersections realize the maximum core mass somewhere passes $\usw$, and then, form a sequence where the core of mass decreases as the total mass increases, in the opposite way to models of single intersection.

As the total mass further increases, the structure line begins with a still smaller $\uoe$ with a smaller core mass, and goes through the region of smaller $V$, eventually to traverse the critical line at a smaller value of $V < \gammad$.
As the structure line goes through very small value of $V \ll 1$, the structure line becomes indifferent to the polytropic index, in particular, in the outer part of larger $U$, responsible for the increment of mass, as seen from eqs.~(\ref{eq:struct_uv}).
As a corollary, the core mass vanishes at the similar total mass, regardless of the polytropic index.

The models of these two sequences combine to form a linear series with the total mass of the planet as a parameter.
Thee structures of these models embody the interior of the configuration of RGB stars, cut out in the middle of the loop structure.
The duplicity of the envelope configuration for a given core is related to the surface conditions, and the maximum core mass occurs after the sequence of models with one intersection terminates and sequence of models with the local minimum of $V$ begins.
However, the precise location, $\usurf^{\crit}$, of the surface of the critical model with the core of maximum mass depends on the polytrope.
For larger polytropic index, the difference of $\usurf^{\crit}$ from $\usw$ decreases since the difference in the slopes between the structure line and loci of constant core mass diminishes.
In particular, for $N = 3$, both slopes coincide so that the maximum core mass occurs for the model at the transition point.
There is the tendency of $\usurf^{\crit}$ decreasing for larger polytropic index;
in Table~\ref{tab:properties_single_model_BH}, $\usurf^{\crit}$ is larger than $\usw$ by as large as 36\% for $N = 1.5$ and both almost agree for $N = 3$.

\subsection{Hill Models with Stiff Polytropes }
\label{sss:hill_stiff,result_of_polytrop}

The behavior of the Hill models on the characteristic plane is illustrated in the right column of Figure~\ref{fig:uv_p1.5+3.0_hill} for the polytrope of $N = 1.5$ and 3.
Since the Hill outer boundary condition allows the value of $\vsurf > N+1$ (for $\qtot > \sqrt{(N+1) \rhodisk/\dmean} \qchar$), the centrally condensed-type solutions which run above the horizontal line are also applicable in addition to those which run below the latter.
These two solutions are both characterized by the power law distributions of density in the inner part surrounding the core, but different in the density distribution in the outer part.
The solution with the inner edge above the horizontal line keeps the value of $V$ increasing outwards monotonically, for which the power-law distribution is broken off while the gravity of the core dominates the structure of whole envelope.
The solution with the inner edge below the horizontal line, on the other hand, first decreases the value of $V$ to have a flatter density distribution, which may leads to predominance of the self-gravity of gas component over the gravity of core.
Then, it takes the minimum of $V$ when crossing the horizontal line, and starts to increase the value of $V$ to reach the surface.

These two types of solutions may give rise to another mode of the duplicity for the configuration of envelopes.
For models with a given core mass, the inner edges of their envelope lie on the loci, specified by the jump condition and on the straight line in the $\log U $-$ \log V$ diagram, as stated above.
In order for two models with the same core mass to exist in either side of horizontal line, the line of the loci of inner edge has to traverse the horizontal line.
They have the point of intersection only for $\uoe < 1/(N + 1)$ and the point of contact is located at $(U,V) = (1/N+1, N)$.
    This imposes an upper bound to the maximum core mass, i.e.,
\begin{eqnarray}
	&&\qcore^{\crit}<N^{3/2}(N+1)^{-3/2N}(\rhocore/3\rhodisk)^{(3-N)/2N}\qchar.\label{eq:qcore_hill_upper}
\end{eqnarray}
Numerically the upper bound is set at $473 \mear$ for $N = 3$, rather close to our result of $\qcore^{\crit} = 344 \mear$, while for $N = 1.5$, it results much larger ($2.4 \times 10^7 \mear$).   

For $\vsurf$ below the horizontal line, the model draws the structure line similar to the Bondi model discussed above, though it is truncated by the condition $\usurf = \rhodisk/\dmean$.
Even after $\vsurf$ moves into the upper side of horizontal line, similar behaviors persist until the surface reaches the critical line.
Once $\vsurf$ reaches above the critical line, the structure line makes a turn on the way, i.e., $U/V$ changes from increasing to decreasing outward while the value of $V$ is kept increasing all the way.
As seen for models of $N = 1.5$, the inner edge is on the upper side of the horizontal line and runs outwards along it before crossing the critical line.
This corresponds to the power-law part of density distribution under predominance of gravity of core over self-gravity of gas component.
   On this diagram, the power-law part is shorter for larger polytropic index, and it reduces essentially to a point on this diagram for $N=3$ because of the proximity of the horizontal line to the critical line.
   For a larger total mass, the structure line comes to cross the critical line at a larger value of $U$ because of larger $\vsurf$.
   Then, the inner edge moves into the region of larger $U$, and hence, for larger $\qcore$ along the horizontal line since the length of structure line along the horizontal line is shorten to satisfy the condition of the radius ratio in eq.~(\ref{eq:radius_condition});
   note that $d \log \rtot = (1/3) d \log \qtot$.
Accordingly, these models constitute the sequence that the core mass increases along with the total mass.

On the other hand, for the model with the inner edge on the lower side of the horizontal line, the structure line first decreases the value of $V$ on the way outward until it hits the minimum on the horizontal line.
The density distribution deviates from the power-low to a flatter distribution.
Then, it makes a turn at the crossing of the critical line and reaches the surface with a rapid decline in density as $V$ increases.
As the total mass increases, the structure line tends to cross the critical line at larger $U/V$ and the horizontal line at a smaller value of $V$, and hence, it draws a larger loop and departs farther from the critical line on the lower side of horizontal line.
   Accordingly, the structure line has to extend into a smaller value of $U_{1e}$ to have smaller core mass in order to satisfy the condition of the radius ratio.
These models with the local minimum of $V$ form a sequence of the core mass decreasing with the total mass, except for those with the inner edge close to the horizontal line.

Models on both sequences bring the inner edge closer to the horizontal line as the core mass increases, finally merging on the horizontal line to give a linear series with the total mass as a parameter.
The maximum core mass is reached for the model with the inner edge of envelope somewhat below, but close to, the horizontal line, as seen in Fig.~\ref{fig:uv_p1.5+3.0_hill}, when the detours around the $V$ minimum from the critical line grows large enough to restrain contribution to radial increment in the condition of the radius ratio below that due to mass increase.
The location of the inner edge of envelope, $\ucrit$ and $\vcrit$, is listed in Table~\ref{tab:properties_single_model_BH} for critical models of various polytropic indices.
The value of $\uoe$ decreases to be much smaller for the smaller polytropic index, by a factor of $\sim 140$ between $N = 1.5$ and 3, which reflects the distance of structure line from the critical line, or the slope of power-law distribution.

The critical model has most of the mass in the middle of interior, different from the Bondi model;
   the mass fraction, contained in the pressure scale height, $\vert d \log M / \log P \vert =  U/V$, takes a maximum at the crossing of the critical line in the middle of structure line.
   And the contribution of the integration of radius ratio comes largely from the internal structure along the horizontal line.
   Namely, the envelope structure is sensitive to the power-law part around the core, and hence, to the value of $\uoe$ and $\voe$ at the inner edge of envelope.
Once $\ucrit$ and $\vcrit$ are derived, the critical core mass directly follows from eq.~(\ref{eq:env_inner_boundary}).
Dependence on parameters other than the polytropic index manifests itself through the characteristic mass, $\qchar$, and the density ratio, $\rhocore / \rhodisk$.
The density at the inner edge of envelope in the critical model is given by $\rhooe^{\crit}/\rhocore = \ucrit/3$.
As typical of the centrally condensed-type solution of stiff polytropes, the density at the inner edge is much smaller than the core density;
   the ratio is strongly decreasing function of the polytropic index, ranging from $\rhooe / \rhocore = 0.03$ to $2.2 \times 10^{-4}$ between $N=3$ and $N=1.5$ and .

As the total mass increases, passing through the critical model, the core mass decreases and the structure in the envelope grows little dependent on the inner core.
   When the structure line comes to cross the critical line at $U/V \gg 1$, then, it approaches to the Emden solution for the given polytrope and the total mass is given by
\begin{equation}
\qtot (N) = \Phi (N) \qchar (\rho_{\rm center} / \rhodisk)^{(3-N)/2N},
\label{eq:mass_emden}
\end{equation}
   where $\Phi (N)$ is the non-dimensional radius and $\rho_{\rm center}$ is the density at the center.
   For $N=3$, the total mass is determined by the polytropic constant indifferent of the central density [$\qtot = 2.9 \times 10^3 \msun$ with $\Phi (3) = 16.14$].
   The central density is determined from the radius condition that
\begin{equation}
\rtot (N) = \Xi (N) R_0 (\rho_{\rm center} / \rhodisk)^{(1-N) /2N} = \rhill
\label{eq:radius_emden}
\end{equation}
   where $\Xi (N)$ is the non-dimensional radius and $R_0 $ is the characteristic radius as given by
\begin{equation}
R_0 = [(1/4 \pi G)(\pdisk / \rhodisk^2)]^{1/2},
\label{eq:char_rad}
\end{equation}
   and numerically by $R_0 = 166 R_\odot$ for out parameters.
   For $N <3$, the total mass that the core mass vanishes increases with the central density.
   For $N=1.5$, where $\Xi (1.5) = 5.777$ and $\Phi (1.5) = 10.73$, the central density is $\rho_{\rm center} / \rhodisk) = 4.7 \times 10^{4}$ and $\qtot (1.5) = 4.2 \times 10^5 \mear$.
   Such larger mass is solely due to the larger thermal energy, or the larger polytropic index, for the smaller polytropic index with a given thermal condition at the surface.

\subsection{Bondi Models with Soft Polytropes}
\label{sss:bondi_soft,result_of_polytrop}

For the polytrope of index $N>3$, the centrally-condensed type solution has the configurations that the inner edge converges to the singular point \citep{Chandrasekhar1939}.
Since the soft polytrope gives rise to a steep inward increase of density, $\uoe$ can be large, as seen from eq.~(\ref{eq:env_inner_boundary}).
In particular, the horizontal line runs in the upper domain of critical line on the left side of singular point, i.e., for $U < (N-3)/(N-1)$, so that the structure line is able to transverse characteristic lines in opposite directions to the stiff polytropes.
This enables the inner edge of envelope to lie in the upper domain of critical line, differently from models of $N \le 3$.


Left panel in Figure~\ref{fig:uv_p4.0_bondi+hill} illustrates the behaviors for the Bondi model of the polytrope with $N=4$.
The inner edge of envelope lies near the horizontal line, on the both side in contrast to Bondi models of $N \le 3$.
As the structure line runs in the upper domain of the critical line, $V/U$ first increases outwards, in contrast to models with the stiff polytrope.
Starting from the upper side, the structure line can enter below the horizontal line and keep decreasing $V$ to attain at the outer boundary line, $\vsurf = \gammad$.
On the way to the outer boundary, it crosses the critical line where $V/U$ turns to decrease.
After that, it crosses the vertical line and $U$ changes from decreasing to increasing.
For a larger total mass, the structure line shifts into the region of larger $U$ as a whole;  it starts from a larger $\usurf$, draws a loop, runs along the horizontal line and further crosses it to arrive at the inner edge.
In order to satisfy the condition of the radius ratio in eq.~(\ref{eq:radius_condition}), the inner edge also moves to larger values of $\uoe$.
These models constitute a sequence with a larger core mass for larger total mass.

As the inner edge approaches the singular point eq.~(\ref{eq:singular_point}) from above, the structure line starts to spiral down and steepens the gradient.
Eventually, when the inner edge arrives at right above the singular point, the slope of the structure line becomes steeper than the slope of loci of the inner edge of the envelope for a fixed core mass.
Beyond the core mass at this point, there is no solution because the structure line shifts downward for a larger total mass while the jump condition moves upward for a larger core mass.
Accordingly, this designates the value, $\ucrit$, of $\uoe$ for the critical model with the maximum core mass as $\ucrit = (N-3)/(N-1)$ in eq.~(\ref{eq:ucrit_soft}).
In comparison with our numerical models, however, we should take into account the increase in the core density with $\uoe$ due to the penetration of gaseous component (see footnote in the text).

As the total mass increases further beyond the critical model, the structure line has to extend into larger $\uoe$, beyond $\ucrit$, in order to satisfy the condition of the radius ratio eq.~(\ref{eq:radius_condition}).
The structure line near the inner edge runs below that of the model with a smaller total mass so that it can intersect only with loci of the inner edge of a smaller core mass, specified by the jump condition.
Accordingly, these models constitute the sequence with the core mass decreasing as the total mass increases.
This new sequence terminates when the surface reaches the horizontal line.
Here models change from the sequence of single intersection with the line of outer boundary condition to the sequence of two intersections, as seen for Bondi models of $N \le 3$.
This occurs at the same value of $\usurf = \usw$ in eqs.~(\ref{eq:switch_point}) and at the same total mass in (\ref{eq:pmass_critmodel_bondi}), but in this case, the core mass takes a local minimum, instead of a maximum for the latter model.
The opposite trend of core mass is due to the difference in the density distribution, as stated in Appendix \S \ref{sec:small_envelope}.
For the soft polytrope of $N>3$, the integration of mass (see, eqs.~(\ref{eq:struct_uv}) and (\ref{eq:radius_condition}) is dominated by the contribution from the inner edge rather than by that from the outer part, different from the case for the stiff polytrope of $N \le 3$.

Furthermore, beyond this point of contact, the sequence begins with models of two intersections, as in the case for Bondi models of $N \le 3$.
As in the case for the latter model, these models also reverse the relation between the core mass and total mass for above models of a single intersection, sharing a part of structure line above $V > \gamma$.
Along this sequence, the inner edge of the envelope traces back its trajectory, but increases the core mass with the total mass, up to the same value as with the first maximum of core mass again.
Accordingly, the value of $\uoe$ and the maximum core mass little differ from those of the former model with the first maximum.

After passing this second maximum, the model continues to make a larger loop for a larger total mass, starting from a smaller $\uoe$ with a smaller core mass, passing through the region of smaller $V$ and attaining at a larger $\usurf$, to have a larger total mass.
As $\uoe$ decreases and $\usurf$ increases, the mass tends to be dominated by the contribution from the outer part, and the total mass of the model tends to be indifferent to the polytropic index as the structure line passes through a region of small $V$.
Eventually, the core mass vanishes for almost the same total mass as in the Bondi model of $N \le 3$.

\subsection{Hill Models with Soft Polytropes }
\label{sss:hill_soft,result_of_polytrop}

Behaviors of Hill models for the polytrope of $N=4$ are illustrated in the right panel of Figure~\ref{fig:uv_p4.0_bondi+hill}.
For the Hill outer boundary condition, solutions that run all through the upper side of the horizontal line are applicable in addition to those which enter into the lower side of it.
This is common to models with Hill models for the polytrope of $N \le 3$.

Hill models share the same structure as Bondi models as long as the models run into the lower side of the horizontal line.
However, it is only those which cross and/or reach the vertical line at $U = \usurf{}_{,Hill}$ that can satisfy the Hill outer boundary condition.
These models include two types according to whether the structure line intersects the line of outer boundary once or twice.
In this case, both types are connected at the model with the outer boundary on the vertical line and have the same tend that the core mass increases with the total mass since its mass increment is determined by the inner edge.

Along with the increase in the core mass, the inner edge of its envelope shifts into the region of a larger $\uoe$.
When it is right above the singular point, the maximum core mass occurs, as in the case of Bondi models.
For the Hill boundary condition, the structure line with a larger total mass shifts upwards into the region of larger $V$ and forms a larger spiral around the singular point, differently from Bondi models.
But it has little effect since structure lines converge near the critical model and reduces the spacing $\delta \log V$ much smaller than $\delta \log \vsurf = (2/3) \delta \log \qtot \simeq (2/3) \delta \log \qcore$ at the surface.
For $N > 3$, therefore, the same criterion for the critical model is applied both to Bondi and Hill models.
The boundary condition may take a part in determining the value of $\vcrit$, or the distance to the critical line, which is again only a small effect since the mass is concentrated to the inner edge of the envelope.

Beyond the critical model, the core mass decreases with the total mass since the inner edge of the envelope goes around the singular point.
When the inner edge spirals in a semicircle and approaches right below the singular point, the variation of core mass with the total mass stagnates because the structure line moves to have the same gradient as the jump condition line for a fixed core mass.
This corresponds to the plateau in Fig.~\ref{fig:solutions_property_mass}.
After passing this point, the model with a larger total mass forms the structure line with a larger loop around the singular point, and the inner edge reaches into a smaller $\uoe$ and a smaller core mass.
As $\uoe$ and core mass decreases, the structure line crosses the horizontal lines at smaller of $V$, so that the total mass converges to a value independent of the polytropic index, as in the case of Bondi models of $N > 3$, but to a larger mass because of different outer boundary condition.


\end{document}